%% file: brick_I.tex
\documentclass[a4paper,fleqn,usenatbib,useAMS]{mnras}

\input{structure.tex}

\title[`The Brick' is not a brick]{`The Brick' is not a \emph{brick}: A comprehensive study of the structure and dynamics of the Central Molecular Zone cloud \brick }

\author[J. D. Henshaw et al.]{J. D. Henshaw,$^{1}$\thanks{Contact e-mail: jonathan.d.henshaw@gmail.com},  A. Ginsburg,$^{2}$ T. J. Haworth,$^{3}$ S. N. Longmore,$^{4}$ J. M. D. Kruijssen,$^{5}$
\newauthor E. A. C. Mills,$^{6}$ V. Sokolov,$^{7}$ D. L. Walker,$^{8,9}$ A. T. Barnes,$^{10}$ Y. Contreras,$^{11}$
\newauthor J. Bally,$^{12}$ C. Battersby,$^{13}$ H. Beuther,$^{1}$ N. Butterfield,$^{14}$ J. E. Dale,$^{15}$ T. Henning,$^{1}$
\newauthor J. M. Jackson,$^{16}$ J. Kauffmann,$^{17}$ T. Pillai,$^{18,19}$ S. Ragan,$^{20}$ M. Riener,$^{1}$ Q. Zhang$^{21}$
\\
$^{1}$ Max Planck Institute for Astronomy, K\"{o}nigstuhl 17, D-69117 Heidelberg, Germany\\
$^{2}$ National Radio Astronomy Observatory, 1003 Lopezville Rd., Socorro, NM 87801, USA\\
$^{3}$ Astrophysics Group, Imperial College London, Blackett Laboratory, Prince Consort Road, London SW7 2AZ, UK\\
$^{4}$ Astrophysics Research Institute, Liverpool John Moores University, Liverpool, L3 5RF, UK\\
$^{5}$ Astronomisches Rechen-Institut, Zentrum f\"{u}r Astronomie der Universit\"{a}t Heidelberg, M\"{o}nchhofstra{\ss}e 12-14, 69120 Heidelberg, Germany\\
$^{6}$ Physics Department, Brandeis University, 415 South Street, Waltham, MA 02453, USA\\
$^{7}$ Max Planck Institute for Extraterrestrial Physics, Gie{\ss}enbachstra{\ss}e 1, 85748, Garching bei M\"{u}nchen, Germany\\
$^{8}$ Joint ALMA Observatory, Alonso de Cordova 3107, Vitacura, Santiago, Chile\\
$^{9}$ National Astronomical Observatory of Japan, 2-21-1 Osawa, Mitaka,Tokyo, 181-8588, Japan\\
$^{10}$ Argelander Institute for Astronomy, University of Bonn, Auf dem H\"{u}gel 71, 53121 Bonn, Germany \\
$^{11}$ Leiden Observatory, Leiden University, P.O. Box 9513, NL-2300 RA Leiden, The Netherlands \\
$^{12}$ Center for Astrophysics and Space Astronomy, Astrophysical and Planetary Sciences Department, University of Colorado, Boulder, CO 80309, USA\\
$^{13}$ National Radio Astronomy Observatory, PO Box 2, Green Bank, WV 24944, USA\\
$^{14}$ Department of Physics, University of Connecticut, Storrs, CT 06269, USA\\
$^{15}$ Centre for Astrophysics Research, University of Hertfordshire, College Lane, Hatfield, AL10 9AB, UK\\
$^{16}$ School of Mathematical and Physical Sciences, University of Newcastle, University Drive, Callaghan NSW 2308, Australia \\
$^{17}$ Hay Stack Observatory, Massachusetts Institute of Technology, 99 Millstone Road, Westford, MA 01886, USA  \\
$^{18}$ Max-Planck-Institut f\"{u}r Radioastronomie, Auf dem H\"{u}gel 69, D53212 Bonn, Germany \\
$^{19}$ Institute for Astrophysical Research, Boston University, 725 Commonwealth Avenue, Boston, MA 02215, USA \\
$^{20}$ School of Physics and Astronomy, Cardiff University, Queen's Buildings, The Parade, Cardiff CF24 3AA, UK\\
$^{21}$ Harvard-Smithsonian Center for Astrophysics, 60 Garden Street, Cambridge, MA, 02138, USA
\vspace{-0.5cm}
}

\date{Accepted 2019 February 4. Received 2019 January 11; in original form 2018 November 29.}

\pubyear{2019}

\begin{document}
\label{firstpage}
\pagerange{\pageref{firstpage}--\pageref{lastpage}}
\maketitle

\begin{abstract}
\input{Tex/Abstract.tex}

\end{abstract}




\input{Tex/Introduction.tex}

\input{Tex/Data.tex}
\input{Tex/Results.tex}

\input{Tex/Discussion.tex}
\input{Tex/Conclusions.tex}
\input{Tex/Acknowledgements.tex}




\bibliographystyle{mnras}
\bibliography{References/references} 



\appendix
\input{Tex/scouse_decomp.tex}

\input{Tex/Methodology.tex}


\bsp	
\label{lastpage}
\end{document}

%% file: structure.tex
\usepackage{float}
\usepackage{graphicx}
\usepackage{times}
\usepackage{amstext}
\usepackage{amsmath}
\usepackage{amssymb}	
\usepackage{natbib}
\usepackage{url}
\usepackage{tabularx}
\usepackage{mathtools}
\usepackage{mathrsfs}
\usepackage{color}
\usepackage{graphics}
\usepackage{multirow}
\usepackage{ar}
\usepackage{multimedia}
\usepackage{media9}
\usepackage{lineno}
\usepackage{upgreek}

\newcommand{\kms}{\,km\,s$^{-1}$} 

\newcommand{\vel}{km\,s$^{-1}$\,pc$^{-1}$}
\newcommand{\scouse}{{\sc scousepy}}

\newcommand{\brick}{G0.253+0.016}

\def \micron{\hbox{$\upmu$m}}

\usepackage[T1]{fontenc}
\usepackage{ae,aecompl}

\usepackage{mathptmx}
\usepackage{txfonts}

%% file: Tex/Abstract.tex
In this paper we provide a comprehensive description of the internal dynamics of \brick \ (a.k.a. `the Brick'); one of the most massive and dense molecular clouds in the Galaxy to lack signatures of widespread star formation. As a potential host to a future generation of high-mass stars, understanding largely quiescent molecular clouds like \brick \ is of critical importance. In this paper, we reanalyse Atacama Large Millimeter Array cycle 0 HNCO $J=4(0,4)-3(0,3)$ data at $3$\,mm, using two new pieces of software which we make available to the community. First, \scouse, a Python implementation of the spectral line fitting algorithm {\sc scouse}. Secondly, {\sc acorns} (Agglomerative Clustering for ORganising Nested Structures), a hierarchical n-dimensional clustering algorithm designed for use with discrete spectroscopic data. Together, these tools provide an unbiased measurement of the line of sight velocity dispersion in this cloud, $\sigma_{v_{los}, {\rm 1D}}=4.4\pm2.1$\,\kms, which is somewhat larger than predicted by velocity dispersion-size relations for the Central Molecular Zone (CMZ). The dispersion of centroid velocities in the plane of the sky are comparable, yielding $\sigma_{v_{los}, {\rm 1D}}/\sigma_{v_{pos}, {\rm 1D}}\sim1.2\pm0.3$. This isotropy may indicate that the line-of-sight extent of the cloud is approximately equivalent to that in the plane of the sky. Combining our kinematic decomposition with radiative transfer modelling we conclude that \brick \ is not a single, coherent, and centrally-condensed molecular cloud; `the Brick' is not a \emph{brick}. Instead, \brick \ is a dynamically complex and hierarchically-structured molecular cloud whose morphology is consistent with the influence of the orbital dynamics and shear in the CMZ.

%% file: Tex/Introduction.tex
\section{Introduction}\label{Section:Introduction}

The lifecycles of molecular clouds and stars are inextricably linked. Molecular cloud evolution drives the formation of the stellar populations which light the Universe and, in turn, feedback from these stars drives the dispersal of the gas clouds from which they are born. It is a self-regulating process which helps to control the evolution of galaxies through cosmic time. 

Developing a complete understanding of molecular cloud evolution requires detailed studies which probe a vast range of physical conditions. While nearby molecular clouds (i.e. those within $\sim500$\,pc of Earth) have been studied in extensive detail over the past decades (see e.g. \citealp{andre_2014} and references therein), only now, with facilities such as the Atacama Large Millimeter Array (ALMA), are we able to target the more extreme ends of this parameter space over an equivalent spatial dynamic range. 

\subsection{Star formation in the Milky Way's Central Molecular Zone}

The Central Molecular Zone (hereafter, CMZ) of the Milky Way (i.e. the central $\sim500$\,pc) contains some of the Galaxy's densest and most massive molecular clouds and star clusters, offering an important window into molecular cloud evolution under extreme physical conditions. The interstellar medium (ISM) conditions found in the CMZ differ substantially from those found in the Galactic disc. Molecular gas densities \citep{guesten_1983b, bally_1987, longmore_2013, rathborne_2014a, mills_2018}, pressures \citep{oka_2001,rathborne_2014, walker_2018}, temperatures \citep{huettermeister_1993, ao_2013, ott_2014, mills_2013, ginsburg_2016, krieger_2017}, and velocity dispersions \citep{bally_1988, shetty_2012, henshaw_2016, kauffmann_2017a} of CMZ clouds, as well as the cosmic ray ionisation rate \citep{oka_2005,yusef-zadeh_2007} and the interstellar radiation field \citep{clark_2013}, can be factors-of-several to orders of magnitude greater than those found in solar-neighbourhood clouds when compared on the same spatial scale. Although the conditions found in the CMZ are therefore often considered to be extreme in the context of the Milky Way, \citet{kruijssen_2013} argue they are comparable to those found in high-redshift galaxies (e.g. \citealp{swinbank_2012}) at the time of peak cosmic star formation rate (around $z\sim2-3$; \citealp{madau_2014}). Consequently, understanding stellar mass assembly in the CMZ may help to provide a representative view of the conditions necessary for star formation at its cosmic peak. 

One currently open question regarding star formation in the CMZ is that despite harbouring a vast reservoir of dense ($\gtrsim~10^{3}$\,cm$^{-3}$) gas ($\sim$a few $10^{7}$\,M$_{\odot}$ or roughly $\sim$5\% of the total molecular gas content of the Milky Way, e.g. \citealp{dahmen_1998}), the estimated star formation rate (SFR) is just $\lesssim0.09$\,M$_{\odot}$\,yr$^{-1}$ \citep{longmore_2013, koepferl_2015, barnes_2017}. This SFR is approximately one order of magnitude below that expected from the observed linear relationship between the SFR and the gas mass above a surface density of $\Sigma_{\rm gas}=116$\,M$_{\odot}$\,yr$^{-1}$ (\citealp{lada_2010, lada_2012}), despite almost all of the molecular gas in the CMZ lying above this threshold (\citealp{longmore_2013,barnes_2017}). This low SFR cannot be explained by incomplete statistical sampling of independent star-forming regions \citep{kruijssen_2014}. Instead, the current underproduction of stars in the CMZ appears to be genuine. 

\input{Figures/Figure_1_CMZ_context.tex}

Numerous possible explanations for this discrepancy were discussed by \citet{kruijssen_2014b}. The authors hypothesised that the low SFR in the CMZ may be due to the high turbulent gas pressure, which would result in an elevated critical density threshold for star formation.\footnote{The SFR of a molecular cloud is determined in turbulent theories of star formation by computing the gas mass fraction above an effective critical density threshold, $\rho_{\rm crit}$. These theories assume that clouds are supersonically turbulent, and that star-forming cores arise as self-gravitating density fluctuations in the turbulent flow. In the models of \citet{krumholz_2005} and \citet{padoan_2011}, $\rho_{\rm crit}\propto \mathcal{M}_{\rm 3D}^{2}$, where $\mathcal{M}_{\rm 3D}$ is the turbulent Mach number, leading to an elevated critical density for star formation with increasing turbulent pressure. Although, as summarised by \citet{federrath_2012}, note that \citealp{hennebelle_2011} instead predict $\rho_{\rm crit}\propto \mathcal{M}_{\rm 3D}^{-2}$.} This led \citet{kruijssen_2014b} to suggest that star formation in the CMZ may be episodic, entering a starburst phase every 10-20 million years. In this episodic picture, turbulent gas flows towards the Milky Way's CMZ along the Galactic bar, providing the fuel for new generations of star formation (as demonstrated in simulations; e.g. \citealp{emsellem_2015, krumholz_2015, sormani_2018}). The key point is that this process takes \emph{time}: time to build up sufficient gas mass such that gravity can overcome the high turbulent pressure and star formation can proceed at a normal rate \citep{krumholz_2015, krumholz_2017}. Previous starburst activity is evident throughout the CMZ. A large population of 24\,\micron \ point sources at negative Galactic longitudes (e.g. \citealp{hinz_2009}) and the young massive clusters known as the Arches and Quintuplet (\citealp{figer_1999, longmore_2014}), may add support to the notion of episodicity. 

Of course, the CMZ is not in a period of complete dormancy. In fact, it hosts some remarkable star-forming complexes, namely Sgr~A, Sgr B1, Sgr~B2, and Sgr~C \citep{gusten_1983, goss_1985, mehringer_1992, mehringer_1993, yusef-zadeh_2009, kendrew_2013, ginsburg_2018}. Where star formation is underway, there is evidence to suggest that it is closely coupled to the orbital dynamics of the gas. \citet{longmore_2013b}, studying the subset of CMZ clouds known as the `dust ridge' \citep{lis_1994}, noted an increase in star formation activity as a function of increasing Galactic longitude along the dust ridge, and argued that these clouds may share a common formation timeline. \citet{longmore_2013b} further postulated that star formation may have been triggered by the tidal compression experienced by the clouds as they pass close ($\sim60$\,pc; \citealp{kruijssen_2015}) to the minimum of the global Galactic gravitational potential located at the position of the central supermassive black hole, Sgr A*. The link between the orbital dynamics of the gas and star formation in the dust ridge molecular clouds is supported by trends in observed star formation activity \citep{immer_2012, barnes_2017, walker_2018, ginsburg_2018} and, less directly, in increasing gas temperatures with increasing Galactic longitude \citep{ginsburg_2016, krieger_2017}. However, the notion of an evolutionary sequence has also been disputed (see e.g. \citealp{kauffmann_2017, simpson_2018}).

\citet{henshaw_2016c} extended the \citet{longmore_2013b} hypothesis following the discovery of several quiescent molecular clouds situated upstream from (but connected in position-position-velocity space to) the dust ridge clouds \citep{henshaw_2016}. Having possibly formed via gravitational instabilities, this portion of the CMZ possibly represents a physically continuous sequence of molecular clouds which we can follow from their formation and on-going assembly through to their subsequent collapse and emergent star formation in the dust ridge. 

Theoretically, this picture is supported by recent hydrodynamical simulations of molecular clouds orbiting the Galactic centre. These simulations demonstrate that many of the observed physical features of CMZ clouds are plausibly controlled by the background gravitational potential and their passage through the orbit's pericentre \citep{kruijssen_2019}. However, it is worth noting that the effect of the potential is dominant here, with the triggering of star formation due to pericentre passages expected to be rare (occurring in only $\sim10-30\%$ of accretion events into the inner CMZ;  \citealp{jeffreson_2018}). Although there are numerous models with differing perspectives on the three dimensional structure and orbital configuration of the CMZ (e.g. \citealp{sofue_1995, sawada_2004, molinari_2011, kruijssen_2015, ridley_2017}),\footnote{See \citealp{henshaw_2016} for a summary of how some of these geometries can either be ruled out or further constrained by observations.} as well as some disagreement on the physical mechanisms driving the flow of material along the Galactic bar and into the CMZ (e.g. \citealp{krumholz_2015, sormani_2018}), there is general agreement that Galactic dynamics play an important role in the regulation of star formation in this environment (e.g. \citealp{kruijssen_2014b, sormani_2019}). 

The aforementioned prominent features are displayed in Fig.~\ref{Figure:brick}, where we show a three-colour image of the CMZ gennerated from \emph{Spitzer} GLIMPSE wavebands. Here, blue is 3.6\,\micron, green is 5.8\,\micron, and red is 8.0\,\micron \ emission. The group of molecular clouds collectively known as the dust ridge are those stretching from \brick \ to Sgr B2.

\subsection{\brick: The prototypical Infrared Dark Cloud}

A key proving ground for understanding star formation in the CMZ is the molecular cloud \brick \ (also, GCM0.253+0.016, G0.216+0.016, M0.25+0.01, M0.25+0.11, or `The Brick'). \brick \ is the first cloud in the dust ridge sequence. With a mass of $\sim10^{5}$\,M$_{\odot}$ and a radius of just $\sim2-3$\,pc, \brick \ is one of the densest and most massive molecular clouds within the Galaxy \citep{lis_1994, longmore_2012, kauffmann_2013, rathborne_2015}. Paradoxically, however, \brick \ shows very few signatures of active star formation \citep{mills_2015} and appears mostly in absorption at 8\,\micron \ (see Fig.~\ref{Figure:brick}). The only direct (and published) evidence for star formation in the cloud comes from a H$_{2}$O maser identified by \citet{lis_1994}.\footnote{Note that there have been claims of ongoing star formation based on more indirect measures. \citet{lis_2001} estimate the internal luminosity of \brick \ to be of the order $\sim2.7\times10^{5}$\,L$_{\odot}$, which they claim is approximately equivalent to that of four B0 zero-age main-sequence stars. Moreover, the presence of emission from warm dust towards the edge of the cloud has been interpreted as being caused by heating from embedded protostars \citep{marsh_2016}. However, these indirect tracers of star formation activity are yet to be supported by independent lines of evidence.} This makes \brick \ one of the only $\gtrsim10^{5}$\,M$_{\odot}$ molecular clouds in the Galaxy, identified thus far, that does not display the signatures of advanced star formation \citep{ginsburg_2012, tackenberg_2012, urquhart_2014, longmore_2017}. The star formation potential of the cloud is therefore far from certain. Despite \brick \ having sufficient mass to form an arches-like cluster, it is not clear if we are observing a cloud on the verge of collapse \citep{longmore_2012, rathborne_2014a, rathborne_2014, rathborne_2015} or if instead the internal turbulent pressure and dynamic surrounding environment will hinder this evolution towards star formation \citep{kauffmann_2013, kauffmann_2017a}. 

Establishing the role of environment on the evolution of \brick \ is vital if we are to understand its fate. Recently, \citet{federrath_2016} performed an investigation into the physical and dynamical state of the cloud, speculating that shearing motions on large scales may be responsible for the dearth of star formation. The authors discuss this in the context of turbulent star formation theory. Simulations indicate that solenoidal motions (i.e. those with a high degree of vorticity) are capable of suppressing the SFR of a molecular cloud by approximately one order of magnitude in comparison to fully compressive modes \citep{federrath_2012}. Combining estimates of the turbulent velocity dispersion and the magnetic field strength, \citet{federrath_2016} conclude that turbulence within the cloud is dominated by solenoidal modes which is the result of the shear on large scales. Highlighting the potential importance of the orbital dynamics, \citet{kruijssen_2015} argue that \brick's recent pericentre passage may be the source of the shear. This argument was supported by recent hydrodynamical simulations of molecular clouds following the \citet{kruijssen_2015} orbit, which show that the observed velocity gradient across \brick \ (e.g. \citealp{rathborne_2015}) is consistent with shear-induced counter-rotation \citep{kruijssen_2019}.

In this Paper, we aim to perform a detailed investigation into the structure and kinematics of \brick, which have thus far often been analysed using moment analysis (\citealp{higuchi_2014, johnston_2014, rathborne_2015, federrath_2016}, although see \citealp{kauffmann_2013}). \citet{henshaw_2016} demonstrated that moment analysis' insensitivity to complex line-of-sight density and velocity structure can result in critical information being missed. We therefore revisit the analysis of the kinematics of \brick \ with the view to categorising and understanding its internal dynamics. In Section~\ref{Section:Data} we describe the data used throughout this paper. In Sections~\ref{results:global} and \ref{results:local} we present our results. In \ref{Section:discussion} we make detailed comparison to previous results in the literature. In \ref{discussion:acorns} summarise our new view of the structure of \brick \ before drawing our conclusions in Section~\ref{Section:Conclusions}.

%% file: Figures/Figure_1_CMZ_context.tex
\begin{figure*}
\begin{center}
\includegraphics[trim = 5mm 18mm 5mm 18mm, clip, width = 0.93\textwidth]{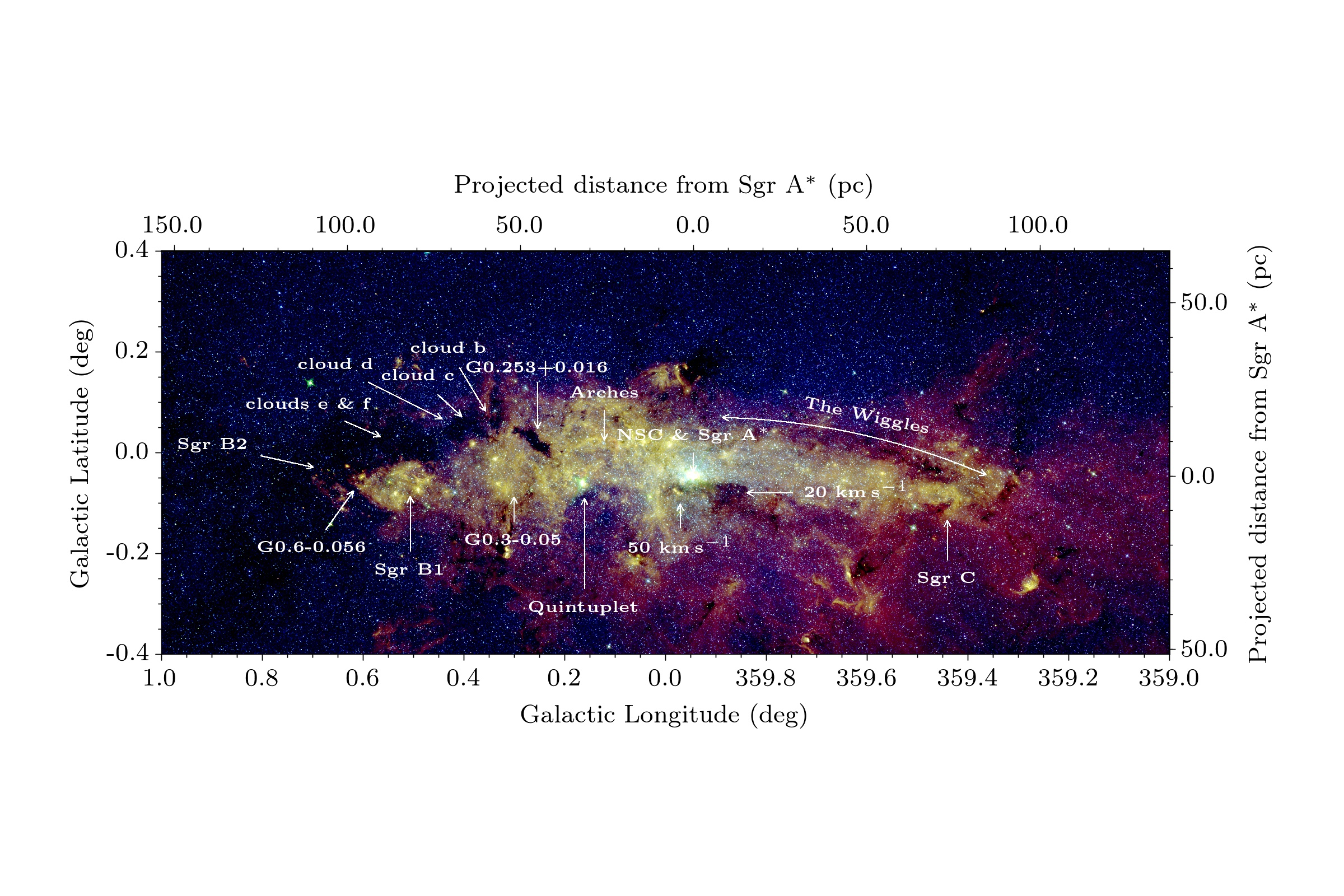}
\vspace{-0.3cm}

\end{center}
\caption{\brick \ in context. A three colour composite image of the CMZ, highlighting some of the most prominent features. All data are from \emph{Spitzer} GLIMPSE \citep{churchwell_2009}. Blue is 3.6\,\micron, green is 5.8\,\micron, and red is 8.0\,\micron \ emission. We highlight the dust ridge clouds (\brick, clouds~`b', `c', `d', `e \& f', and Sgr B2), additional well-known and studied molecular clouds (Sgr C, and the 20\,\kms \ and 50\,\kms \ clouds), star forming complexes (Sgr B1, G$0.6-0.056$, and G$0.3-0.05$), young massive clusters (the Arches and Quintuplet), the location of the velocity oscillations identified in \citet{henshaw_2016, henshaw_2016c}, and finally, the location of the nuclear star cluster and Sgr A$^{*}$. \brick \ can be clearly identified as a strong extinction feature against the bright mid-IR emission arising from the Galactic centre. }
\label{Figure:brick}
\end{figure*}

%% file: Tex/Data.tex
\section{Data}\label{Section:Data}

This paper makes use of the ALMA Early Science Cycle 0 Band 3 observations of G0.253+0.016 originally presented in \citet{rathborne_2014, rathborne_2015}. The ALMA 12m observations cover the full $3^{'}\,\times\,1^{'}$ extent of the cloud using a 13 point mosaic. The correlator was configured to use four spectral windows in dual-polarization mode centred at 87.2, 89.1, 99.1, and 101.1 GHz, each with 1875 MHz bandwidth and 488 kHz (1.4-1.7\kms) channel spacing. Because the data was Hanning smoothed by default by the ALMA correlator in Cycle 0, the spectral resolution of the data is 3.4\,\kms \ \citep{rathborne_2015}. The spatial resolution of the observations is $1.7^{"}$. This corresponds to a physical spatial resolution of $\sim0.07$\,pc assuming a distance to the Galactic centre of $8.34 \pm 0.16$\,kpc \citep{reid_2014}, which we adopt throughout this work, assuming that \brick \ is at an equivalent distance. 

The ALMA dataset provided data cubes for 17 different molecular species. \citet{rathborne_2015} studied each of these in detail, making a statistical comparison with the available continuum data (these data were combined with single-dish data provided by the \emph{Herschel} Space Observatory). Measuring the 2-D cross-correlation coefficients, the authors were able to look for similarities between the molecular species and the dust continuum (used here as a proxy for density). The strongest correlations were found between NH$_{2}$CHO, HNCO, CH$_{3}$CHO. Out of these species we select the HNCO 4(0,4)--3(0,3) transition (rest~freq.~$\approx~87.925$\,GHz) as our primary tracer of the kinematics since it is bright and extended. HNCO is often spatially extended towards galactic centres (e.g. \citealp{dahmen_1997, meier_2005, jones_2012}), and has proved fruitful for tracing the gas kinematics on both large ($\sim$ pc; \citealp{henshaw_2016}) and small ($\sim0.1$\,pc; \citealp{federrath_2016}) scales. The ALMA data were combined with single-dish data available from the Millimetre Astronomy Legacy Team 90 GHz Survey (MALT90; \citealp{foster_2011, jackson_2013}) obtained with the Mopra 22m telescope. For further information regarding the data reduction and image processing we refer the reader to \citet{rathborne_2015}.

%% file: Tex/Results.tex
\section{A global look at the kinematics of \brick}\label{results:global}

\subsection{SCOUSEPY decomposition of the ALMA HNCO data}\label{results:scouse}

Our kinematic decomposition of the ALMA HNCO data is performed using a newly-developed Python implementation of the Semi-Automated multi-COmponent Universal Spectral-line fitting Engine ({\sc scouse}), first presented in \citet{henshaw_2016}.\footnote{{\sc scousepy} is publicly available for download here: \url{https://github.com/jdhenshaw/scousepy}. Alternatively, the original IDL implementation can be downloaded here: \url{https://github.com/jdhenshaw/scouse}.} \scouse \ is a semi-automated routine used to fit large quantities of complex spectroscopic data in an efficient and systematic way. The procedure followed by \scouse \ is discussed in detail by \citet{henshaw_2016}, but we highlight the key points here.

Briefly, the \scouse \ fitting procedure can be broken down into several stages. \scouse \ first identifies the spatial region over which it will perform the fitting. This can be tailored by the user to target localised regions (in both position and velocity), or to target data above a specified noise threshold. The philosophy behind this step is to minimise workload. For example, although the \brick \ HNCO data contains $>3\times10^{5}$ pixels, we masked all spectra whose peak flux is below 0.03\,mJy\,beam$^{-1}$. The unmasked region is (approximately) comparable to that studied by \citet{federrath_2016}, who employed a H$_{2}$ column density threshold for their study of 5$\times10^{22}$\,cm$^{-2}$. 

\scouse \ then breaks up the map into small areas, referred to as Spectral Averaging Areas (SAAs), and extracts a spatially-averaged spectrum from each. In the new Python implementation, the user has the option to refine the size of the spectral averaging area depending on the local complexity of the line profiles. To gauge the complexity of a spectrum a very simplistic metric is used. We compute the difference in velocity between the intensity-weighed average velocity (i.e. moment 1; $v_{1}$) to the velocity of the channel containing the peak emission in the spectrum ($v_{peak}$). The idea is that for a simple, singly-peaked, symmetric line profile the difference between these two quantities $\Delta v_{\rm m}\equiv|v_{1}-v_{peak}|\sim 0$. Alternatively, $\Delta v_{\rm m}$ will be $>0$ for a highly asymmetric line profile. This is demonstrated in Fig.~\ref{Figure:cov} located in Appendix~\ref{SCOUSE}. The map is then divided up into different sized SAAs, where the smallest areas contain spectra with a high degree of complexity. 

The refinement of the SAA size leads to higher quality fits overall, particularly for large and complex datasets, because of the greater accuracy of the input guesses supplied to the automated fitting procedure. Moreover, having many overlapping SAAs (of potentially different sizes) provides a variety of models to any given pixel, enabling \scouse \ to make an informed choice about which is the best-fitting solution. 

The spatially averaged spectra extracted from each SAA are then manually fitted by the user. Fitting is performed interactively using {\sc pyspeckit},\footnote{{\sc pyspeckit} can be downloaded here: \url{https://github.com/pyspeckit/pyspeckit}.} whose extensible framework facilitates the modelling of a variety of line profiles (including Gaussian, Voigt, and Lorentzian profiles, as well as hyperfine structure fitting). Specifically, for the ALMA HNCO data, we assume that the spectra can be decomposed into individual Gaussians. This assumption is reasonable given the lack of line wings in the spectral profiles as well as the likelihood that the HNCO emission is optically thin (we quantify this statement further in \S~\ref{discussion:thick}).

\input{Figures/Figure_2_PPV_scousepy.tex}

Best-fitting solutions to the SAAs are then supplied to the fully-automated fitting procedure that targets all of the individual spectra contained within each region. This process is controlled by a number of tolerance levels. For a full description of the tolerances see \citet{henshaw_2016}. In summary, we fixed the following tolerance criteria during our search: (i) all detected components must have a flux density which is greater than three times the local noise value ($T_{1}=3.0$; \citealp{henshaw_2016}); (ii) each Gaussian component must have a full-width-at-half-maximum (FWHM) line-width of at least one channel ($T_{2}=1.0$);\footnote{It should be noted that this leads to the detection of unresolved velocity components. Often these components are necessary for a good fit to the remaining spectral components, and so we choose to fit them. However, as we will discuss later, these components are removed for the clustering analysis (see \S~\ref{results:local}). } (iii) for two Gaussian components to be considered distinguishable, they must be separated by at least half of the FWHM of the narrowest of the two ($T_{5}=0.5$). The remaining two tolerance levels ($T_{3}$ and $T_{4}$) restrict the degree to which the parameters describing the velocity components can deviate from their closest matching counterparts in the SAA spectrum. We set both of these tolerance levels to 3.0. As in \citet{henshaw_2016} the final best-fitting solution for each pixel is that which has the smallest value of the (corrected) Akaike Information Criterion (AICc; \citealp{akaike_1974}).

The statistical information regarding the \scouse \ fitting of the \brick \ can be found in Table~\ref{Table:global_stats} which can be found in Appendix~\ref{SCOUSE}. To summarise, a total of 2355 SAAs were manually fitted. This resulted in best-fitting solutions to 133065 out of a total 315219 pixels (note the total here includes those pixels that were masked during stage 1 of the fitting process), and a total of 457264 velocity components. Multiple component fits are required to describe the spectral line profiles over a significant ($\sim96\%$) portion of the map. These large values indicate the complexity of the velocity structure.

\subsection{Centroid velocities: Ubiquitous velocity oscillations, cloud substructure, and velocity gradients}\label{results:vlsr}

The result of the fitting procedure is displayed in Fig.~\ref{Figure:ppv}. This image is a 3-D PPV diagram highlighting the distribution of HNCO gas throughout \brick. Each data point represents the $\{l,\,b,v\}$ coordinates of an individual Gaussian component extracted by \scouse. The colour (light to dark) of each data point encodes the peak flux density of each spectral component. 

The velocity structure of the cloud is clearly complex. The most striking features of Fig.~\ref{Figure:ppv} are the vertical velocity oscillations appearing in the gas distribution appearing across a range of spatial scales. These oscillatory gradients are reminiscent of those first identified on larger scales in \citet{henshaw_2016,henshaw_2016c}, and suggest that such gradients are a common feature of the interstellar medium in the CMZ. However, unlike those analysed in detail by \citet{henshaw_2016c}, which display a characteristic amplitude ($\sim3.7\,\pm\,0.1$\,\kms) and wavelength ($\sim22.5\,\pm\,0.1$\,pc), the \brick \ oscillations appear to be more stochastic. This will be explored further in a future publication (Henshaw et al., in preparation). 

\input{Figures/Figure_3_v_scousepy.tex}

Further, one notices two large scale, dominant features that appear to merge (caution: in PPV-space) towards the southern portion of the cloud. The first appears at a velocity of $\sim35-50$\,\kms. The second shows a distinct velocity gradient increasing in velocity from $\sim0$\,\kms \ in the north and appears to merge in PPV-space\footnote{We stress that this does not necessarily indicate a merger of structure in physical space.} with the first feature at a velocity of $\sim30$\,\kms \ towards the south of the cloud. Many studies have described the prominent velocity gradient observed across \brick \ (e.g. \citealp{higuchi_2014, johnston_2014, rathborne_2015}). Most recently, it has been cited as evidence for the rotation induced by the orbital dynamics of the CMZ \citep{federrath_2016}, which was argued from a theoretical perspective by \citet{kruijssen_2015}, and further quantified using hydrodynamical simulations \citep{kruijssen_2019}. In this picture, as a cloud makes its closest approach to the bottom of the Galactic gravitational potential well, the side of the cloud closest to the central potential accelerates with respect to the far-side, inducing shear, and causing the cloud to counter-rotate with respect to its orbital motion. 

We can estimate the velocity gradient across \brick \ using the intensity-weighted velocity field provided by the first order moment
\begin{equation}
v_{\rm m1}=\frac{\sum_{n=i}^{N}S_{\nu}(v_{i})v_{i}}{\sum_{n=i}^{N} S_{\nu}(v_{i})}
\label{Eq:m1}
\end{equation}
where $S_{\nu}(v_{i})$ is the flux density at a velocity channel $v_{i}$ and $N$ is the number of channels. Following \citet{federrath_2016}, we compute this over a velocity range of $0-45$\,\kms \ and clip all data below 3$\sigma_{\rm rms}$. The velocity gradient is estimated as a fit to all $\{l,b,v\}$ data points assuming that the velocity field is well approximated by a first-degree bivariate polynomial (e.g. \citealp{goodman_1993, henshaw_2016})
\begin{equation}
v = v_{0}+\mathscr{G}_{v_{l}}\Delta l+\mathscr{G}_{v_{b}}\Delta b.
\label{Equation:gradient}
\end{equation}
Here, $v_{0}$ is the systemic velocity of the mapped region, $\Delta l$ and $\Delta b$ are the offset Galactic longitude and latitude values (expressed in radians), and $\mathscr{G}_{v_{l}}$ and $\mathscr{G}_{v_{b}}$ are free-parameters in the least squares fit and refer to the magnitudes of the velocity gradients in the $l$ and $b$ directions, respectively (in \kms\,rad$^{-1}$). The magnitude of the velocity gradient ($\mathscr{G}$), and its direction ($\Theta_{\mathscr{G}}$), are then estimated using:
\begin{align}
\mathscr{G} \equiv |\mathscr{G}_{v_{l,b}}| = \frac{(\mathscr{G}_{v_{l}}^{2}+\mathscr{G}_{v_{b}}^{2})^{1/2}}{D},
\label{Equation:grad_calc}
\end{align}
and
\begin{equation}
\Theta_{\mathscr{G}} \equiv {\rm tan}^{-1}\bigg(\frac{\mathscr{G}_{v_{l}}}{\mathscr{G}_{v_{b}}}\bigg),
\label{Equation:dir_calc}
\end{equation}
whereby $D$ is the distance to the cloud in pc (see \S~\ref{Section:Data}). For the velocity gradient, $\mathscr{G}_{v_{\rm m1}}$, we find $4.0$\,km\,s$^{-1}$\,pc$^{-1}$ ($\mathscr{G}_{v_{\rm m1}}=9.7$\,km\,s$^{-1}$arcmin$^{-1}$). 

The computed velocity gradient is consistent with that reported by \citet{federrath_2016}, $\mathscr{G}_{v_{\rm m1}}=3.9$\,km\,s$^{-1}$pc$^{-1}$ ($\mathscr{G}_{v_{\rm m1}}=9.5$\,km\,s$^{-1}$arcmin$^{-1}$), where the slight difference is most likely due to the slight difference in the intensity-threshold used.\footnote{Here we have simply used an intensity threshold cut in HNCO whereas \citealp{federrath_2016} make a cut based on the continuum-derived column density (see \S~\ref{results:scouse}). Differences between our results derived from moment analysis and those of \citet{federrath_2016} will therefore propagate throughout any comparisons made in this work. However, we note that the differences are inconsequentially small.} Moreover, this is similar to, albeit slightly larger than the value derived from single-dish MALT90 data $\mathscr{G}_{v_{\rm m1}}=3.1$\,km\,s$^{-1}$pc$^{-1}$ ($\mathscr{G}_{v_{\rm m1}}=7.43$\,km\,s$^{-1}$arcmin$^{-1}$; \citealp{rathborne_2014a}).\footnote{This value actually differs from that reported by \citet{rathborne_2014a}, which has been corrected due to a conversion error (see also \citealp{kruijssen_2019}).} Despite this general agreement with other observational work, each of these derived-gradients is considerably smaller than the 20\,km\,s$^{-1}$pc$^{-1}$ value quoted by \citet{higuchi_2014}, who compute the gradient using the full range of velocities which are spatially coincident with \brick \ (see below and Fig.~\ref{Figure:vlsr}). However, as discussed in \citet{henshaw_2016}, the southern portion of \brick \ spatially overlaps with portions of the CMZ gas stream at velocities of $\sim70$\,\kms. This gas, according to our best understanding of the 3-D geometry of the gas distribution in the CMZ, is physically unassociated with the cloud. Finally, the velocity gradient derived from the intensity-weighted velocity field is also similar to, but larger than that extracted from simulations of molecular clouds following the \citet{kruijssen_2015} orbit, $\mathscr{G}_{v}=2.4$\,km\,s$^{-1}$pc$^{-1}$ ($\mathscr{G}_{v}=5.9$\,km\,s$^{-1}$arcmin$^{-1}$; \citealp{kruijssen_2019}), where this gradient is driven by shear. 

In the same way as described above, we can also compute the velocity gradient using the information available from our \scouse \ decomposition. Here, we find $\mathscr{G}=4.3$\,km\,s$^{-1}$\,pc$^{-1}$ ($10.5$\,km\,s$^{-1}$arcmin$^{-1}$) in a direction $\Theta_{\mathscr{G}}=-151.7\degr$ east of north. This value is consistent with, albeit larger than, the other observational derivations (see above). This discrepancy is likely a result of the fact that we are utilising all of our \scouse \ measured velocities and ignoring the complex structure presented in Fig.~\ref{Figure:ppv}. We will revisit this topic in \S~\ref{results:local}.

To emphasise the difference between moment analysis and spectral decomposition, we plot in the left-hand panel Fig.~\ref{Figure:vlsr} a histogram of the \scouse \ data in blue and the first order moment in green. The \scouse \ velocity data covers the range $-35.0\,\,{\rm km\,s^{-1}}<v<87.7\,{\rm km\,s^{-1}}$ (note that these extremes may themselves not be associated with \brick) and has a mean of $\langle v\rangle\sim24.96\,\pm\,0.03$\,\kms \  (median $=25.78$\,\kms; standard deviation $=16.8$\,\kms), where the uncertainty here refers to the standard error of the mean. In the right-hand panel we plot the centroid velocity of all the identified velocity components as a function of their peak flux density (we over plot the point density as contours). Both panels of Fig.~\ref{Figure:vlsr} illustrate that the \scouse \ data can be split into 4 (possibly 5) main features. In the histogram there are peaks at $\sim3$\,\kms, $\sim16$\,\kms, $\sim31$\,\kms, and $\sim40$\,\kms \ (and a smaller peak at $\sim80$\,\kms), each of which is clearly evident in the $S_{\nu}-v_{\rm LSR}$ plane in the right-hand panel. Some of the multiplicity observed in both panels of Fig.~\ref{Figure:vlsr} may be a result of the velocity gradients observed across the dominant features seen in Fig.~\ref{Figure:ppv} (in the same way that the double-peaked feature in the moment 1 histogram seen in green in the left panel has been interpreted as a signature of rotation; \citealp{federrath_2016}). 

The above analysis demonstrates that although intensity-weighted average quantities may encode important information about the bulk gas dynamics throughout \brick, Figs.~\ref{Figure:ppv} and \ref{Figure:vlsr} clearly show that these quantities miss significant detail in the structure and kinematics of the cloud. Therefore, while the kinematics may be interpreted as displaying the hallmarks of rotation, our \scouse \ decomposition indicates that a single-component model (i.e. a singular, coherent and rotating cloud), may be too simplistic in describing the complexity of \brick's dynamics, and that complex line-of-sight structure is present (we will discuss our interpretation of the cloud structure further in \S~\ref{Section:discussion}). 

\subsection{Velocity dispersions and estimated (line-of-sight) turbulent Mach numbers}\label{results:disp}

\input{Figures/Figure_4_disp_scousepy.tex}

In the left hand panel of Fig.~\ref{Figure:disp} we show the distribution of 1-D line-of-sight velocity dispersions, $\sigma_{v_{los}, {\rm 1D}}$, measured with \scouse. Dispersions range between $0.8\,{\rm km\,s^{-1}}<\sigma_{v_{los}, {\rm 1D}}<23.1\,{\rm km\,s^{-1}}$ (the 25th and 75th percentile are $2.9$\,\kms \ and $5.6$\,\kms, respectively), with a mean value $\langle\sigma_{v_{los}, {\rm 1D}}\rangle=4.4\,{\rm km\,s^{-1}}$ (median $=4.0$\,\kms) and a standard deviation of $2.1\,{\rm km\,s^{-1}}$. The standard error of the mean is of the order $\sim10^{-3}$\,\kms.\footnote{Note that the velocity components at the lower end of the distribution are unresolved (the spectral resolution is $\Delta v_{\rm res}=3.4$\,\kms). We allowed these components in the \scouse \ decomposition to improve the overall quality of the fit. This affects approximately $\sim2.5\%$ of the data. However, they are removed from the analysis in \S~\ref{results:local}. } The distribution is skewed towards higher values (with a skewness of $\sim1$). The skew can also be seen in the right hand panel, where we show the velocity dispersion as a function of the peak flux density. In the left hand panel of Fig.~\ref{Figure:disp} we also show the velocity dispersion as derived using moment analysis, for comparison. The second order moment is given by 
\begin{equation}
\sigma_{v_{\rm m2}}=\bigg[\frac{{\sum_{n=i}^{N}S_{\nu}(v_{i})(v_{i}-v_{\rm m1})^{2}}}{\sum_{n=i}^{N} S_{\nu}(v_{i})}\bigg]^{1/2}
\label{Eq:m2}
\end{equation}
where $S_{\nu}(v_{i})$ is the flux density at a velocity $v_{i}$ and $v_{\rm m1}$ refers to the first order moment (Equation~\ref{Eq:m1}). The second order moment values are distributed about a mean value of $\sigma_{v_{\rm m2}, {\rm 1D}}=11.1\,{\rm km\,s^{-1}}$ and have standard deviation of $4.3\,{\rm km\,s^{-1}}$. This latter value is consistent with that derived by \citet{rathborne_2014a} [see their Figure 11]. The mean velocity dispersion is more than a factor of 2 greater than that extracted using \scouse \, which is due to the presence of multiple velocity components in the data. 

Our \scouse-measured mean velocity dispersion differs significantly from the value reported by \citet{federrath_2016}. However, \citet{federrath_2016} perform a fundamentally different measurement. These authors instead use the standard deviation of centroid velocities. This represents a measurement of the dispersion of line-of-sight velocities across the plane of the sky (which we label $\sigma_{v_{\rm m1}, {\rm 1D}}$;  as their value is derived from moment analysis) rather than along the line of sight, as is measured (directly) with \scouse. Consequently, their measured value of $\sigma_{v_{\rm m1}, {\rm 1D}}$ is a factor of $\sim2$ larger than our \scouse-derived mean dispersion, $\sigma_{v_{los}, {\rm 1D}}$. Repeating their analysis using the first order moment we find $\sigma_{v_{\rm m1}, {\rm 1D}}=9.1\,{\rm km\,s^{-1}}$, which is close to the value quoted in \citet{federrath_2016}, $\sigma_{v_{\rm m1}, {\rm 1D}}=8.8\,{\rm km\,s^{-1}}$. By comparison, if we take the standard deviation of all centroid velocity measurements made by \scouse \ we find $16.8\,{\rm km\,s^{-1}}$, clearly indicating the dominance of multiple velocity components.

In an attempt to isolate the turbulent velocity dispersion (i.e. motions which are exclusively associated with turbulence), \citet{federrath_2016} subtracted the observed large-scale velocity gradient from the intensity-weighted average velocity field. This yields a value of $\sigma_{v_{\rm m1}, gs, {\rm 1D}}=3.9\,{\rm km\,s^{-1}}$ (where the subscript `$gs$' stands for `gradient-subtracted'). As discussed in \S~\ref{results:vlsr}, given the complex distribution of centroid velocities that is evident in Fig.~\ref{Figure:ppv}, it is unclear whether the velocity gradient observed in the intensity-weighted velocity field can be exclusively attributed to the ordered motion of the cloud. Velocity gradients derived from an intensity-weighted average velocity field may be exaggerated by independent clouds or sub-clouds situated along the line-of-sight, each of which has its own independent velocity gradient. Moreover, it is also unclear, on a pixel-by-pixel level, to what extent the intensity-weighted average velocity field (and by extension $\sigma_{v_{\rm m1}, gs, {\rm 1D}}$) is influenced by the presence of multiple velocity components in the HNCO data. That is to say that different regions within the cloud do not have the same number of components (as can be inferred from Fig.~\ref{Figure:cov}) and so the first order moment will be affected differently as a function of position. Therefore the subtraction of a singular velocity gradient from an intensity-weighted velocity field should be approached with caution.

We convert our velocity dispersions, $\sigma_{v_{los}, {\rm 1D}}$, measured on the scale of the synthesised beam ($0.07$\,pc; \S~\ref{Section:Data}), into an estimate of the turbulent Mach number, $\mathcal{M}_{\sigma_{v_{los}}, {\rm 3D}}$ using \citep{henshaw_2016}
\begin{equation}
\mathcal{M}_{\sigma_{v_{los}}, {\rm 3D}}\approx \sqrt{3}\frac{\sigma_{v_{\rm turb}, {\rm 1D}}}{c_{s}}=\sqrt{3}\bigg[\bigg(\frac{\sigma_{v_{los}, {\rm 1D}}}{c_{s}}\bigg)^{2}-\bigg(\frac{\overline\mu_{\rm p}}{\mu_{\rm obs}}\bigg) \bigg]^{1/2}
\label{Eq:mach}
\end{equation}
where $\sigma_{v_{\rm turb}, {\rm 1D}}$, in the centre of this equation refers to the 1-D turbulent velocity dispersion measured along the line-of-sight, which we estimate by subtracting the contribution of thermal motions from the observed 1-D line-of-sight velocity dispersion in quadrature. The isothermal sound speed is given as $c_{\rm s}=(k_{\rm B}T_{\rm kin}/{\overline\mu}_{\rm p} m_{\rm H})^{0.5}$, for a gas with kinetic temperature, $T_{\rm kin}$, and mean molecular mass, $\overline{\mu}_{\rm p}~=~2.33$\,amu ($k_{\rm B}$ and $m_{\rm H}$ are the Boltzmann constant and the mass of atomic hydrogen, respectively), and $\mu_{\rm obs}$ is the molecular mass of the observed molecule (43\,amu in the case of HNCO). Assuming a fixed temperature of 60\,K \citep{ginsburg_2016, krieger_2017}, $c_{s}=0.46$\,\kms. Plugging these values into Equation~\ref{Eq:mach}, we derive a mean Mach number of $\langle\mathcal{M}_{\sigma_{v_{los}}, {\rm 3D}}\rangle~=~16.45\,\pm~\,0.01$ (the 25th and 75th percentile are $10.7$ and $21.0$, respectively), where the uncertainty here reflects the standard error of the mean. This should be taken as an upper bound on the level of turbulent motion, since our assumptions do not take into account the contribution from coherent motions or substructure within the ALMA beam, we assume a uniform temperature, and because of the relatively coarse spectral resolution of the observations $\Delta v_{\rm res}=3.4$\,\kms. These factors could combine to result in us overestimating the velocity dispersion and therefore the Mach number throughout the cloud. 

This analysis, and the subsequent reduction in measured velocity dispersions and estimated Mach numbers in comparison to other techniques, adds to mounting evidence for the identification of narrow lines in CMZ clouds \citep{kauffmann_2013, kauffmann_2017a}.\footnote{Although the ALMA cycle 0 dataset used here has insufficient spectral resolution to confirm the identification of the more extreme cases ($<1$\,\kms) of narrow velocity dispersions presented by \citet{kauffmann_2017a}.} This in itself should not come as a surprise, given the increasing spatial resolution of the aforementioned observations. However, despite this, our \scouse-derived velocity dispersions are broader than those predicted by the observationally-derived, steep velocity dispersion-size relationships of the CMZ. Using $\sigma=(\sigma_{0}/{\rm km\,s^{-1}})(r/{\rm pc})^{\zeta}$, where $\sigma_{0}$ is the absolute scaling of the velocity dispersion and $\zeta$ is the slope, we can predict the magnitude of the velocity dispersions measured on $0.07$\,pc scales (representative of the ALMA synthesised beam), from the relationships derived by \citet{shetty_2012} and \citet{kauffmann_2017a}. Using $\{\sigma_{0}, \zeta \}=\{2.8\,{\rm km\,s}^{-1}, 0.64\}$ \citep{shetty_2012} and $\{5.5\,{\rm km\,s}^{-1}, 0.66\}$ \citep{kauffmann_2017a}, velocity dispersions of the order $\sim0.5$\,\kms \ and $\sim1.0$\,\kms, respectively, are predicted. These are factors of $\sim9$ and $\sim4$ narrower than those measured from our \scouse \ decomposition, respectively. 

The fact that our mean measured velocity dispersion of $\langle\sigma_{v_{los}, {\rm 1D}}\rangle=4.4\,{\rm km\,s^{-1}}$ is fully resolved by ALMA ($2[2{\rm ln}(2)]^{1/2}\langle\sigma_{v_{los}, {\rm 1D}}\rangle/\Delta v_{\rm res}>3$, where $\Delta v_{\rm res}$ is the spectral resolution), could indicate that, in contrast to the derived relationships of \citet{shetty_2012} and \citet{kauffmann_2017a}, velocity dispersions $\lesssim1$\,\kms \ are not dominant on (projected) $\sim0.07$\,pc scales throughout \brick. This could imply a shallower velocity dispersion-size relationship. However, this comparison comes with the caveat that although our dispersion measurements are taken on projected scales of the ALMA synthesised beam ($\sim0.07$\,pc), we do not know the extent of the cloud along the line-of-sight. Although this is also true of both the \citet{shetty_2012} and \citet{kauffmann_2017a} studies, the discrepancy between our measured, and the predicted, velocity dispersions could instead indicate that the depth of the cloud is much greater than the projected spatial extent over which the measurements are taken.

Quantifying both the absolute scaling of non-thermal motions measured at a given spatial scale as well as how the magnitude of non-thermal motions varies as a function of spatial scale throughout the CMZ is of critical importance to understanding star formation in this environment (see \S~\ref{Section:Introduction}). A steep velocity dispersion-size relationship in the CMZ, if confirmed, may have profound implications for how molecular clouds in this environment begin to build their stellar mass.\footnote{The shape of the stellar Initial Mass Function, or more specifically, the turnover in the IMF may be closely tied to the sonic length, which is the scale below which thermal or magnetic support dominates over turbulence (see e.g. \citealp{offner_2014}, and references therein).} Therefore, higher spatial and spectral resolution observations, those which are capable of resolving the sound speed in the molecular gas ($\sim0.46$\,\kms \ for 60\,K gas), are first required to confirm if the turnover in the \scouse \ histogram in the left hand panel of Fig.~\ref{Figure:disp} is real, and secondly, to fully characterise the gas motions on small spatial scales throughout \brick. 

\section{A detailed study of \brick's kinematic substructure}\label{results:local}

\subsection{ACORNS decomposition of the ALMA HNCO data}\label{results:acorns_decomposition}

To date, analyses of the gas kinematics of G0.253+0.016 have predominantly relied on techniques such as moment analysis \citep{rathborne_2015, federrath_2016}, and dendrograms \citep{kauffmann_2013}. The former technique is beneficial as it is simple and fast to implement. It returns information on the pixel scale and is an intuitive way of taking a `first look' at spectroscopic data. However, as is clearly demonstrated in \S~\ref{results:global}, detail is easily lost when using moment analysis. Conversely, the latter technique is beneficial in that complex line-of-sight structure is accounted for as the algorithm seeks to build a hierarchy of structure, which can be represented graphically in the form of a dendrogram (see e.g. \citealp{rosolowski_2008}). However, kinematic information is provided in the form of intensity-weighted average quantities relating to each structure. Further work is therefore required if one is interested in how those kinematic quantities vary with position within a given structure on the pixel scale. There was previously no publicly-available code whose primary function is to extract hierarchical information from spectroscopic data, but which simultaneously retains the pixel scale information needed to study variation in the kinematics throughout each member of the hierarchy. 

Our solution to this problem is the development of a new analysis tool, written in Python, named {\sc acorns} (Agglomerative Clustering for ORganising Nested Structures).\footnote{{\sc acorns} is publicly available for download here: \url{https://github.com/jdhenshaw/acorns}.} {\sc acorns} is based on a technique known as hierarchical agglomerative clustering, whose primary function is to generate a hierarchical system of clusters within discrete data. Although {\sc acorns} was designed with the analysis of discrete spectroscopic position-position-velocity (PPV) data in mind (rather than uniformly spaced data cubes), clustering can be performed in n-dimensions, and the algorithm can be readily applied using information in addition to PPV measurements. For a full description of the {\sc acorns} algorithm see Appendix~\ref{Methodology}. 

In the following sections we use {\sc acorns} to further characterise the velocity structure of the cloud. We perform the {\sc acorns} decomposition only on the most robust spectral velocity components extracted by \scouse. We define `robust' as all velocity components whose peak flux density is greater than $\sim5\times$ the typical measured rms noise value\footnote{This is performed on a pixel-by-pixel basis. The mean rms value is $\langle\sigma_{\rm rms}\rangle=0.8$\,mJy\,beam$^{-1}$.} and whose velocity dispersion is greater than $\sim1.4$\,\kms \ (this corresponds to a FWHM of $\sim3.4$\,\kms, which is a single resolution element). The selected data constitute $\sim92\%$ of the total dataset extracted by \scouse \ (420398 kinematic measurements).

\input{Figures/Figure_5_dendrogram_acorns.tex}
\input{Figures/Figure_6_PPV_acorns.tex}
\input{Figures/Figure_7_int_acorns.tex}

For the clustering, we set the minimum radius of a cluster\footnote{Note that here and throughout this paper the term `cluster' is used in the statistical sense to refer to an agglomeration of data points.} to be $1.2^{''}$, which is $\sim10\%$ larger than the semi-major axis of the ALMA synthesised beam. This is to ensure that all identified clusters are spatially resolved. In addition to spatial information we also include velocity information in the clustering. For two data points to be classified as `linked' we specify that the euclidean distance between the points and the absolute difference in both their measured centroid velocity and velocity dispersion can be no greater than $1.2^{''}$ and 3.4\,\kms, respectively. In summary, these constraints are selected because they reflect our observational limitations. 

During the initial phase of the clustering a total of 1152 clusters were identified, representing $\sim97\%$ of the subsample selected above.\footnote{Note that using a linking length of 1.7\,\kms \ for both the centroid velocity and velocity dispersion (i.e. a single channel), respectively, changes the results only slightly. In this case, the total number of clusters identified is 1231 and these clusters contain $\sim95\%$ of the data. This doesn't however, affect any of the conclusions of this work.} Having fixed these parameters for the initial development of the hierarchy, we then relaxed all linking lengths (position, velocity, velocity dispersion) by 50\% to further develop the clusters. Our final dataset contains 1182 clusters, accounting for $\sim98\%$ of all data. 

As with any hierarchical system of clusters, the result can be displayed graphically as a dendrogram (see e.g. \citealp{rosolowski_2008}). In Fig.~\ref{Figure:dendrogram} we display the resultant {\sc acorns} hierarchy for \brick. To avoid confusion in star formation nomenclature, we drop the statistical terminology of `cluster' and instead expand on the nomenclature typically used in describing dendrograms (see e.g. \citealp{houlahan_1992}). We refer to the hierarchical system presented in Fig.~\ref{Figure:dendrogram} as the forest, which itself contains numerous trees. Each tree may then be further subdivided into branches or leaves in a hierarchical fashion (trees with no hierarchical substructure are also classed as leaves). In the case of \brick, the forest consists of a total of 195 trees. The forest is dominated by 4 trees; \#3, \#22, \#85, and \#98 (highlighted in red, blue, green, and yellow, respectively). These 4 trees contain over 50\% of all data. In Fig.~\ref{Figure:ppv_acorns} we display these trees in PPV space (as in Fig.~\ref{Figure:ppv}). As can be clearly seen in this figure, these trees are associated with the dominant features which are evident in Fig.~\ref{Figure:ppv} and discussed in \S~\ref{results:vlsr}. Given the enormity of the dataset, we focus on these dominant trees for the remainder of our analysis. For simplicity, we henceforth refer to the trees as A (red), B (blue), C (green), and D (yellow).

\subsection{Peak intensity distributions}\label{results:int_acorns}

\subsubsection{Tree features: Localised peaks, arcs, and shocks}\label{results:trees}

In Fig.~\ref{Figure:intfield} we display the spatial distribution of peak flux density for each of the main trees to give an impression of their physical structure. While the trees appear to follow the overall distribution and curvature of the cloud, which is commonly observed on large scales in dust continuum maps (see e.g. \citealp{johnston_2014, rathborne_2015}), our analysis has also revealed a lot of small scale structure in the gas distribution. 

Trees B and C, appearing in blue and green in Fig.~\ref{Figure:ppv_acorns} and in the top right and bottom left of Fig.~\ref{Figure:intfield}, respectively, are the most prominent of the identified trees. Together they dominate the physical appearance of \brick, accounting for $\sim34\%$ of the data (which is roughly distributed evenly between them). A cursory visual comparison of the two trees in Fig.~\ref{Figure:intfield} suggests that the HNCO emission is brighter throughout tree C, on average. This can be inferred from Figure\,\ref{Figure:dendrogram}, where C has a greater number of leaves that have greater peak flux density than those associated with B. 

Qualitatively, the small scale peaks of emission (identified as leaves by {\sc acorns}) in tree C show a similar spatial distribution to those observed in the corresponding 3\,mm dust continuum image presented by \citet{rathborne_2015} and displayed in Fig.~\ref{Figure:continuum}. There is, however, a notable exception. The green circle in Fig.~\ref{Figure:intfield} denotes the location of the H$_{2}$O maser identified by \citet{lis_1994}. This coincides with a `hole' in the emission associated with tree C and we will discuss this further in \S~\ref{results:sf}.

Another prominent feature evident in Fig.~\ref{Figure:intfield} is the `C'-shaped arc structure associated with tree B (top-right panel of Fig.~\ref{Figure:intfield}). The arc was originally discovered with ALMA as a prominent feature traced by sulphur monoxide (specifically the SO~[$v=0,\,3(2)-2(1)]$ transition).\footnote{These data were taken as part of the same ALMA cycle 0 dataset as that used in this paper: ADS/JAO.ALMA\#2011.0.00217.S. The spatial ($\sim$1.9\,arcseconds) and spectral ($\sim3.4$\,\kms) resolutions are therefore approximately equivalent to the HNCO data presented here.} \citet{higuchi_2014} characterise the arc as being associated with a number of emission peaks (both in the dust continuum and SO), some of which show broad velocity dispersions (of the order 30-40\,\kms) as well as strong velocity gradients. Despite these relatively extreme values, the right hand panel of \citet{higuchi_2014}'s Fig.~2 (which displays the second order moment map), shows that most of the emission associated with the arc has velocity dispersions up to $\sim10$\,\kms. \citet{mills_2015} later confirmed that the arc is observed in other molecular species and transitions, identifying it clearly in the (peak) emission maps of NH$_{3}$ transitions from (1,1) up to (7,7). Although the presence of the `C'-shaped arc was therefore noted in previous studies, {\sc acorns} provides the first evidence that the arc is coherent in both (projected) space and velocity. 

Tree D (bottom-right panel of Fig.~\ref{Figure:intfield}) resides at the interface of trees B and C in terms of velocity (see Figure\,\ref{Figure:ppv_acorns} and \S~\ref{results:vlsr_acorns}). This tree is associated with a linear feature referred to as the `tilted bar' by \citet{mills_2015}, contains the bulk of the brightest clumps seen in NH$_{3}$ (3, 3) and a multitude of `class {\sc i}', collisionally excited, and shock tracing CH$_{3}$OH masers and maser candidates. The `tilted bar' is also evident in \citet{johnston_2014}'s Fig.~14 which displays the integrated flux line ratio of different H$_{2}$CO transitions. Radiative transfer analysis suggests that this region shows elevated gas temperatures (\citealp{johnston_2014}), consistent with \citet{mills_2015}. Moreover, this region is observed to exhibit enhanced emission from shocked and warm ($>140$\,K) gas tracers (e.g. SiO (5-4) and H$_{2}$CO; Kauffmann et al. in preparation). These features are complemented by the HNCO emission, which is very bright throughout the tree and follows a linear feature running perpendicular to the major axis of \brick. This region of the cloud has previously been cited as a potential location for cloud-cloud collisions \citep{johnston_2014} and the linear feature that is observed may be the result of large-scale shocks \citep{mills_2015}. We will return to this discussion in \S~\ref{Section:discussion}.

Finally, tree A overall has fewer regions of bright emission than the others despite showing a lot of substructure. This is evident in Figure\,\ref{Figure:dendrogram}, where the tree is seen to exhibit a complex dendrogram. In larger-scale single-dish observations of \brick \ emission at the low velocities associated with A extends further north of the cloud in the direction of the dust ridge cloud `b' (see e.g. \citealp{lis_2001}), whose mean velocity is measured to be $\sim3.4$\,\kms \ \citep{henshaw_2016}. This extension is also evident in dust continuum observations (see e.g. \citealp{immer_2012}). 

\subsubsection{Star formation within \brick}\label{results:sf}

\input{Figures/Figure_8_brick_core_acorns.tex}

In the previous section we noted that there is a lack of emission in tree C at the only (currently) confirmed location of ongoing star formation in \brick.\footnote{Note that recently, two additional H$_{2}$O masers have recently been discovered. One further to the north of the cloud at 70\,\kms \ and another $\sim10^{''}$ to the south of the maser identified by \citet{lis_1994} at a velocity of 28.4\,\kms \ \citep{lu_2019}. } To investigate whether or not there is a true absence of emission at this location, we first of all inspected the best-fitting solutions extracted using \scouse. A cursory inspection indicates that there are several velocity components at this location. We then further explored the {\sc acorns} hierarchy for any trees that spatially overlap with the H$_{2}$O maser and are located at the `appropriate' velocity (\citealp{lis_1994} quote velocities of 32.1\,\kms \ and 41.6\,\kms \ for the maser). Using these criteria we identified two trees (\#32 and \#108). We then identified all leaves which spatially coincide with the gap in emission associated with tree C. 

We identify a centrally-concentrated leaf associated with the first of these two trees that fits this criteria. It has a mean centroid velocity of $\sim$42.0\,\kms \ and a velocity dispersion of $\sim2.8$\,\kms. Given the spectral resolution of 3.4\,\kms \ this is in satisfactory agreement with the velocity of the H$_{2}$O maser identified by \citet{lis_1994}. In Fig.~\ref{Figure:continuum} we plot the 3\,mm continuum map first presented by \citet{rathborne_2014}. Overlaid on this image we display the contoured outline of the tree (blue). In the inset image we zoom in on the 3\,mm dust continuum peak (red contours and background) which is associated with the maser emission identified by \citet{lis_1994}. Comparing the ALMA dust continuum with our {\sc acorns} decomposition, we find that the {\sc acorns} leaf (cyan contours) does not trace the main dust continuum peak, but instead traces an extension of this peak observed towards the south. 

To further investigate this, we compare our results with new high resolution ($\sim0.13\,\arcsec$) ALMA band 6 observations towards the maser region (Walker et al. in preparation). Using a combination of dust continuum observations and CH$_{3}$CN emission we find that there is evidence for line emission associated with the 3\,mm dust continuum peak at $\sim42-43$\,\kms, consistent with the velocity of tree\,\#32. The reason for the lack of a line emission peak in our 3\,mm HNCO data is currently unclear and further investigation at high-angular resolution and with molecular line tracers that probe different critical densities and excitation conditions are necessary. Nevertheless, there is evidence for a small compact continuum source which coincides with the extension in emission seen in the 3\,mm data presented in Fig.~\ref{Figure:continuum}  (D. L. Walker, private communication), and therefore our {\sc acorns} leaf. 

\subsection{Gas kinematics}

\subsubsection{Centroid velocities: non-Gaussian Velocity PDFs and velocity gradients}\label{results:vlsr_acorns}

In Fig.~\ref{Figure:velpdf} we plot velocity probability density functions (PDFs) of the {\sc acorns} trees. In laboratory experiments of incompressible turbulence the PDF of the velocity field is often very nearly Gaussian (see e.g. \citealp{anselmet_1984}). This has also been demonstrated in numerical simulations of turbulence \citep{lis_1996,klessen_2000, federrath_2013}. Results from observations of the interstellar medium however, have been mixed and largely show some deviation from pure Gaussian behaviour (e.g. \citealp{miesch_1999,ossenkopf_2002, federrath_2016}). To assess this we fit a normal distribution to the centroid velocities measurements associated with each tree and also compute the higher order moments (skewness and kurtosis) of the distributions. The first four central moments of a dataset (in our case $v$) with $N$ elements are:
\begin{align}
{\rm mean} = \langle v \rangle = \frac{1}{N} \sum_{n=i}^{N} v_{i} \\
{\rm dispersion} = \sigma_{v_{pos}, {\rm 1D}} = \sqrt{\langle[v_{i} - \langle v \rangle]^{2}\rangle} \label{Eq:mom2}\\
{\rm skewness} = \mathcal{S} = \frac{\langle [v_{i} - \langle v \rangle]^{3}\rangle}{\sigma_{v_{pos}, {\rm 1D}}^{3}} \\
{\rm kurtosis} = \mathcal{K} = \frac{\langle [v_{i} - \langle v \rangle]^{4}\rangle}{\sigma_{v_{pos}, {\rm 1D}}^{4}}.
\end{align}
Note that the dispersion in Equation~\ref{Eq:mom2} is a measurement of the dispersion of centroid velocities in the plane of the sky measured across the trees, which we denote $\sigma_{v_{pos}, {\rm 1D}}$ (this will be discussed further in \S~\ref{results:disp_acorns}; cf. $\sigma_{v_{\rm m1}, {\rm 1D}}$ in \S~\ref{results:disp}). The skewness and kurtosis are measures of the symmetry and flatness of a distribution, respectively. Negative skewness indicates that the distribution is skewed to the left and a positive skewness the opposite. A Gaussian distribution has a kurtosis of 3. A value larger than 3 implies that the distribution has prominent tails, and therefore rarer, high-amplitude events occur more frequently than would be expected for purely Gaussian behaviour. A value less than 3 implies the opposite. 

The trees are mostly well separated in velocity (as can also be seen in Fig.~\ref{Figure:ppv_acorns}) with mean velocities of $\langle v \rangle=\{2.9, 16.5, 33.1, 37.0\}$\,\kms \ for trees A, B, D, and C, respectively. Note however, that this is not an explicit requirement of {\sc acorns}. For example, trees C and D are more closely related in velocity but are identified as distinct due to their differing velocity dispersions (their median velocity dispersions are separated by $\sim2$\,\kms; see \S~\ref{results:disp_acorns}). 

Each of the trees shows a slightly skewed distribution of centroid velocity. Trees B, C, and D are negatively skewed while A is positively skewed. In terms of the kurtosis, A and D have similar values of $\mathcal{K}\sim3.9$ indicating that the tails of the distribution are more prominent than those expected from a purely Gaussian distribution. Conversely, A has $\mathcal{K}\sim2.5$. Finally, B has a kurtosis value of $\mathcal{K}\sim3.0$. Despite all clusters having $\mathcal{S}\lesssim|0.4|$ and $2.5<\mathcal{K}<3.9$, the centroid velocities of the clusters are statistically inconsistent with Gaussian distributions based on the computation of the D'Agostino ($p\approx0.0$, which combines the skewness and kurtosis of the distribution; \citealp{dagostino_1990}) and the Anderson-Darling statistics ($p\approx0.0$; \citealp{anderson_1952}). 

\input{Figures/Figure_9_velocity_pdf_acorns.tex}
\input{Figures/Figure_10_v_acorns.tex}

It has been argued that deviation from Gaussianity can occur when systematic or ordered motions are present within the velocity field. \citet{federrath_2016} recently argued that the large-scale velocity gradient observed across \brick \ contributes to producing a non-Gaussian velocity PDF. After subtraction of the systematic motions from the velocity field, \citet{federrath_2016} state (following visual inspection of the data) that the velocity PDF is in excellent agreement with a Gaussian profile, and used this as a method to decouple the contribution of turbulent gas motions from the observed velocity dispersion. 

It is worth noting that despite appearing consistent with a Gaussian profile, the gradient-subtracted velocity field derived from intensity weighted mean velocities (see \S~\ref{results:vlsr}) also produces a non-Gaussian distribution, in a statistical sense. We examine the gradient-subtracted velocity field for the moment 1 map and find: $\langle v \rangle=0.0$\,\kms \ (note this is because the gradient has been subtracted); $\sigma_{v_{\rm m1}, gs, {\rm 1D}} = 4.0$\,\kms; $\mathcal{S}=-0.24$; $\mathcal{K}=3.0$. As with the {\sc acorns} trees, the null hypothesis that the distribution of velocities is drawn from a Gaussian distribution can be rejected following the computation of the D'Agostino and Anderson-Darling statistics ($p$-values\,$\approx0.0$). However, with many physical processes at work within the interstellar medium, deviations from Gaussianity are unsurprising \citep{klessen_2000}. Moreover, \citet{federrath_2016} clearly acknowledge that there are residual deviations from their Gaussian fit. These deviations, the authors argue, are most likely due to a combination of noise in the data, the excitation conditions of HNCO, and the fact that small scale systematic motions may still be present in the data. 

In Fig.~\ref{Figure:velfield} we plot the velocity fields of the four {\sc acorns} trees. Velocity gradients are clearly evident in the data. Using the methodology outlined in \S~\ref{results:vlsr} we compute velocity gradients for each tree. We find $\mathscr{G}=\{1.1, 2.1, 1.9, 4.2\}$\,km\,s$^{-1}$pc$^{-1}$ (corresponding to $\mathscr{G}=\{2.7, 5.0, 4.6, 10.2\}$\,km\,s$^{-1}$arcmin$^{-1}$) in directions $\Theta_{\mathscr{G}}=\{-82.6\degr, -141.9\degr, -129.6\degr, -31.2\degr\}$ east of north for trees, A, B, C, and D, respectively. The magnitude of the velocity gradients of trees B and C are more consistent with those derived from simulations of molecular clouds following the \citet{kruijssen_2015} orbit ($2.4$\,km\,s$^{-1}$pc$^{-1}$; \citealp{kruijssen_2019}; cf. \S~\ref{results:vlsr}). We display the magnitude and direction of these gradients as arrows in Fig.~\ref{Figure:velfield}. 

\input{Figures/Figure_11_disp_histo_acorns.tex}

\subsubsection{Velocity dispersions: plane of the sky vs. line-of-sight velocity fluctuations}\label{results:disp_acorns}

In this section we focus on the velocity dispersions of the {\sc acorns} trees. The standard deviation of centroid velocities estimated above (Equation~\ref{Eq:mom2}) provides an estimate for $\sigma_{v_{pos}, {\rm 1D}}$ for each tree. For trees A, B, C, and D we measure $\sigma_{v_{pos}, {\rm 1D}} = \{3.5, 5.2, 4.6, 4.5\}$\,\kms, respectively. 

If we recompute the dispersions after subtracting a 2-D velocity plane constructed from the velocity field of each tree (cf. the linear model in Equation~\ref{Equation:gradient} and the gradients displayed in Fig.~\ref{Figure:velfield}), we find for A, B, C, and D $\sigma_{v_{pos, gs}, {\rm 1D}} = \{3.3, 4.1, 3.1, 4.1\}$\,\kms, respectively (where the subscript $gs$ stands for `gradient-subtraction'). Accounting for these large-scale systematic motions leads to a reduction of $\sim17\%$ in the total dispersion of centroid velocities, in contrast to the $\sim56$\% reduction inferred by \citet{federrath_2016}. This indicates that although large-scale systematic motions, if they are indeed systematic, may contribute to the observed dispersion in the plane of the sky velocity, they do not dominate.  

In Fig.~\ref{Figure:disp_acorns} we plot histograms of $\sigma_{v_{los}, {\rm 1D}}$ for each of the {\sc acorns} trees. For A, B, C, and D we find $\langle \sigma_{v_{los}, {\rm 1D}} \rangle = \{5.3, 4.9, 4.0, 5.8\}$\,\kms, respectively (where the angle brackets indicate that we have taken the mean value over all \scouse \ measurements associated with each cluster). As is evident in Fig.~\ref{Figure:disp_acorns}, the distributions are skewed and so we report median velocity dispersions of $\{5.2, 4.5, 3.8, 5.8\}$\,\kms. In the bottom panels we plot the cumulative histograms of the velocity dispersions. A two-sample Kolmogorov-Smirnov test for each of the six unique parings of the four tree samples reveals that none of the samples are drawn from the same distribution, indicating that there are statistical differences between the clusters in terms of their measured line-of-sight velocity dispersions. The peak of the distribution for D is, for example, shifted rightwards from those of B and C indicating broader velocity dispersions on average. This can be seen in Fig.~\ref{Figure:sigfield}, where we have plotted the spatial distribution of velocity dispersions throughout each tree. 

It is notable that taking the ratio of the line-of-sight and plane of the sky velocity dispersions yields $\sigma_{v_{los}, {\rm 1D}} /\sigma_{v_{pos}, {\rm 1D}} = \{1.5, 0.9, 0.9, 1.3\}$ for trees A, B, C, and D, respectively. On average this is $\langle \sigma_{v_{los}, {\rm 1D}} /\sigma_{v_{pos}, {\rm 1D}} \rangle=1.2\,\pm\,0.3$, where the uncertainty here reflects the standard deviation. We speculate that this isotropy in the line-of-sight velocity distribution and the line-of-sight fluctuations in the centroid velocity in the plane of the sky may encode information about the cloud geometry. Namely, that the line-of-sight extent of the cloud components are approximately equivalent to that in the plane of the sky. This could perhaps explain some of the discrepancy between our measured velocity dispersions and those predicted from the steep velocity dispersion-size relationship derived for the CMZ (see \S~\ref{results:disp}). However, we hasten to add that this result would need to be tested rigorously with numerical simulations.

\input{Figures/Figure_12_disp_acorns.tex}

%% file: Figures/Figure_2_PPV_scousepy.tex
\begin{figure*}
\begin{center}
\includegraphics[trim = 0mm 5mm 0mm 10mm, clip, width = 0.85\textwidth]{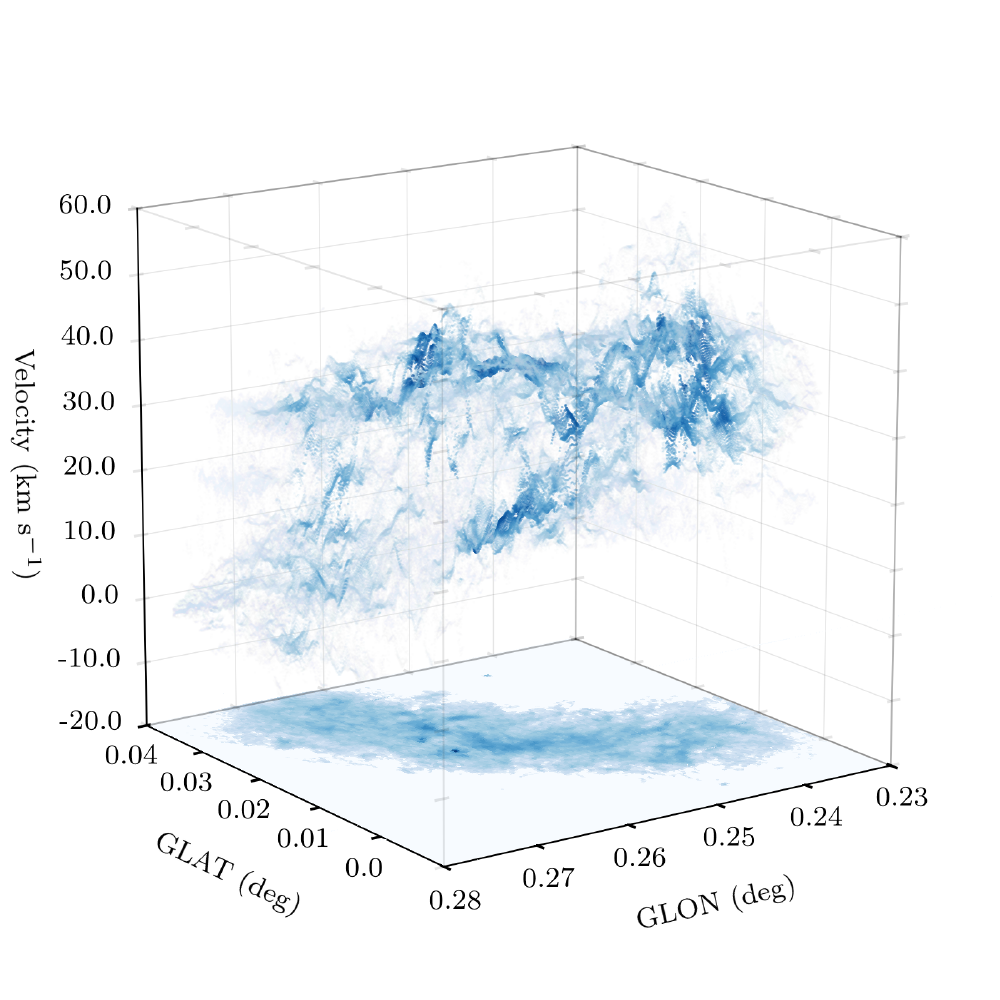}
\end{center}
\caption{A PPV image of \brick. Each data point denotes the location and centroid velocity of a Gaussian component identified in HNCO emission in the ALMA data and extracted using \scouse. The colour of each data point (from light to dark) is proportional to the peak intensity of the corresponding spectral component. Note that only the data between -20.0\,\kms \ and 60.0\,\kms \ are shown. Emission outside of this velocity range, although spatially coincident with \brick \ may not be associated with the cloud itself. The full extent of the data can be seen in Figure~\ref{Figure:vlsr}. At the base of the plot, we plot the 3\,mm dust continuum emission first presented in \citet{rathborne_2015}. This figure highlights the kinematic complexity of \brick. }
\label{Figure:ppv}
\end{figure*}

%% file: Figures/Figure_3_v_scousepy.tex
\begin{figure*}
\begin{center}
\includegraphics[trim = 15mm 0mm 10mm 5mm, clip, width = 0.8\textwidth]{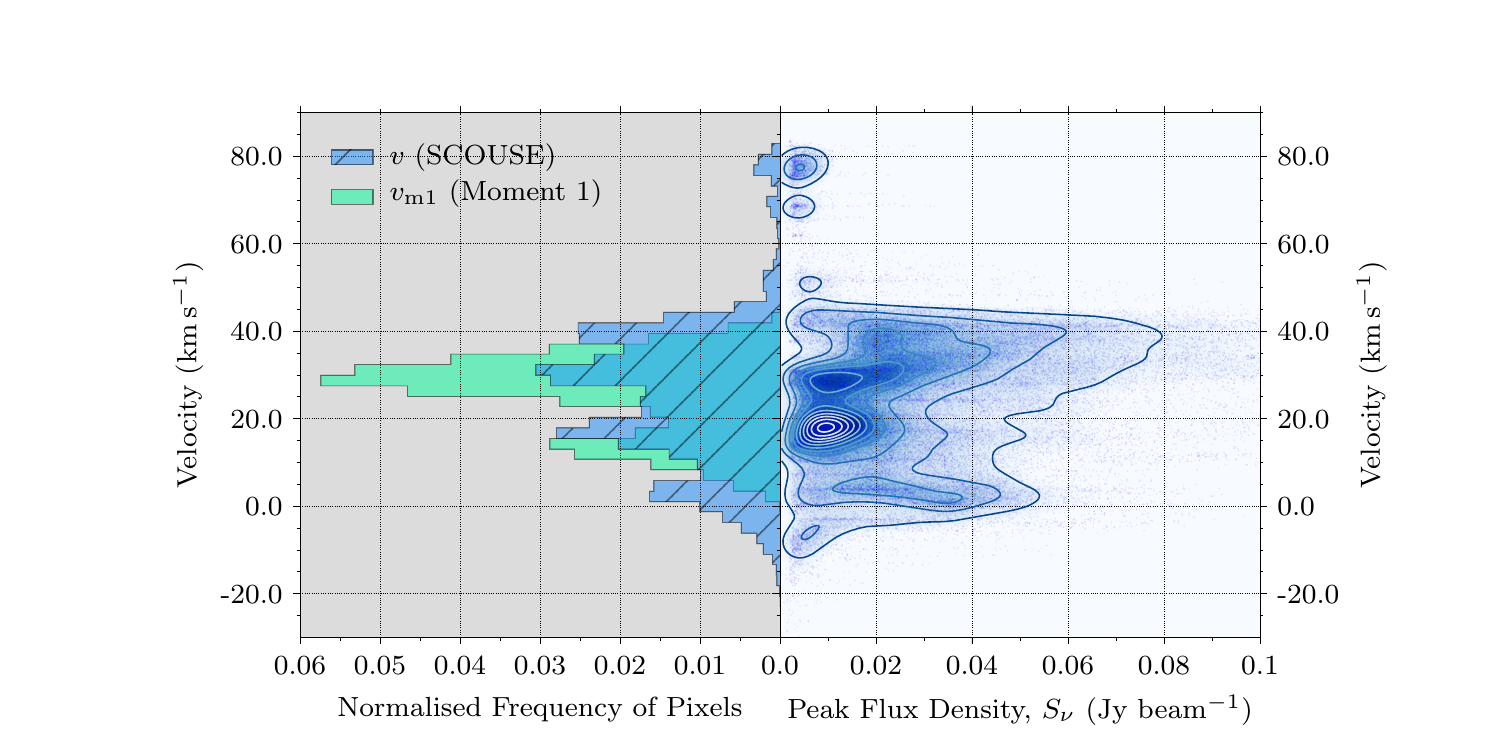}
\vspace{-0.5cm}
\end{center}
\caption{Left: A histogram of the centroid velocities extracted using {\sc scousepy} ($v$, blue) compared with the intensity-weighted average velocities extracted using moment analysis ($v_{\rm m1}$, green). The latter histogram shows a double-peaked profile which has previously been interpreted as a signature of cloud rotation. Right: {\sc scousepy} centroid velocities, $v$, as a function of peak flux density, $S_{\nu}$. The contours reflect the point density. Note that we have truncated the x-axis in order to show the main structure in the $S_{\nu}-v$ plane (peak flux densities actually go up to $\sim0.4$\,Jy\,beam$^{-1}$). The {\sc scousepy} decomposition displays significantly more structure over a broader distribution of velocities than that derived from moment analysis.  }
\label{Figure:vlsr}
\end{figure*}

%% file: Figures/Figure_4_disp_scousepy.tex
\begin{figure*}
\begin{center}
\includegraphics[trim = 15mm 0mm 10mm 5mm, clip, width = 0.8\textwidth]{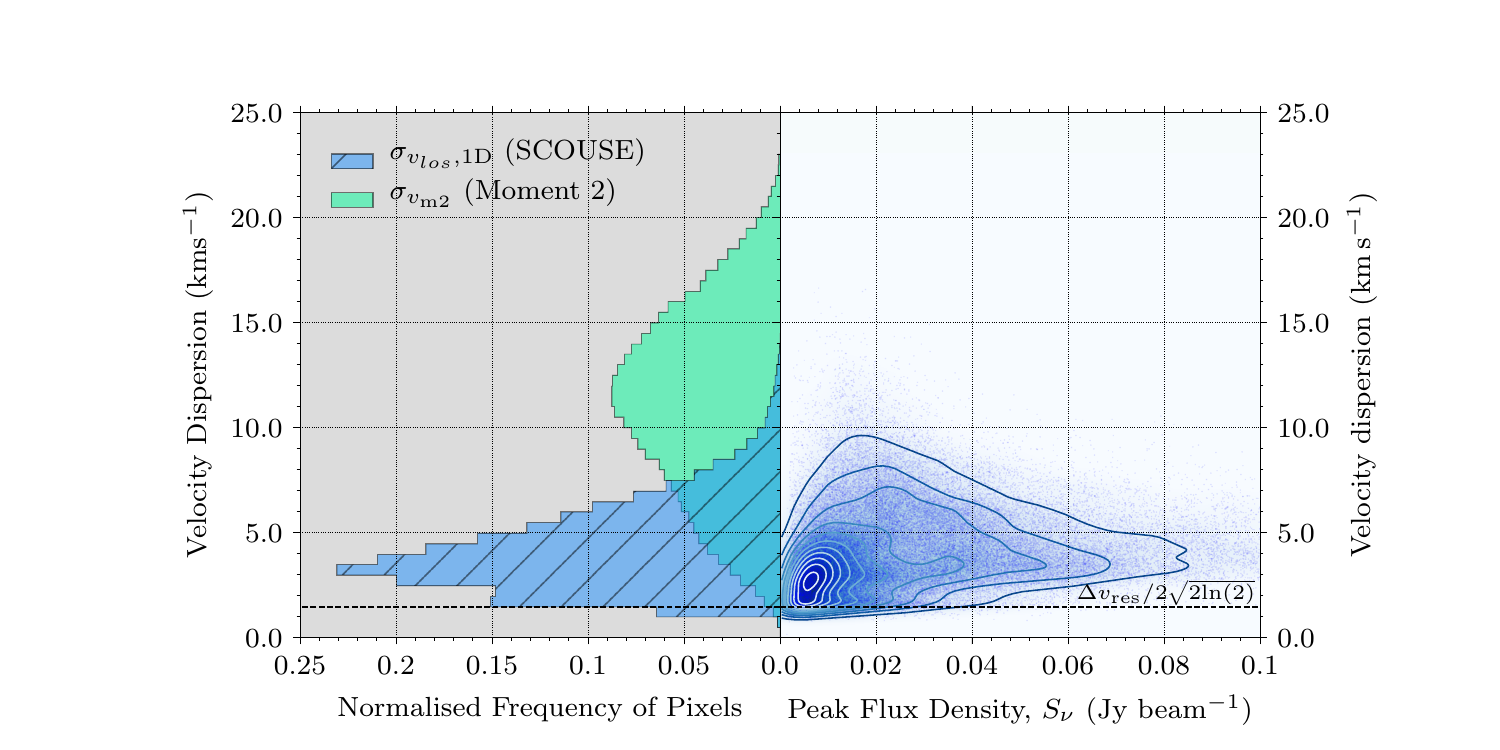}
\vspace{-0.5cm}
\end{center}
\caption{Left: A histogram of the velocity dispersions extracted using {\sc scousepy} ($\sigma_{v_{los}, {\rm 1D}}$, blue) compared with the intensity-weighted average velocity dispersion extracted using moment analysis ($\sigma_{v_{\rm m2}}$, green). Right: {\sc scousepy} Velocity dispersion as a function of peak flux density, $S_{\nu}$. The contours reflect the point density. Note that we have truncated the x-axis in order to show the main structure in the $S_{\nu}-\sigma_{v_{los}, {\rm 1D}}$ plane (peak intensities actually go up to  $\sim0.4$\,Jy\,beam$^{-1}$ ). The horizontal line is located at $\Delta v_{\rm res}/2\sqrt{2{\rm ln}(2)}$, where $\Delta v_{\rm res}=3.4$\,\kms \ (the spectral resolution). Note that the tolerance level input during stage 3 of the {\sc scousepy} fitting procedure was half of this value to ensure good fits to the data (see \S~\ref{results:scouse} and \citealp{henshaw_2016} for further discussion on the input tolerance values for \scouse). As can be seen, due to the presence of multiple velocity components, moment analysis overestimates the velocity dispersions on average by over a factor of 2 compared to {\sc scousepy}.  }
\label{Figure:disp}
\end{figure*}

%% file: Figures/Figure_5_dendrogram_acorns.tex
\begin{figure*}
\begin{center}
\includegraphics[trim = 2mm 6mm 10mm 2mm, clip, width = 0.98\textwidth]{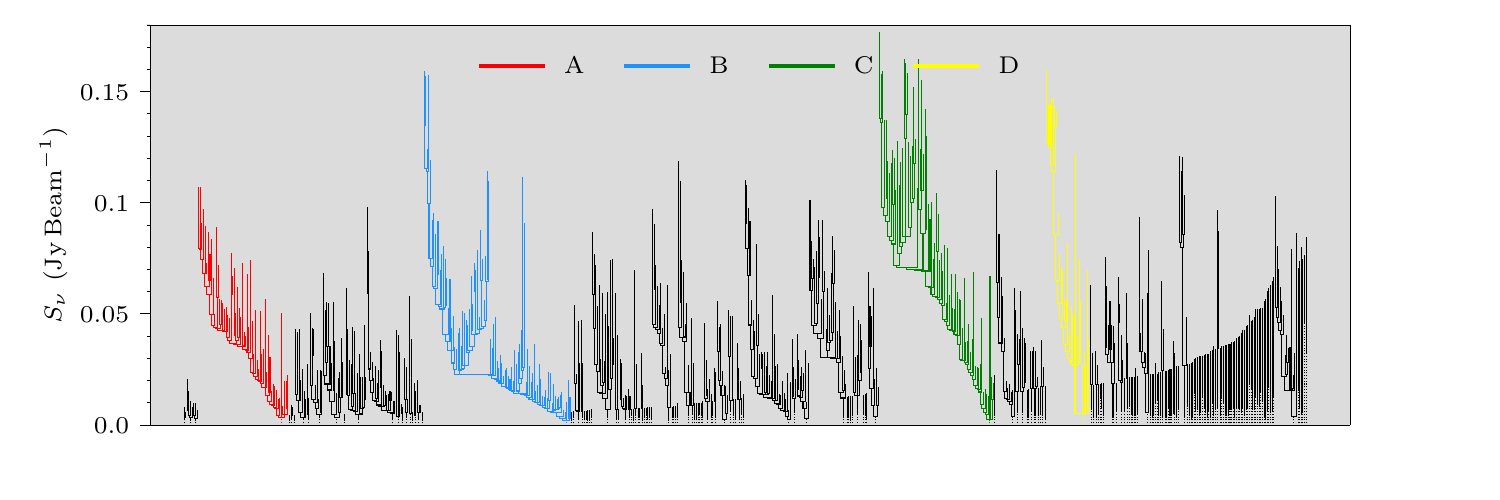}
\end{center}
\caption{{\sc acorns} clustering displayed graphically as a dendrogram. Here we display the full forest of clusters. Each tree in the forest can be further subdivided into branches and leaves in a hierarchical fashion (see \S~\ref{Methodology:acorns} for a full description of the method and nomenclature). The forest, comprising a total of 195 trees, is dominated by 4 trees; A, B, C, and D (highlighted in red, blue, green, and yellow, respectively). Together they comprise $>50\%$ of all data. After these first 4 trees there is a factor of $\sim2$ drop in the next tree's percentage contribution to the total dataset.}
\label{Figure:dendrogram}
\end{figure*}

%% file: Figures/Figure_6_PPV_acorns.tex
\begin{figure*}
\begin{center}
\includegraphics[trim = 0mm 5mm 0mm 10mm, clip, width = 0.85\textwidth]{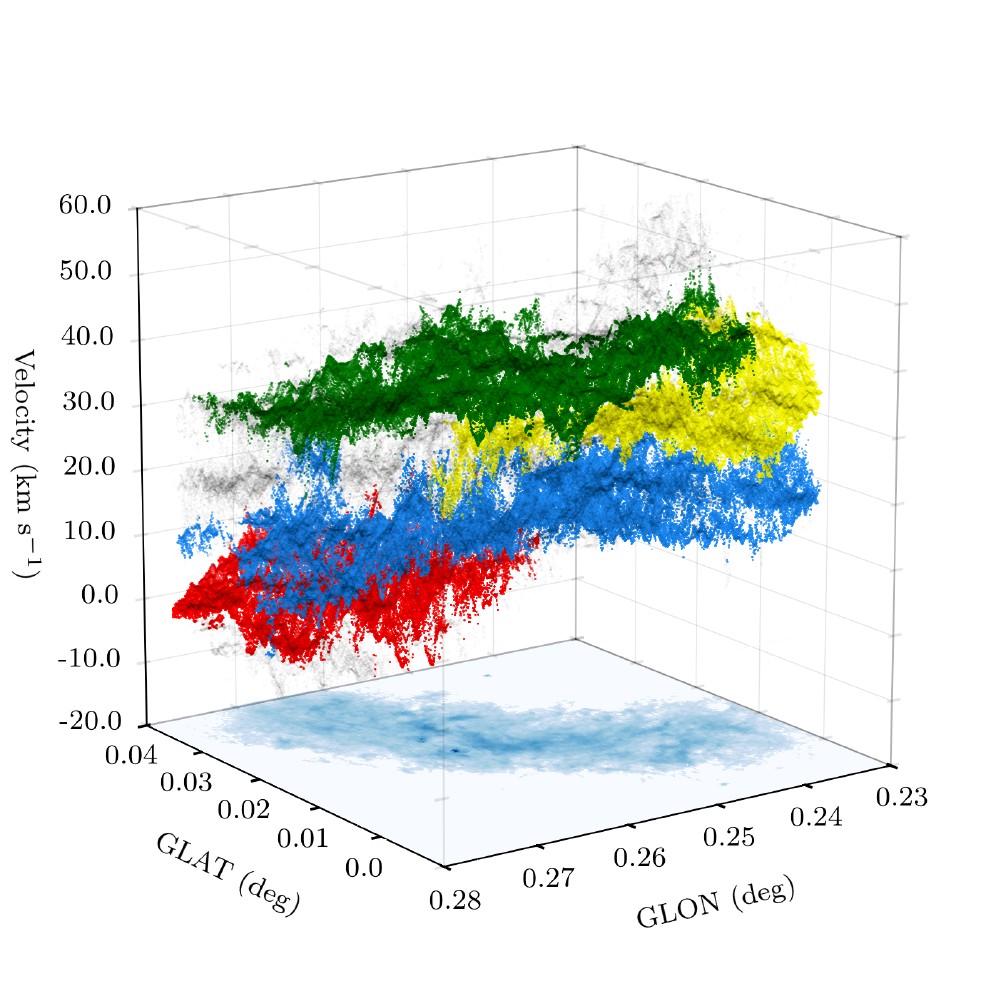}
\end{center}
\caption{In this image we highlight the dominant {\sc acorns} trees in PPV space. Each colour refers to a different tree in the forest: A (red), B (blue), C (green), and D (yellow) [see \S~\ref{results:acorns_decomposition} and Figure~\ref{Figure:dendrogram}]. These 4 (out of 195) trees contain $>50\%$ of all data. The full dataset is included in this image as small black data points. The image at the base of the plot is equivalent to that presented in Figure~\ref{Figure:ppv}. }
\label{Figure:ppv_acorns}
\end{figure*}

%% file: Figures/Figure_7_int_acorns.tex
\begin{figure*}
\begin{center}
\includegraphics[trim = 10mm 15mm 10mm 13mm, clip, width = 0.8\textwidth]{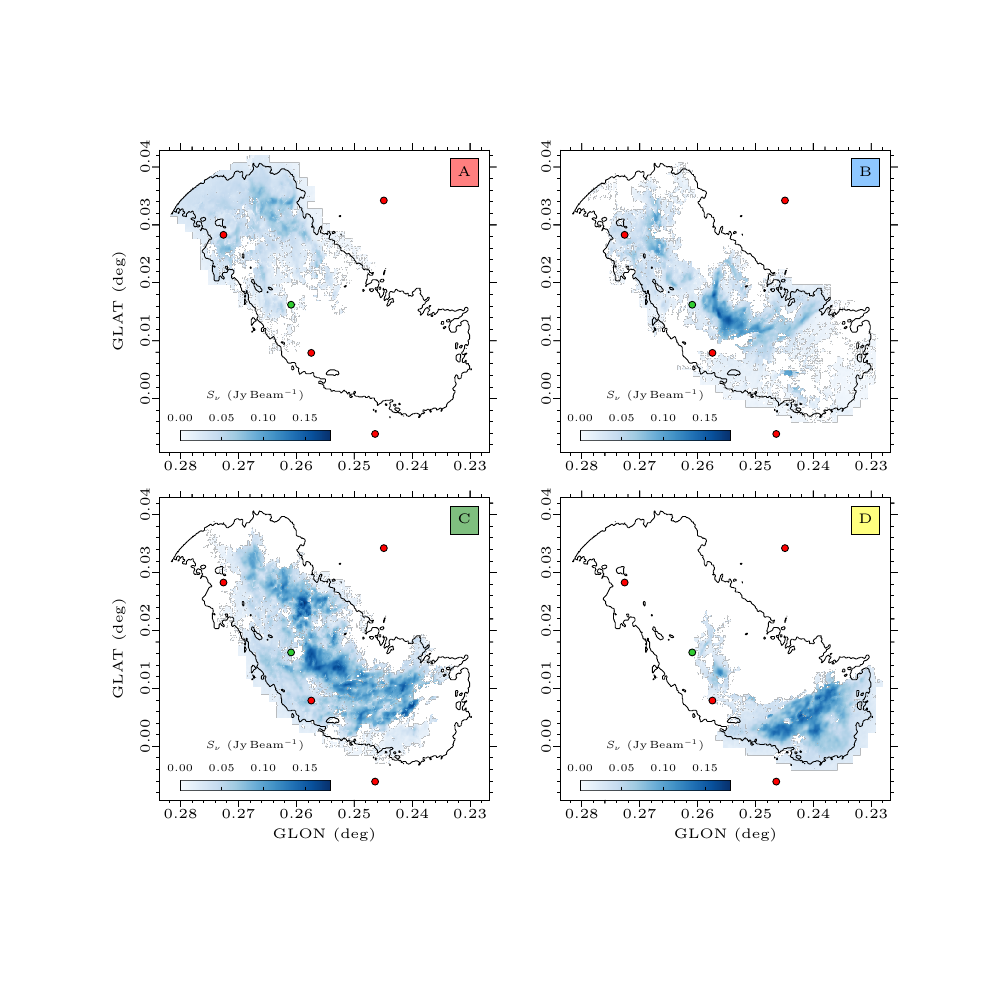}
\end{center}
\vspace{-0.4cm}

\caption{Peak flux density distributions of the main {\sc acorns} trees associated with \brick. The coloured squares in the top right-hand corner of each image refer to the colouring used in Figures~\ref{Figure:dendrogram} and \ref{Figure:ppv_acorns}. The black contour highlights the boundary of the cloud as defined during the masking described in \S~\ref{results:scouse}. The green circle indicates the location of the H$_{2}$O maser identified by \citet{lis_1994}. Note that this coincides with a hole in the intensity distribution of tree C (green). This is discussed further in \S~\ref{results:int_acorns}. Red circles indicate the locations of compact radio continuum sources \citep{rodriguez_2013}. }
\label{Figure:intfield}
\end{figure*}

%% file: Figures/Figure_8_brick_core_acorns.tex
\begin{figure}
\begin{center}
\includegraphics[trim = 2mm 0mm 0mm 12mm, clip, width = 0.45\textwidth]{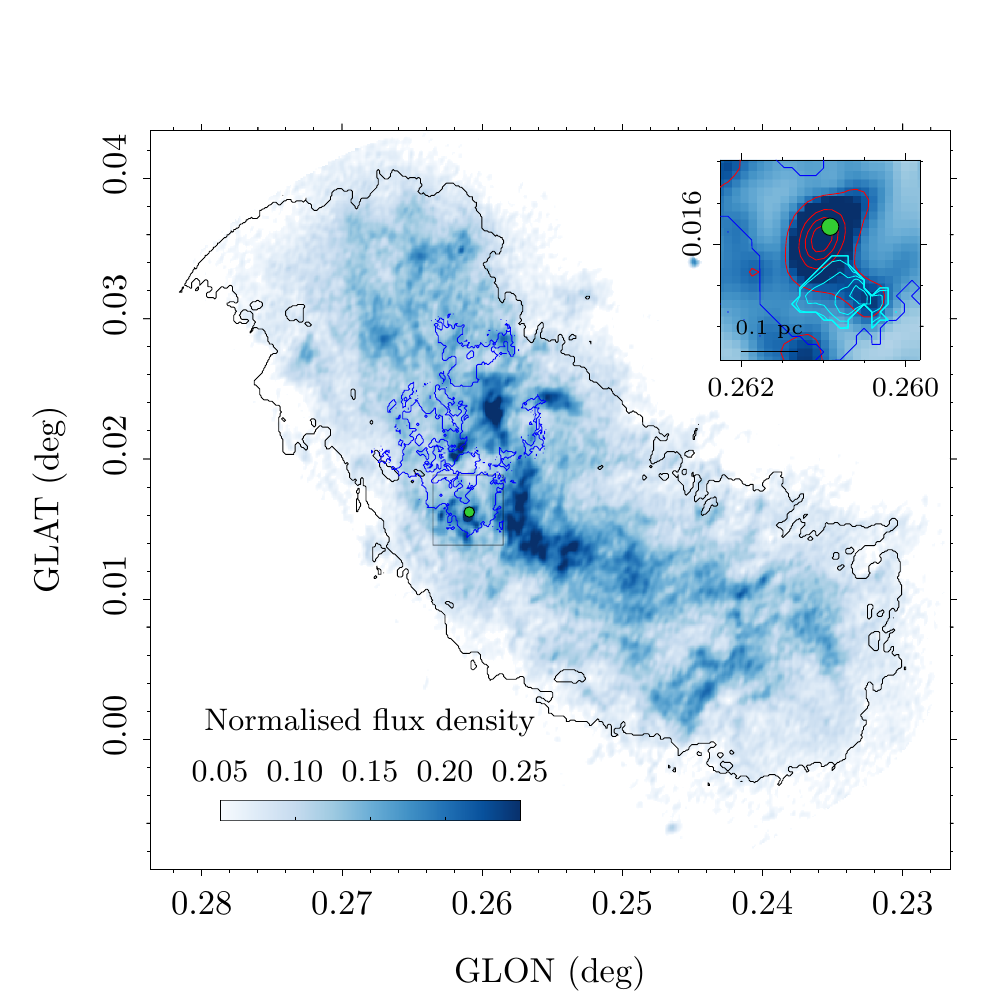}
\end{center}
\vspace{-0.4cm}

\caption{ALMA 3\,mm dust continuum observations of \brick \ in colour scale \citep{rathborne_2014}. The colour scale has been normalised against the peak emission and intentionally truncated in order to emphasise the more diffuse emission. The black contour is equivalent to that presented in Fig.~\ref{Figure:intfield}. Overlaid in blue is the outline of tree\,\#32 (see \S~\ref{results:sf} for details). In the inset image we zoom in on the dust continuum peak associated with the H$_{2}$O maser identified by \citet{lis_1994}. In red contours we display the 3\,mm dust continuum (from $[0.2,0.4,0.6,0.8]\times$ the peak emission $\sim0.004$\,Jy\,beam$^{-1}$). The blue contour indicates the outline of the tree. In cyan contours we show the leaf which is closest to the bright 3\,mm peak seen in dust continuum. The filled green circle in both the main image and the inset indicates the location of the H$_{2}$O maser identified by \citet{lis_1994}.  }
\label{Figure:continuum}
\end{figure}

%% file: Figures/Figure_9_velocity_pdf_acorns.tex
\begin{figure}
\begin{center}
\includegraphics[trim = 12mm 15mm 10mm 10mm, clip, width = 0.48\textwidth]{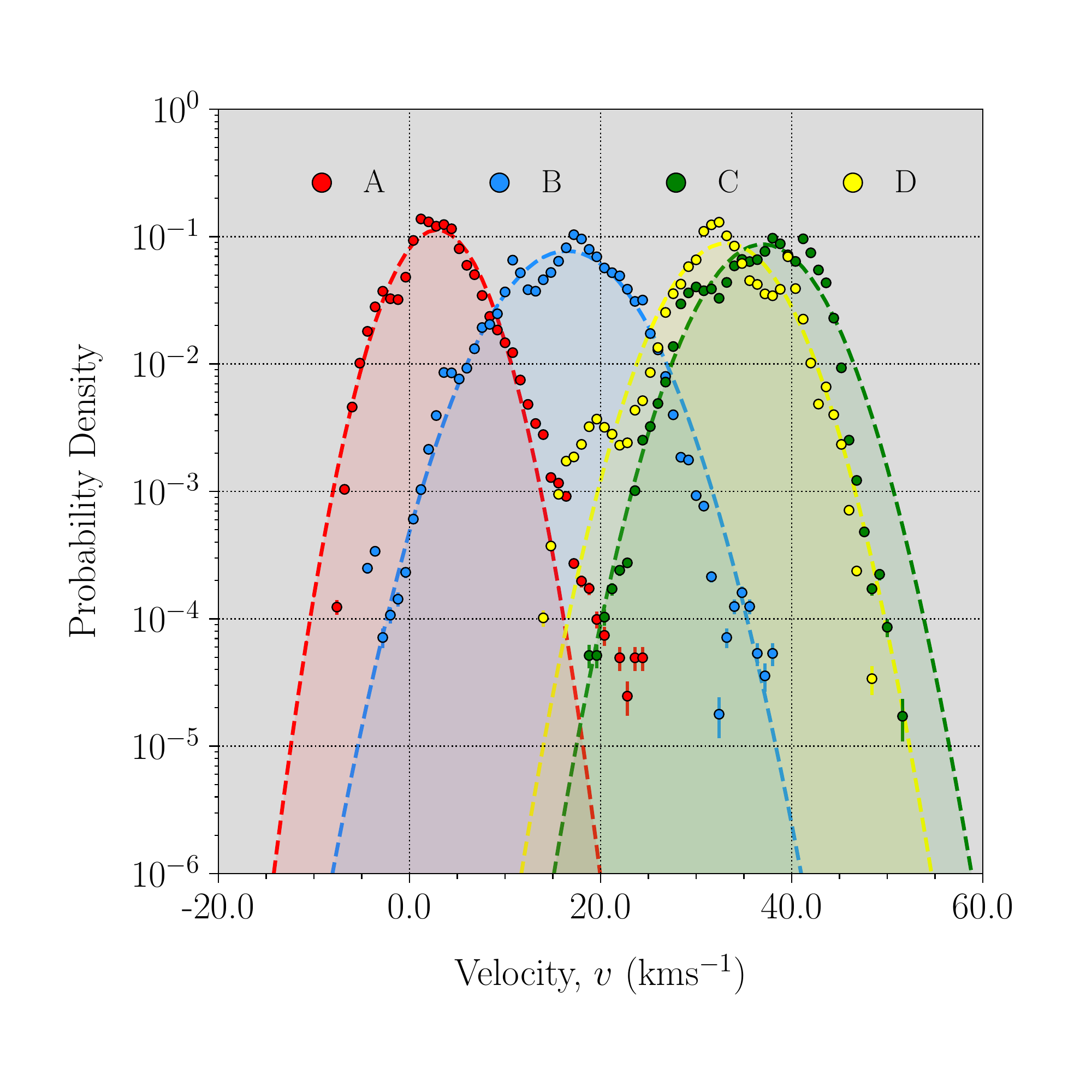}
\end{center}
\caption{Centroid velocity probability density functions (PDFs) of the main {\sc acorns} trees A (red), B (blue), C (green), and D (yellow) [cf. Figure~\ref{Figure:dendrogram}]. Dashed lines are normal distributions fitted to the data. Despite the velocity PDFs appearing broadly consistent with Gaussian distributions (the profiles have a mean kurtosis value of $\langle \mathcal{K} \rangle \sim 3.3$), there are statistically significant deviations from Gaussianity.    }
\label{Figure:velpdf}
\end{figure}

%% file: Figures/Figure_10_v_acorns.tex
\begin{figure*}
\begin{center}
\includegraphics[trim = 10mm 15mm 10mm 13mm, clip, width = 0.8\textwidth]{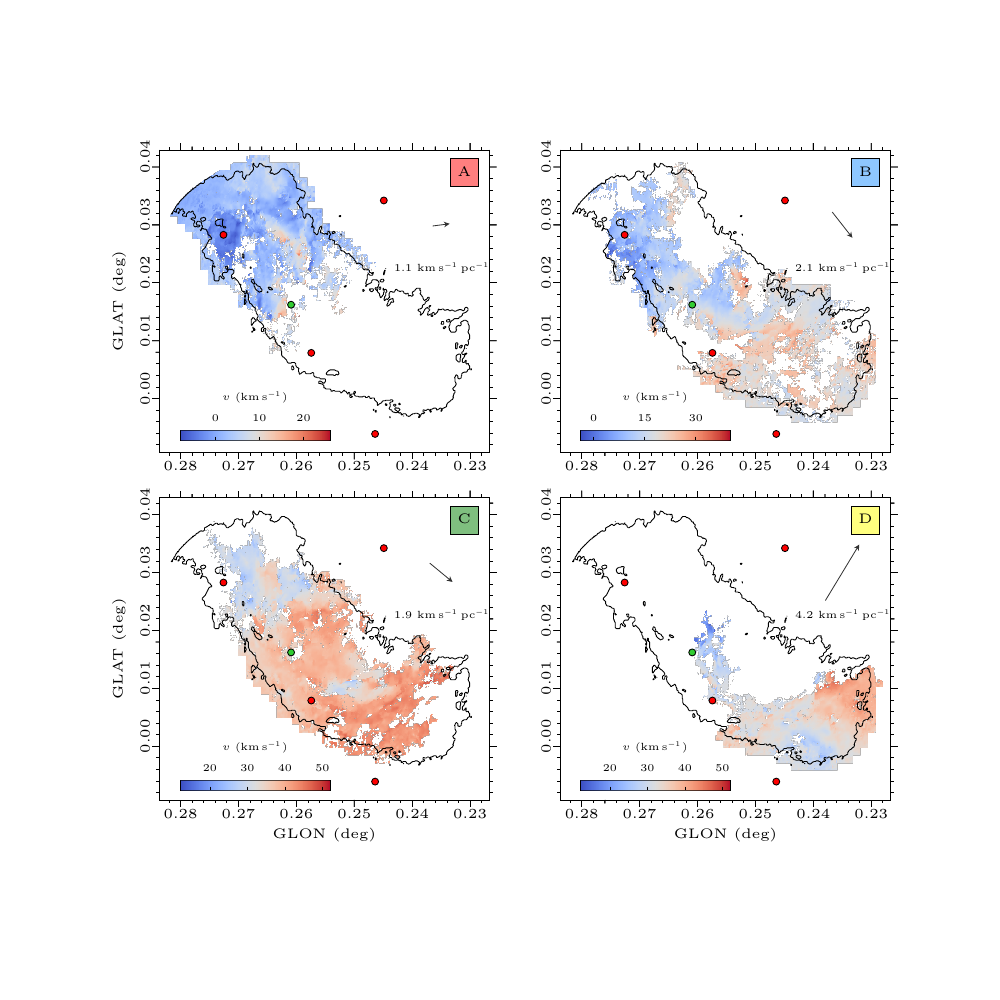}
\end{center}
\vspace{-0.4cm}

\caption{Equivalent to Figure~\ref{Figure:intfield} but for the centroid velocities, $v$, measured throughout each {\sc acorns} tree. The size and direction of the arrow in each plot represents the magnitude and direction of the velocity gradient across each tree (pointing in the direction of increasing velocities). }
\label{Figure:velfield}
\end{figure*}

%% file: Figures/Figure_11_disp_histo_acorns.tex
\begin{figure}
\begin{center}
\includegraphics[trim = 14mm 20mm 14mm 16mm, clip, width = 0.4\textwidth]{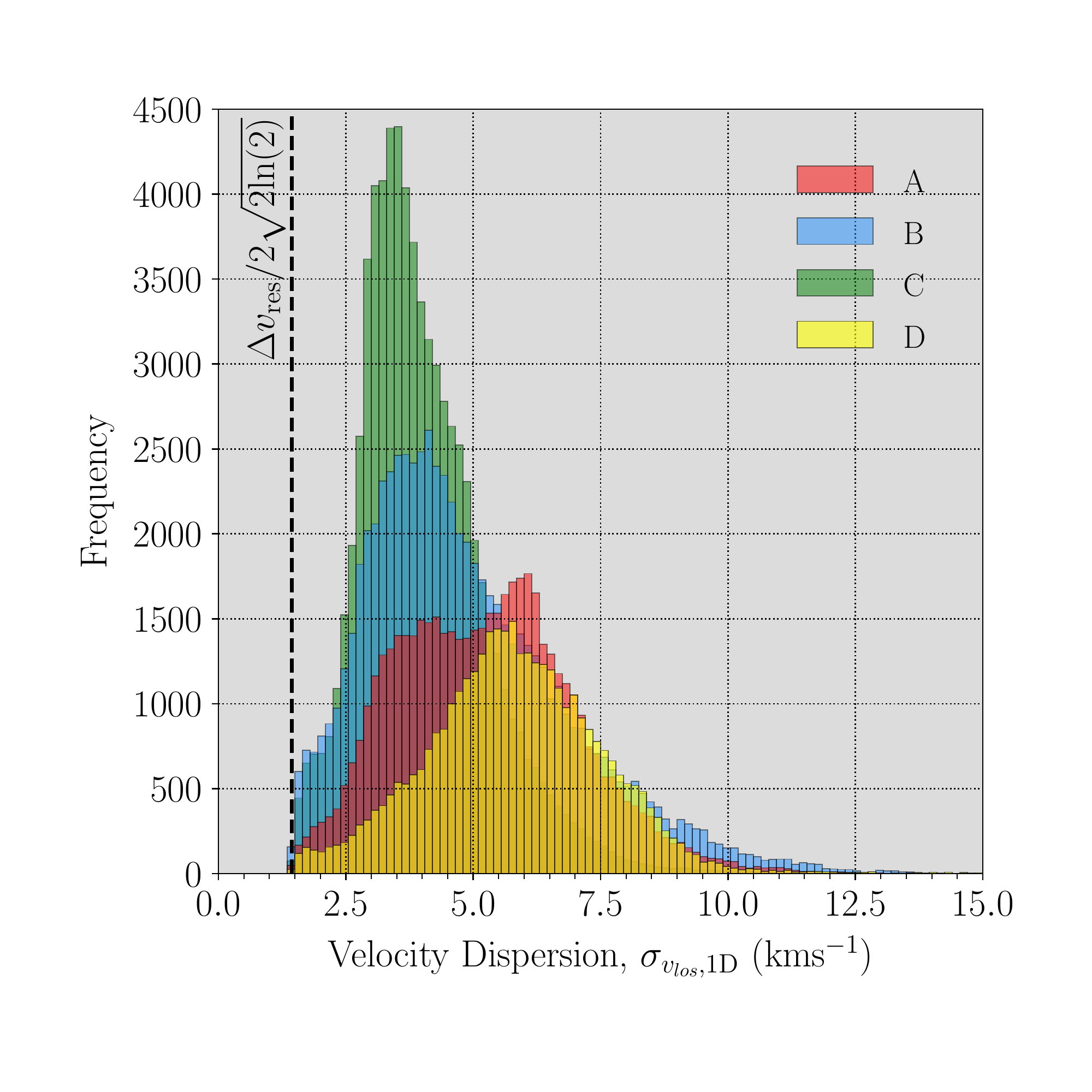}
\includegraphics[trim = 14mm 20mm 14mm 16mm, clip, width = 0.4\textwidth]{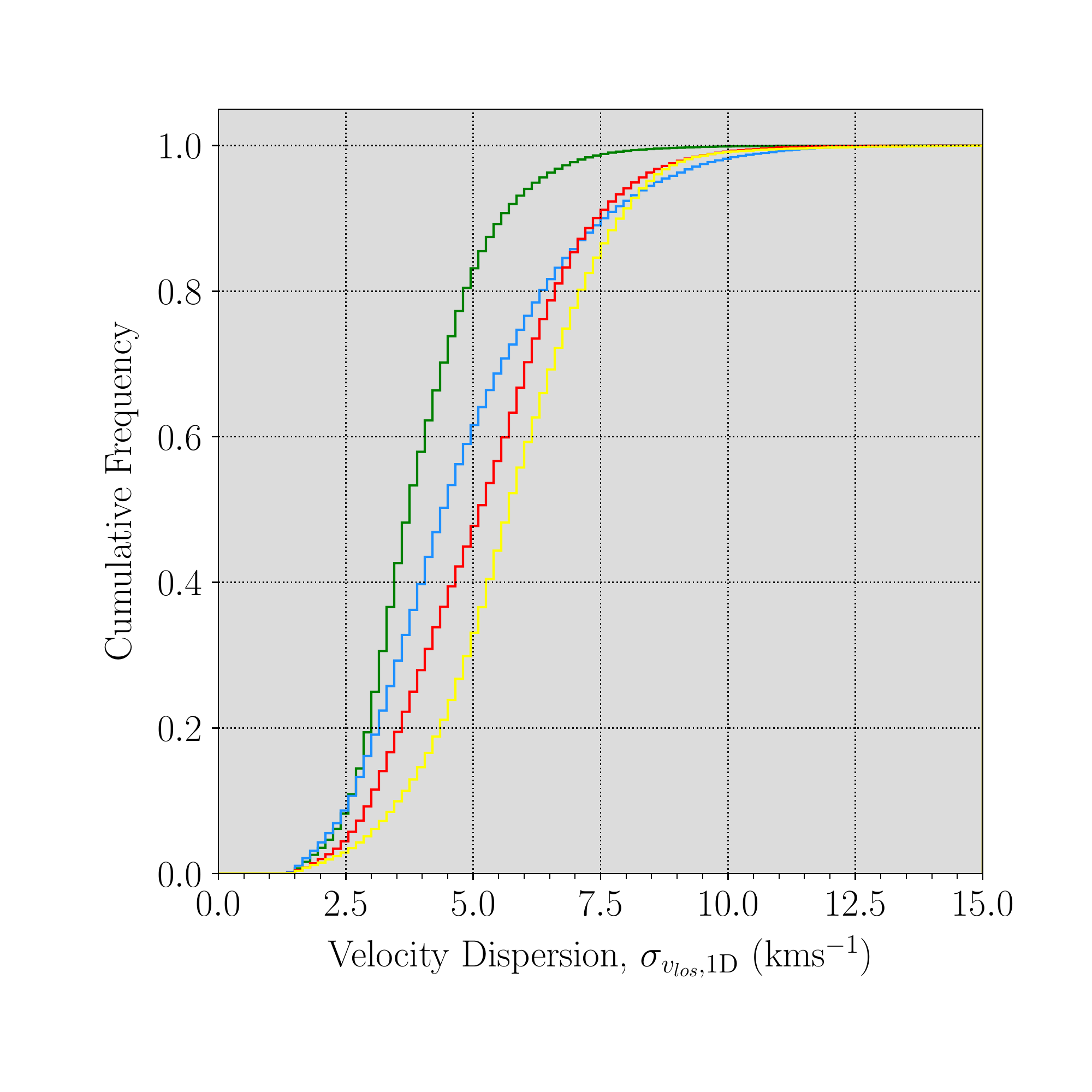}
\end{center}
\caption{Top: Histograms of the velocity dispersion, $\sigma_{v_{los}, {\rm 1D}}$, for the main {\sc acorns} trees (coloured histograms; see Figure\,\ref{Figure:dendrogram}). The vertical line is located at $\Delta v_{\rm res}/2\sqrt{2{\rm ln}(2)}$, where $\Delta v_{\rm res}=3.4$\,\kms \ (the spectral resolution). Bottom: Cumulative histograms of the velocity dispersions for the trees.  }
\label{Figure:disp_acorns}
\end{figure}

%% file: Figures/Figure_12_disp_acorns.tex
\begin{figure*}
\begin{center}
\includegraphics[trim = 10mm 15mm 10mm 13mm, clip, width = 0.8\textwidth]{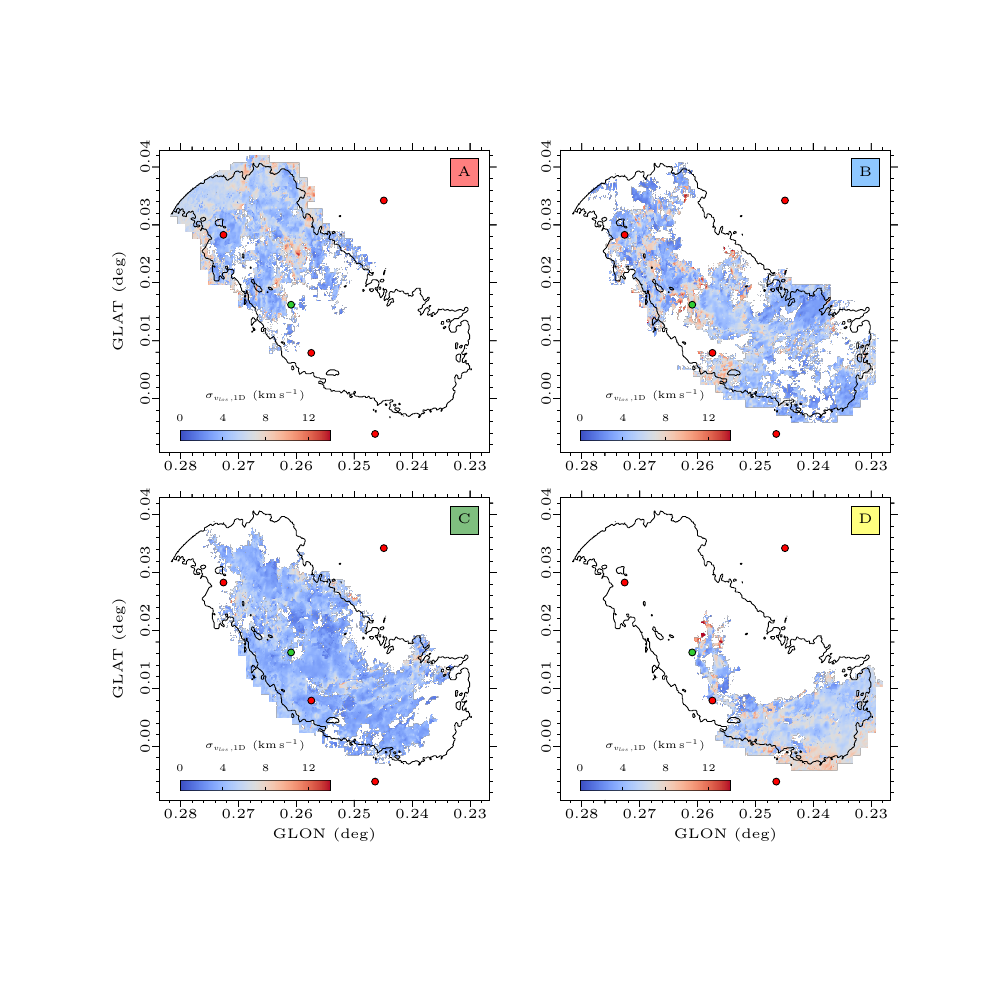}
\end{center}
\vspace{-0.4cm}

\caption{Equivalent to Figure~\ref{Figure:intfield} but for the velocity dispersion, $\sigma_{v_{los}, {\rm 1D}}$, measured throughout each {\sc acorns} tree.}
\label{Figure:sigfield}
\end{figure*}

%% file: Tex/Discussion.tex
\section{Is `the Brick' \emph{really} a brick? A centrally-condensed molecular cloud vs. multiple colliding (sub-)clouds}\label{Section:discussion}

\brick's moniker, `the Brick', reflects both its shape on the plane of the sky and the fact that we see it in silhouette against the bright Galactic mid-infrared background at the Galactic Centre (see Fig.~\ref{Figure:brick} and \citealp{longmore_2012}). However, the analysis presented in \S~\ref{results:global} and \S~\ref{results:local} reveals substantial and complex substructure in both position and velocity, consistent with prior analyses which identified cores, filaments, and other coherent features \citep{bally_2014, higuchi_2014, johnston_2014, rathborne_2014, mills_2015, rathborne_2015, federrath_2016}. In the following sections we discuss the current understanding of the structure of \brick \ both in terms of the kinematic analysis presented in this work and in the global context of the CMZ. 

\subsection{`The Brick': \brick \ as a centrally-condensed molecular cloud}

Using single-dish observations from the MALT90 survey (\citealp{foster_2011, jackson_2013}), \citet{rathborne_2014a} presented a study of the structure of \brick. One of the distinctive features noted by the authors was the presence of multiple velocity components associated with \brick \ (much like in Fig.~\ref{Figure:ppv}). \citet{rathborne_2014} presented two possible explanations for the presence of multiple velocity components in \brick: i) that \brick \ is a single, coherent, centrally-condensed cloud with depletion in its cold interior; ii) that the two velocity components reflect two clumps colliding. \citet{rathborne_2014a} favour the former of the two scenarios, which we assess in this section. In \S~\ref{discussion:collision} we discuss the cloud collision scenario. 

\subsubsection{Scenario 1a: Optically thin lines: \brick \ is a centrally-condensed molecular cloud with depletion in its cold interior}\label{discussion:bakedalaska}

The first interpretation was coined the `Baked Alaska' model by \citet{rathborne_2014a} and was conceived in an attempt to explain the profiles of molecular emission lines observed throughout \brick. Conceptually, it is easiest to think of the Baked Alaska model as an adjustment to the classic blue-shifted infall model (see e.g. \citealp{evans_1999, smith_2012} for intuitive diagrammatic explanations). In this idealised picture, an asymmetric self-absorbed line profile occurs in emission lines with high opacity due to the inside-out collapse of a molecular cloud or a core. If the core exhibits a density and temperature gradient such that the excitation temperature increases inwards, emission from the centre can be absorbed by the low-excitation outer envelope, producing a double peaked emission line profile with an emission dip at the centroid velocity of the core. The blue asymmetry (i.e. where the blue peak appears brighter than the red peak) is due to the high excitation point in the red peak being obscured by the lower excitation point (as only the $\tau=1$ surface is observed). Consequently, a double-peaked line profile with a blue asymmetry in an optically thick line can often be interpreted as a signature of collapse. 

In 5 positions selected by \citet{rathborne_2014a}, the line profiles of the optically thick species (e.g. HCO$^{+}$, HCN, N$_{2}$H$^{+}$) showed redshifted asymmetry (i.e. the opposite of the blue-shifted infall model). \citet{rathborne_2014a} argue that one way to create a redshifted asymmetry would be to invoke the same model, but with a cloud that is externally heated (as is observed in dust emission in \brick; \citealp{longmore_2012}) such that the excitation temperature actually decreases towards the centre (hence the name `Baked Alaska'). A schematic explanation of this idea is presented in Figure\,9 of \citet{rathborne_2014a}.\footnote{Note that another way to create a redshifted asymmetry would be to invoke expansion motions rather than collapse.}

Key to the interpretation of collapse in the aforementioned idealised model is that optically thin tracers peak at the location of the self-absorbed dip in emission in optically thick tracers (see e.g. \citealp{contreras_2018}, for a recent example). This is crucial because multiple spectral components in optically thin lines may simply indicate the presence of additional cloud components along the line-of-sight. 

Herein lies the problem with \brick: the lines that are believed to be optically thin (e.g. H$^{13}$CO$^{+}$, HN$^{13}$C) also show a double peak towards the cloud interior. \citet{rathborne_2014a} argue that a plausible explanation for the double peaked optically thin lines is that, if the lines are not optically thick, there must be severe, parsec-scale, chemical gas depletion of molecules in the cloud's high density and low temperature interior. One proposed line of evidence in favour of the aforementioned scenario is that there is an observed anti-correlation between the dust column density and the integrated intensity of various molecular lines toward the centre of the cloud. This gave rise to the interpretation that \brick \ is a single, coherent, centrally-condensed cloud with depletion in its cold interior. 

In a later publication, \citet{rathborne_2015}, reported a tendency for molecular transitions with higher excitation energies and critical densities to peak toward the centre of the cloud, consistent, they argue, with a cloud with a dense interior. Fig.~8 of \citet{rathborne_2015} shows that PV diagrams of the emission associated with a variety of different molecules, including C$_{2}$H, SiO, HN$^{13}$C, H$^{13}$CN, H$^{13}$CO$^{+}$, HCC$^{13}$CN, SO, NH$_{2}$CHO, CH$_{3}$CHO, and H$_{2}$CS, have similar profiles to that shown in PPV space in Fig.~\ref{Figure:ppv} (i.e. two dominant features separated by $\sim$20\,\kms \ in the north of the cloud that merge in velocity towards to south). 

However, while 2 out of the 17 molecules discussed by \citet{rathborne_2015} do display some emission towards the centre of the cloud (CH$_{3}$CHO and NH$_{2}$CN), it is not extended and it does not peak exclusively in the central region. Rather, the emission qualitatively follows that of the other molecular transitions, but with a small peak towards the centre. Moreover, data from independent studies illustrate that the $\sim$20\,\kms \ gap between the dominant PPV features observed in Fig.~\ref{Figure:ppv} is not populated with emission from nitrogen-bearing species such as N$_{2}$H$^{+}$ \citep{pound_2018}, which are less susceptible to freeze-out at high densities \citep{bergin_2007}.

The fact that the difference in velocity between the dominant components is largest towards the north of the cloud may also be problematic for this scenario. First, when observed at higher resolution, lines that are presumed to be optically thin show multiply-peaked line profiles towards the north and south of the cloud (cf. the singly-peaked profiles in the schematic diagram presented by \citealp{rathborne_2014a}). Secondly, the greatest velocity difference is observed towards the north of the cloud, where we find trees A, B, and C. In the context of widespread depletion, this would necessitate either a density or temperature gradient in \brick. Furthermore it would suggest that either the highest density, or alternatively, lowest temperatures, are observed in the north of the cloud (where the absolute difference in the velocity peaks is the greatest; $\sim35$\,\kms; \S~\ref{results:vlsr_acorns}). Studies of the dust continuum, and therefore the inferred H$_{2}$ column density towards \brick \ show no evidence for such a density gradient \citep{longmore_2012, johnston_2014, rathborne_2015}. Additionally, although the highest temperatures ($>150$\,K) in \brick \ are found towards the south of the cloud (i.e. towards tree D), warm gas temperatures ($80-100$\,K) are also found in the north (and generally distributed throughout; \citealp{ginsburg_2016, krieger_2017}). There is no clear and monotonic trend in decreasing gas temperature towards the north of the cloud. 

It is worth noting that probably the strongest case for complete depletion of molecules within an individual cloud core (although it is yet to be confirmed) comes from \citet{cyganowski_2014}. However, this occurs on $<1000$\,AU scales where densities and temperatures are estimated to be $>10^{9}$\,cm$^{-3}$ and $\lesssim20$\,K, respectively. Although dust temperatures within \brick \ are of the order $\sim20$\,K \citep{longmore_2012}, the gas temperatures are actually considerably higher (of the order $\gtrsim60$\,K; \citealp{ginsburg_2016, krieger_2017}), consistent with the gas and dust not being thermally coupled at the derived cloud density of $\sim10^{4}$\,cm$^{-3}$ \citep{clark_2013}. Therefore without detailed chemical modelling it is currently difficult to reconcile the concept of parsec-scale depletion throughout the interior of a singular, coherent, and centrally-condensed cloud with the absence of either an increasing density gradient or a decreasing temperature gradient towards the northern portion of \brick \ (as would be required to create the PPV profile observed in Fig.~\ref{Figure:ppv}). 

\subsubsection{Scenario 1b: Optically-thick lines: \brick \ is a centrally-concentrated cloud whose interior dynamics are masked due to high optical depth}\label{discussion:thick}

Another conceivable scenario is that the lines which are often considered to be optically thin (e.g. H$^{13}$CO$^{+}$, H$^{13}$CN, HN$^{13}$C), are actually optically thick. If this is the case then the double peaked profile in these lines may simply arise from self-absorption, with the individual peaks representing the outer `shell' of the cloud at the $\tau=1$ surface. 

\input{Figures/Figure_13_RT.tex}

We assess the possibility of the HNCO $J=4(0,4)-3(0,3)$ line being optically thick using radiative transfer modelling. We adopt a kinetic temperature of 60\,K, \citep{ginsburg_2016} and a fixed turbulent line width of 4.4\,\kms \ (i.e. $\langle \sigma_{{v_{los}}, {\rm 1D}} \rangle$). We treat the molecular abundance and gas number density as free parameters, though the best estimate of the average number density is $10^{4}$\,cm$^{-3}$ \citep{federrath_2016} and the assumed canonical HNCO abundance is $10^{-9}$ (the typical abundance found towards dense cores, including those in the CMZ, by \citealp{churchwell_1986} and \citealp{zinchenko_2000}). 

We perform radiative transfer calculations using both the large velocity gradient (LVG) approximation and a 3-D model evaluated on a 1-D grid. The LVG approximation assumes that each emitting position in the cloud can only be absorbed by adjacent material, since more distant material is doppler shifted out of the emission line profile. For the geometric model, we consider a uniform density sphere of fixed radius 2.35\,pc \citep[to give a diameter, $4.7$\,pc, consistent with][]{federrath_2016} evaluated on a 1-D grid.\footnote{Note that the LVG calculation also assumes spherical symmetry.} We employ the NLTE statistical equilibrium solver in the Monte Carlo radiation transport code \textsc{torus} \citep{rundle_2010}, which is similar to that of \citet{hogerheijde_2000}. This approach accounts for the 3-D structure of the cloud by assuming spherical symmetry. The level populations are computed in each cell using either LTE or NLTE assumptions. In LTE the level populations are trivially calculated analytically using the Boltzmann distribution. The NLTE level populations are calculated iteratively. They are initialised to LTE conditions, then ray tracing is performed to determine the radiation field and recalculate the level populations. This process is repeated until level populations converge. To estimate the brightness temperature and optical depth, a ray at the line centre is traced through the centre of the sphere along the observers line of sight. All of the material is assumed to be centred on the same rest velocity with a constant $4.4$\,\kms \ turbulent line width.

The resulting grid in the ray tracing approach, both in LTE and NLTE, is given in the upper two panels of Figure \ref{Figure:tau}. The single white point represents the canonical HNCO abundance and derived mean density of \brick. The lower left panel of Fig.~\ref{Figure:tau} shows the NLTE result in the LVG \textsc{radex} calculations. In this panel the colour bar also represents the brightness temperature distribution and the black dotted contour in each plot denotes the region where $\tau=1$. 

In the ray tracing models, there is no component of the parameter space that is both optically thick and has a low enough brightness temperature to be consistent with the observed $T_{B}$ distribution extracted using \scouse \ throughout \brick \ (see the bottom-right hand panel). In the LVG models, there is a very small region of the parameter space where a solution is possible (hatched contour; $\tau>1$ and $T_{B}<5$\,K, where this latter condition represents three standard deviations from the mean \scouse-derived brightness temperature $\sim1.75$\,K). However, the abundance of HNCO would have to be enhanced above the value observed towards dense cores by \cite{churchwell_1986} and \citet{zinchenko_2000} by at least 1-2 orders of magnitude.

The above analysis comes with the caveats that our calculations assume spherical symmetry, as well as a uniform abundance, density, and temperature. For more realistic conditions, there may be localised regions within \brick \ where the line becomes optically thick. However, we conclude that, in the absence of independent evidence for an extremely elevated HNCO abundance, the line is likely to be optically thin throughout the majority of the cloud. Even if the abundance is highly elevated, there appears to be only specific, unlikely geometric and kinematic structures of the cloud consistent with the line being thick. We therefore conclude that the HNCO line can be used as a reliable tracer of the gas dynamics of \brick, where its emission is widespread (as it is throughout the CMZ; \citealp{henshaw_2016}).

\subsection{Cloud-cloud collision hypothesis: \brick \ is a superposition of two molecular gas clouds undergoing a collision}\label{discussion:collision}

\subsubsection{Scenario 2a: \brick \ formed following a cloud-cloud collision}\label{discussion:higuchi}

It has been argued that collisions between either atomic or molecular gas clouds in the ISM may play a role in both the formation and/or agglomeration of clouds, and in the triggering of star formation events, particularly high-mass star and star cluster formation (see e.g. \citealp{dobbs_2014} and references therein). Hence there is considerable interest in categorising the observational characteristics of such phenomena. However, inferring cloud-cloud collisions from observations is challenging. Numerical simulations can give important insight to some of the characteristics of cloud collisions \citep{inoue_2013, haworth_2015, haworth_2015b}, however, often these characteristics are not unique. 

\citet{higuchi_2014} invoked cloud-cloud collisions as a possible formation mechanism for \brick. The authors identified the presence of a shell (radius $\sim1.3$\,pc) within \brick, in addition to large velocity gradients ($\sim~20$\,\vel) and broad velocity dispersions (of the order 30-40\,\kms). Comparing with simulations of cloud-cloud collisions \citet{higuchi_2014} conclude that the shell structure may have been caused by the collision between two clouds of different mass and radii, resulting in the formation of a dense cloud which we now observe as \brick. 

The shell structure identified is that which we identify as the `C'-shaped arc belonging to tree B in \S~\ref{results:int_acorns}. Our kinematic analysis reveals that the arc is exclusively associated with tree B. The fact that this feature only accounts for a small fraction of the total HNCO emission observed throughout \brick \ (roughly $\sim17$\% of all fitted components) indicates that it is unlikely a relic of the cloud formation process. While we can not rule out the possibility that \brick \ has formed via a cloud-cloud collision, based on our combined \scouse \ and {\sc acorns} decomposition, we dispute that the presence of the arc is residual evidence of the formation process of the cloud as a whole as hypothesised by \citet{higuchi_2014}. More generally, it is unclear whether cloud-cloud collisions occur frequently enough, and on a short enough timescale, for them to be a dominant physical mechanism in the formation of clouds \citep{jeffreson_2018, jeffreson_2018a}. Instead, it has recently been suggested that large-scale instabilities may provide a plausible mechanism for the formation of massive and dense molecular clouds in the CMZ, both in observations \citep{henshaw_2016c} and in simulations \citep{sormani_2018}. 

\subsubsection{Scenario 2b: \brick \ is currently undergoing a cloud-cloud collision}\label{discussion:johnston}

The concept of a cloud-cloud collision in \brick \ is not new. It was first proposed by \citet{lis_1998b} (and further expanded by \citealp{lis_2001}) as a possible explanation for both the presence of multiple line-of-sight velocity components and the observed widespread emission from shocked gas tracers (see also \citealp{kauffmann_2013}). \citet{lis_2001} argued that the collision occurs between a molecular gas component observed at $\sim20$\,\kms \ (cf. tree B) and another at $\sim40$\,\kms \ (cf. tree C). 

\citet{rathborne_2014a} postulated that for a cloud collision one may expect to observe two velocity components and a central zone of hot and shocked gas at the collision interface. The authors point out that while multiple velocity components are indeed observed in the dense gas tracers in single-dish observations, the same is true for those tracing hot and shocked gas. The hot and shocked gas tracers (such as SiO) are not isolated to a single region within the cloud. Instead they have a similar distribution and kinematic profile to the optically thin gas tracers. In the absence of a specific collision region, \citet{rathborne_2014a} conclude that the single cloud interpretation is more consistent with their observations (\S~\ref{discussion:bakedalaska}). 

Using high-spatial resolution interferometric observations however, \citet{johnston_2014} identified the presence of shocked gas tracers and elevated gas temperatures towards the southern portion of the cloud. This emission spatially coincides with our tree D. In investigating the kinematics, the authors noted that there is an additional velocity component, situated at $\sim70$\,\kms \ which is spatially coincident with the emission from shocked gas. These two velocity components `connect' in PV space, which led \citet{johnston_2014} to suggest that they may be interacting. 

This latter possibility was discussed by \citet{henshaw_2016}, who compared the observed kinematics of the CMZ with three different geometries aiming to describe the three-dimensional structure of the CMZ. \citet{henshaw_2016} concluded, albeit using much coarser spatial resolution observations ($1'\sim2.4$\,pc) than \citet{johnston_2014}, that the component observed at $\sim70$\,\kms \ is unlikely to be associated with \brick. The emission from the $\sim70$\,\kms \ component is morphologically distinct from that of \brick \ (despite overlapping in projection), and is more extended (with projected extent $>150$\,pc), appearing to connect to the molecular clouds closest in projection to Sgr A$^{*}$ (i.e. the 20\,\kms \ and 50\,\kms \ clouds). In each of the model geometries discussed by \citet{henshaw_2016}, the $70$\,\kms \ component is unrelated to \brick. Given the observational evidence that is currently available, we therefore conclude that \brick \ and the $70$\,\kms \ velocity component are most likely spatially distinct and non-interacting.

Despite the aforementioned discrepancy with the 70\,\kms \ component, we can not rule out the possibility that interaction between sub-clouds within \brick. The location of elevated gas temperatures and shocked gas emission identified by \citet{johnston_2014} is spatially coincident with our tree D, which sits at the interface of trees B and C (caution: in PPV space). Indeed, our analysis shows that this location in tree D displays an enhancement of HNCO emission (referred to as the `tiled bar' in \citealp{mills_2015}; cf. the bottom right hand panel of Fig.~\ref{Figure:intfield}). Moreover, Fig.~\ref{Figure:disp_acorns} demonstrates that velocity dispersions ($\sigma_{v_{los}, {\rm 1D}}$) measured within tree D are on average greater than those measured throughout the other identified components. This could indicate that the interaction of sub-structure within \brick \ may play an important role in setting the internal dynamics of the cloud as well as its appearance in shocked gas tracers (see also \citealp{lis_2001, kauffmann_2013}).

\section{The ACORNS view of \brick: `The Brick' is not a \emph{brick}}\label{discussion:acorns}

The kinematic analysis presented in \S~\ref{results:global} and \S~\ref{results:local} provides new and unique insight into the structure of \brick \ and the physical processes that are important (or unimportant) in shaping its appearance. The discussions presented in \S~\ref{discussion:bakedalaska} and \S~\ref{discussion:thick} enable us to conclude that, globally, emission from the HNCO $J=4(0,4)-3(0,3)$ transition $\sim3$\,mm is both likely to be optically thin and not widely depleted. Consequently HNCO is likely a reliable tracer of the internal structure and dynamics of the cloud. Our interpretation is therefore that, rather than a single, coherent, centrally-condensed cloud with depletion in its cold interior, \brick \ is a complex, hierarchically-structured molecular cloud exhibiting an intricate network of velocity components situated along the line-of-sight. `The Brick' is not a \emph{brick}.

Despite the aforementioned interpretation, one should always approach the connection between PPV space and true physical 3-D space with caution (as demonstrated by e.g. \citealp{beaumont_2013, clarke_2018}). However, both the arc (top right hand panel in Fig.~\ref{Figure:intfield}) and the `tilted bar' (bottom right) have both been identified in earlier works on \brick, in a variety of molecular lines \citep{higuchi_2014,mills_2015}. {\sc acorns} has uniquely provided the first evidence that these features: i) were also present in datasets in which they had previously not been identified, but were simply masked by the kinematic complexity of the data; and ii) are coherent in both (projected) space, velocity, and velocity dispersion. A key result of our analysis therefore is that {\sc acorns} has blindly identified structures that appear to be both physically meaningful and statistically different from one another, evident through their morphologically distinct emission features (Fig.~\ref{Figure:intfield}) as well as their differing internal dynamics (Figs.~\ref{Figure:vlsr} and \ref{Figure:disp_acorns}). 

So what is shaping the structure of the cloud? It is likely that \brick \ is a product of its complex and dynamic environment. A key result of recent hydrodynamical simulations is that the small scale cloud structure of \brick \ is consistent with the cloud being sculpted by the Galactic dynamics of the CMZ \citep{dale_2019, kruijssen_2019}, but see also the simulations of \citealp{sormani_2018}, where gas clouds are clearly influenced by orbital dynamics). A side-by-side comparison between the dust continuum observations presented in Fig.~\ref{Figure:continuum} \citep{rathborne_2014} and simulated ALMA observations of clouds orbiting the Galactic centre gives good qualitative agreement in terms of global morphology and the complex spatial structure of \brick \ (\citealp{kruijssen_2019}, see their Fig. 6). These simulations demonstrate that high column densities, global velocity gradients, flattened cloud morphology, and inclination on the plane of the sky all naturally occur as a result of the influence of the background gravitational potential and shearing motions induced by eccentric orbits. 

\input{Figures/Figure_14_brick_context.tex}

In addition to the large-scale orbital dynamics that may shape the cloud structure, there may be further external factors that play a significant role in shaping the structure and evolution of \brick. In Figure\,~\ref{Figure:brick_annotate}, we show a zoom of the three-colour \emph{Spitzer} GLIMPSE image in Fig.~\ref{Figure:brick}, however, here we highlight some of the main features in \brick's surrounding environment. As can be seen, the cloud overlaps in projection with the prominent supernova remnant G0.30+0.00 (also, G000.3+00.0, G0.33+0.04, G0.4+0.1; red ellipse; \citealp{kassim_1996,larosa_2000}). Additionally, \citet{ponti_2015} identify another supernova remnant candidate, G0.224+0.032,\footnote{Note this is labelled as G0.224-0.032 in \citet{ponti_2015}.} located directly to the (Galactic) west of \brick \ (dashed green ellipse). The high extinction of the cloud means that the soft X-ray emission is partially obscured by the cloud. Nevertheless, \citet{ponti_2015} argue that the properties of G0.224+0.032 are consistent with those of a supernova remnant, but that the true size and energy are difficult to estimate due to the obscuration.

Also indicated in Figure~\ref{Figure:brick_annotate} are the positions of massive stars located towards \brick. The filled cyan points indicate the locations of Paschen $\alpha$ emitting sources obtained with the Hubble Space Telescope/Near-Infrared Camera and Multi-Object Spectrometer and Multi-Object Spectrometer (HST/NICMOS) identified by \citet{dong_2011} and the filled red circles indicate the locations of Wolf-Rayet stars, O supergiants, and B supergiants, obtained by \citet{mauerhan_2010}. \citet{dong_2011} argue that the majority of these sources are most likely evolved massive stars ($M_{*}>7$\,M$_{\odot}$) with strong stellar winds. The source locations are categorised into four different groups: i) \& ii) those associated with the young massive clusters the Arches and Quintuplet; iii) those located with the nuclear star cluster; iv) and field sources outside the main clusters. Although the footprint of the observations does not include \brick, there is a considerable number of field massive stars spread throughout the observed region. Feedback from such massive stars has the potential to influence the molecular gas in this environment. Indeed, it has been argued that the O4-6 supergiant, which is situated immediately to the (Galactic) east of \brick \ (that which lies within the boundary of the red ellipse in Fig.~\ref{Figure:brick_annotate}), may be responsible for the ionisation of the exterior of the cloud in this direction \citep{mills_2015}.

Although projection effects may play a role in determining whether or not these features indeed influence the structure of \brick; the fact remains that \brick \ displays complex internal dynamics, both in terms of velocity gradients and supersonic velocity dispersions, as well as elevated gas temperatures, and a prevalence of emission from tracers of shocked gas. The complex interplay of these large-scale (e.g. Galactic dynamics) and comparatively small-scale (e.g. feedback) effects may all contribute in sculpting the physical structure of \brick, and therefore its star formation potential. 

%% file: Figures/Figure_13_RT.tex
\begin{figure*}
\begin{center}
\includegraphics[trim = 10mm 5mm 10mm 6mm, clip, width = 0.48\textwidth]{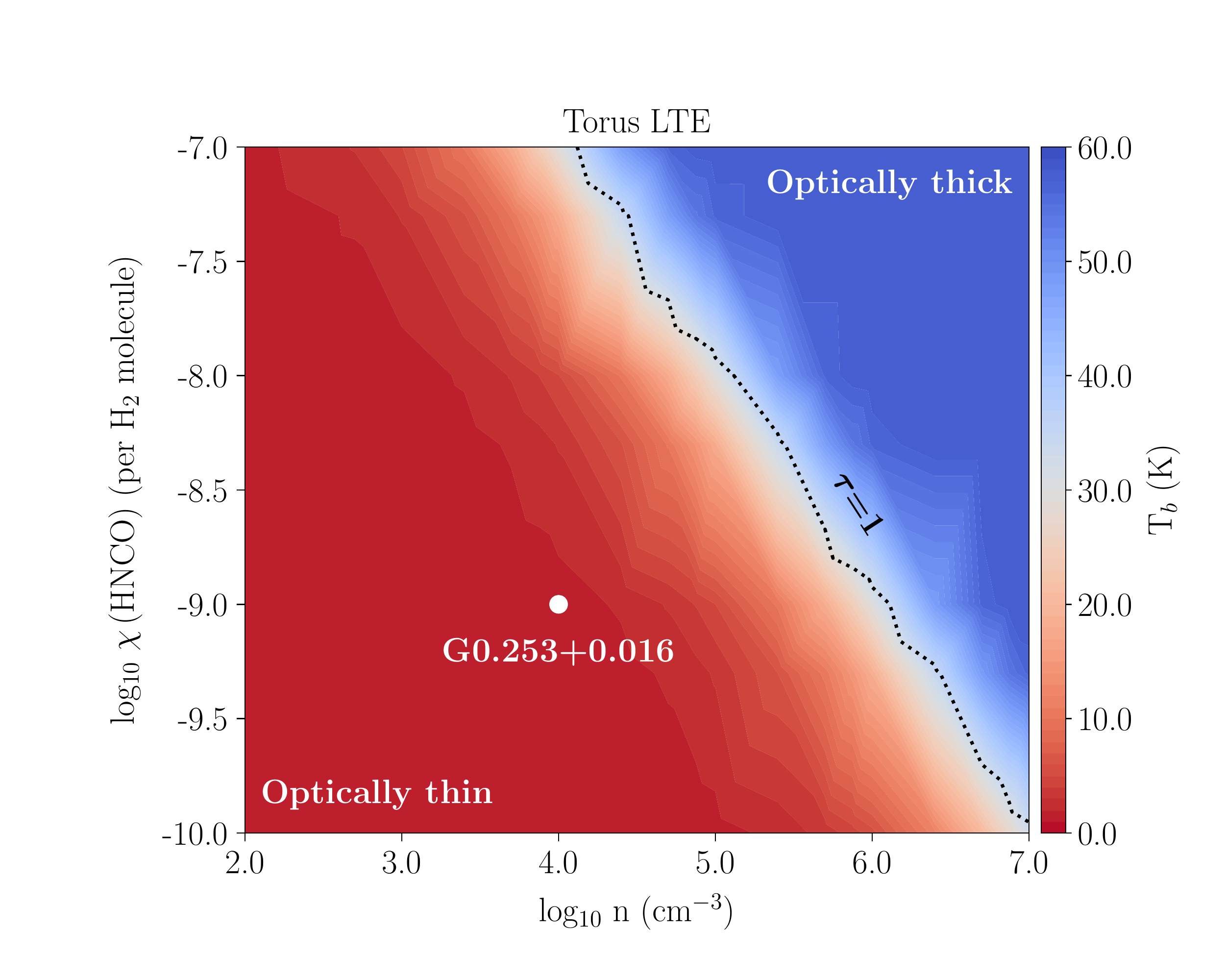}
\includegraphics[trim = 10mm 5mm 10mm 6mm, clip, width = 0.48\textwidth]{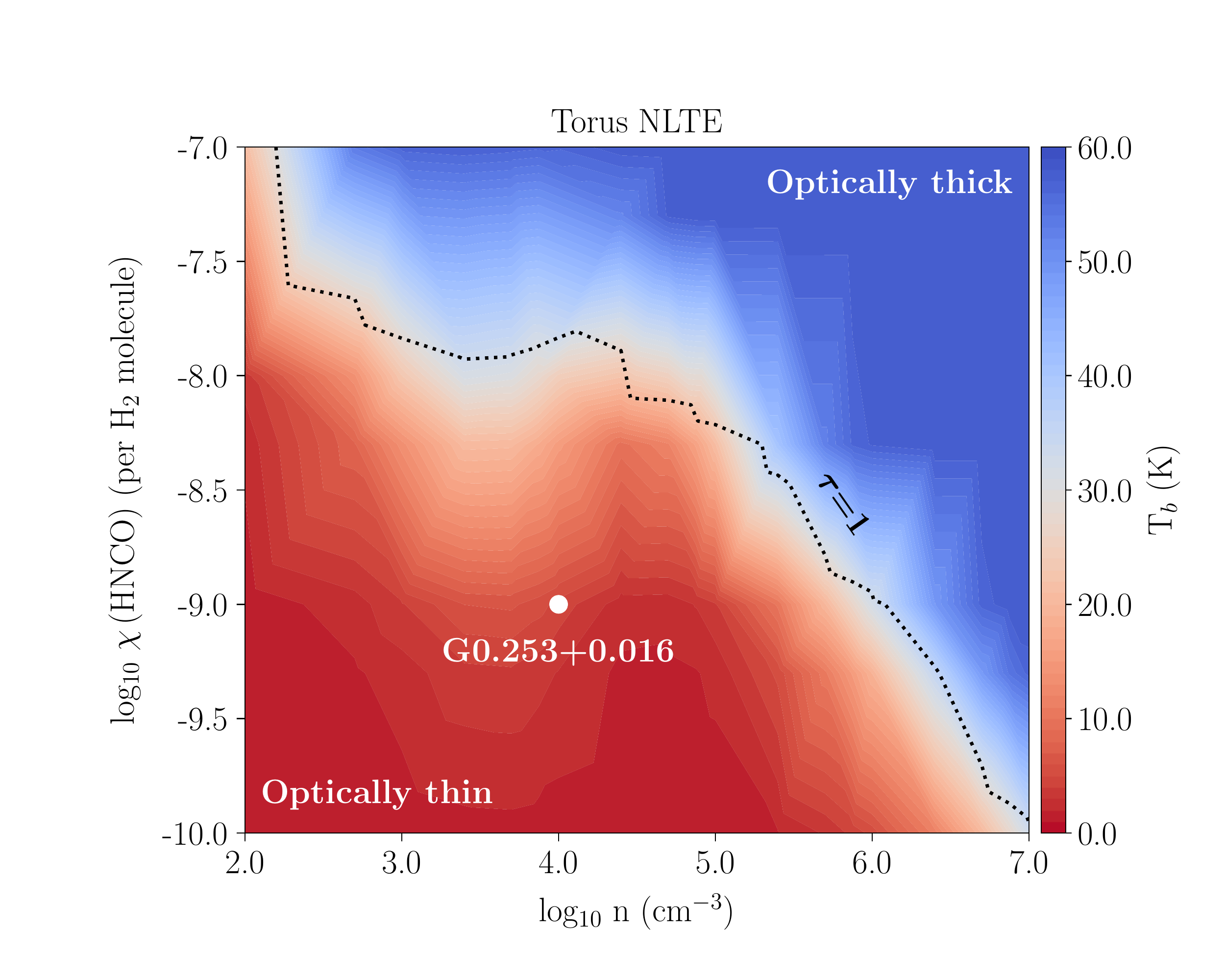}
\includegraphics[trim = 10mm 5mm 10mm 6mm, clip, width = 0.48\textwidth]{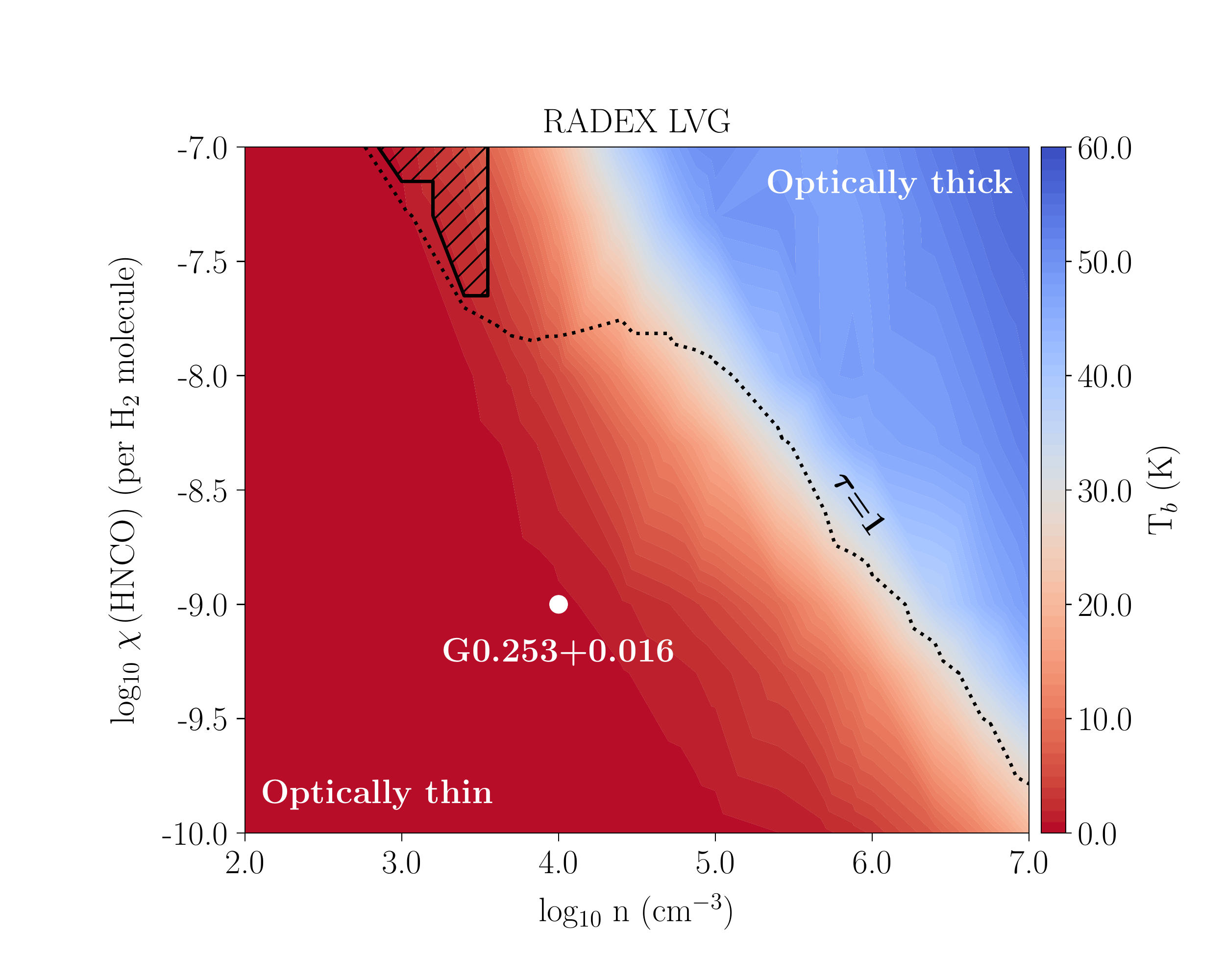}
\includegraphics[trim = 10mm 5mm 10mm 6mm, clip, width = 0.48\textwidth]{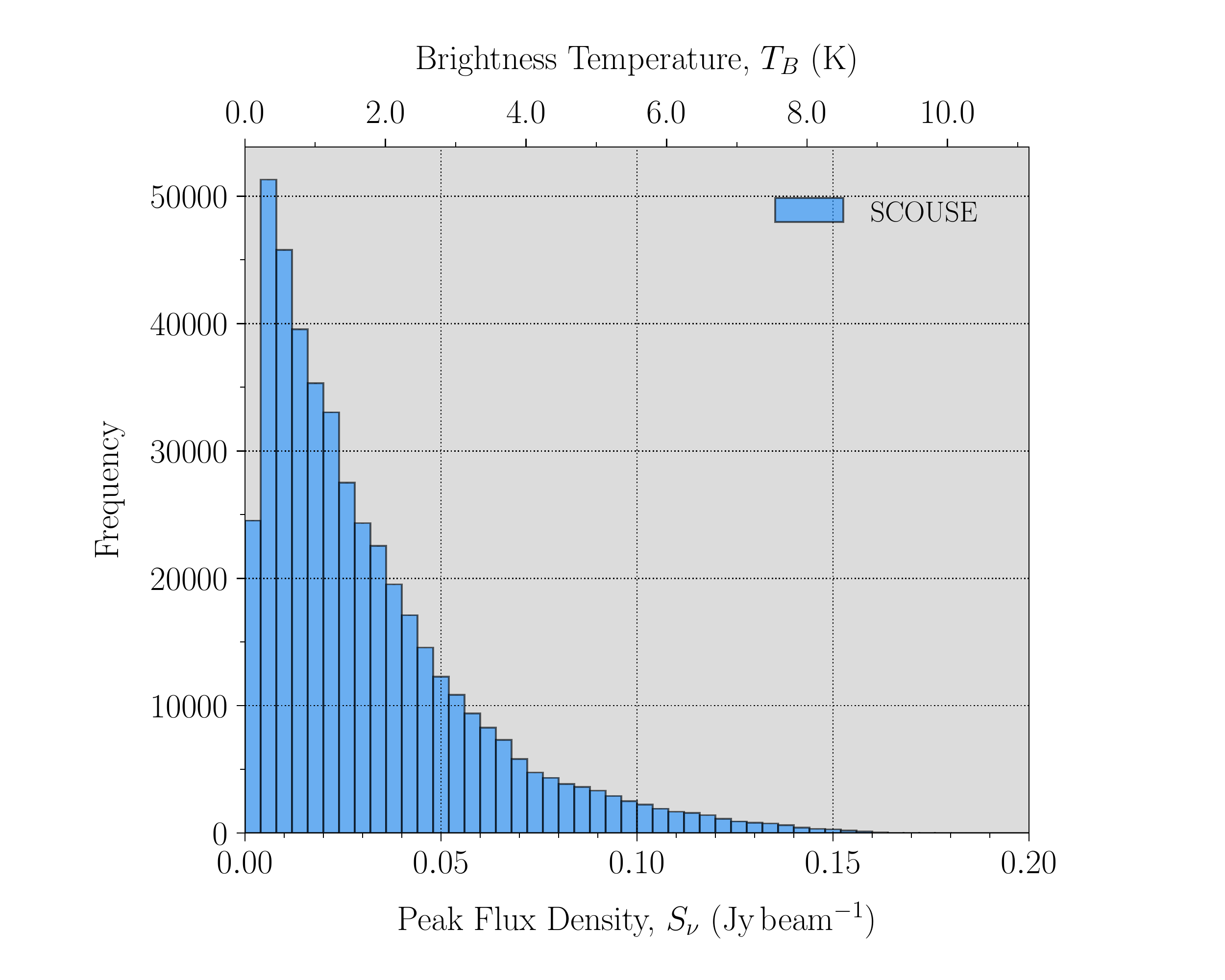}
\end{center}
\caption{Top panels: The brightness temperature as a function of molecular abundance and the number density. The left and right panels follow the ray tracing approach using {\sc torus} \citep{rundle_2010} in LTE and NLTE respectively. The dotted black contour denotes an optical depth of unity and the white point represents the likely conditions in \brick. Bottom left: The lower panel employs an LVG approach computed with {\sc radex} \citep{vandertak_2007}. The hatched region denotes the region in which the medium is optically thick and the brightness temperature is less than 5\,K. Bottom right: A histogram of the peak flux density, $S_{\nu}$, of all spectral components extracted using {\sc scousepy}. The x-axes are given in both Jy\,beam$^{-1}$ (bottom) and K (top) using a conversion factor of 55.8\,K\,(Jy\,beam$^{-1}$)$^{-1}$.}
\label{Figure:tau}
\end{figure*}

%% file: Figures/Figure_14_brick_context.tex
\begin{figure}
\begin{center}
\includegraphics[trim = 1mm 4mm 0mm 4mm, clip, width = 0.45\textwidth]{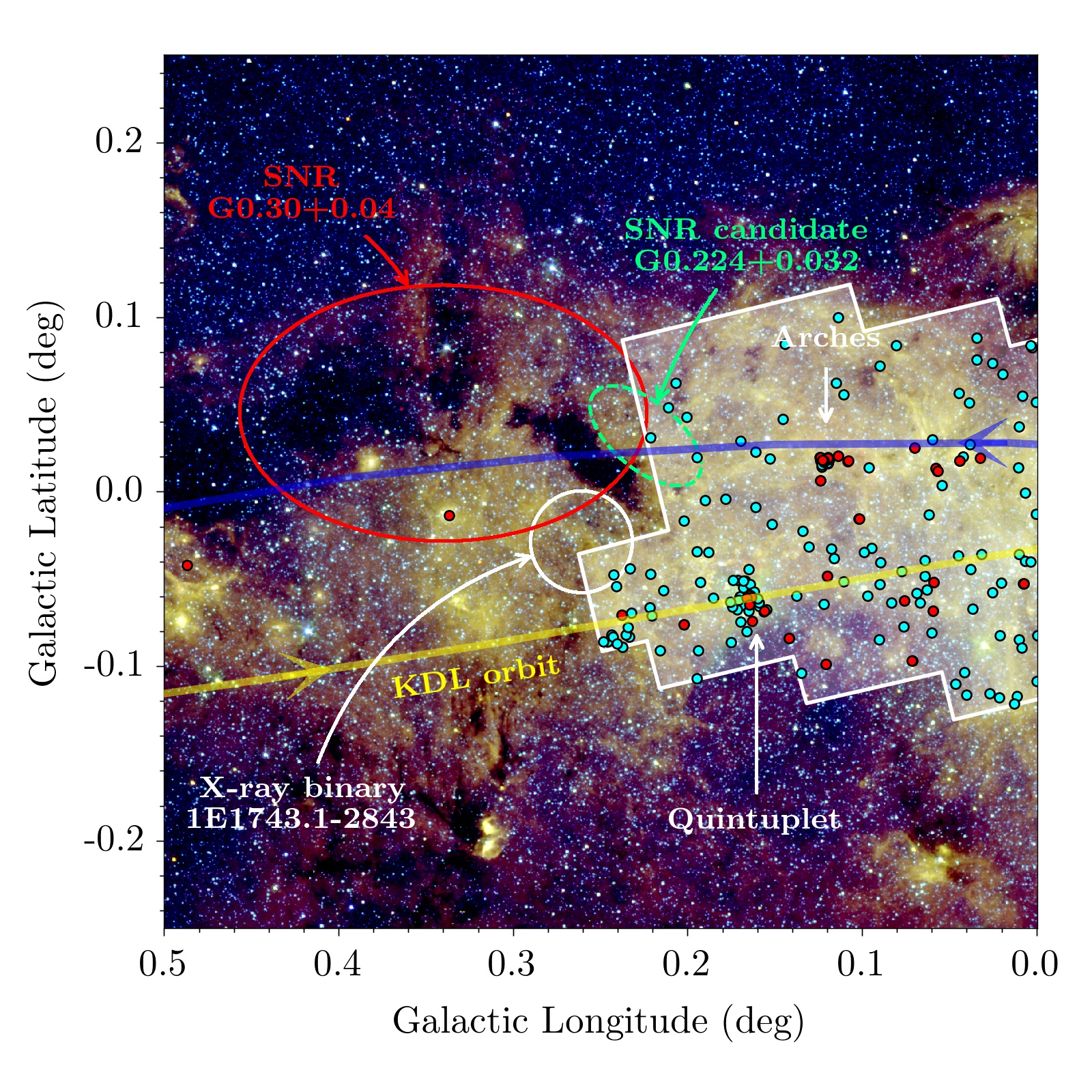}
\end{center}
\caption{A close up view of the immediate environment of \brick. The background image is equivalent to Figure~\ref{Figure:brick}, but here we display some of the additional external factors which may play a significant role in the evolution of \brick. The red ellipse highlights the prominent supernova remnant, ${\rm G}0.30+0.04$ \citep{kassim_1996,larosa_2000} and the dashed green ellipse is a supernova remnant \emph{candidate}, ${\rm G}0.224+0.032$ \citep{ponti_2015}. The white circle highlights the location of X-ray binary $1{\rm E} 1743.1-2843$ \citep{porquet_2003}. The size of the circle corresponds approximately to the spatial extent of the emission observed with $XMM-Newton$ (see \citealp{ponti_2015}). Filled cyan points indicate the locations of Paschen $\alpha$ emitting sources obtained with the HST \citep{dong_2011}. The white shaded region shows the footprint of the HST observations. The filled red circles indicate the locations of Wolf-Rayet stars, O supergiants, and B supergiants, obtained by \citep{mauerhan_2010}. Finally, the blue (near side) and yellow (far  side) lines indicate the orbital model of the CMZ derived by \citet{kruijssen_2015}, with the arrows depicting the direction of gas motion.  }
\label{Figure:brick_annotate}
\end{figure}

%% file: Tex/Conclusions.tex
\section{Conclusions}\label{Section:Conclusions}

We have performed a comprehensive study of the dynamics and physical structure of \brick. To facilitate this study we have developed two pieces of software, both of which we make available to the community. The first, \scouse, is a redevelopment of the spectral line fitting algorithm first presented by \citet{henshaw_2016}. The second, {\sc acorns}, is a hierarchical clustering algorithm designed specifically for use with discrete data such as that output by \scouse. Combined, these algorithms have helped us to develop a new view of \brick. Our main conclusions are summarised below.
\begin{enumerate}
\item We have performed a full kinematic decomposition of the HNCO ALMA data, quantifying and measuring the spectral lines despite their well-known complexity. Globally, the kinematic structure of \brick \ appears to show two dominant features in position-position-velocity space, one situated at $\sim35-50$\,\kms \ and another which ranges from $\sim0-30$\,\kms. Both features have global velocity gradients in the north-south direction (in equatorial coordinates or north east-south west in Galactic) following the major axis of the cloud. However, the magnitude of the velocity gradient across the latter feature is about a factor of $\sim2$ greater than that across the former. This presents a more complex picture than that of a singular cloud exhibiting the hallmarks of rotation as has been suggested in previous works (e.g. \citealp{rathborne_2014a,federrath_2016}).\\
\item A striking feature of our \scouse \ decomposition is the `wiggly' nature of the kinematic substructure. Oscillatory velocity gradients appear ubiquitously throughout the cloud. However, unlike those identified on larger scales \citep{henshaw_2016c}, which display a characteristic wavelength and amplitude, these oscillations appear more stochastic. We will quantify these oscillations further in a future publication (Henshaw et al. in preparation).\\
\item Velocity dispersions measured along the line-of-sight (extracted directly from spectral line fitting) are a factor of $\sim2$ below those derived from moment analysis due to the presence of multiple velocity components identified within the spectra. On average we measure $\langle\sigma_{v_{los}, {\rm 1D}}\rangle = 4.4$\,\kms, with a standard deviation of $2.1$\,\kms. Assuming a fixed temperature of $\sim60\,$K, this translates into a Mach number estimate of $\mathcal{M}_{\sigma_{v_{los}}, {\rm 3D}}\sim16.5$. Although these velocity dispersions are broader than those predicted from the steep linewidth-size relationships of \citet{shetty_2012} and \citet{kauffmann_2017a}, these results add to mounting evidence for the existence of narrow ($\lesssim$ a few \kms) lines on small spatial scales in CMZ clouds. \\
\item $\sim98\%$ of the \scouse \ decomposition data are clustered using {\sc acorns}. We find that the dynamics are dominated by four main features containing $>50\%$ of the data.\\
\item There are important differences between the four main hierarchical structures (referred to as `trees'). The dominant tree (C), situated at a mean velocity of $\langle v \rangle\sim37.0$\,\kms, is most similar to the intensity distribution observed in dust continuum observations giving the cloud its physical appearance as we observe it on the plane of the sky. Tree B ($\langle v \rangle\sim16.5$\,\kms) exhibits a prominent arc shaped feature which has been noted in previous studies \citep{higuchi_2014, mills_2015}. Out of the two smaller trees, D ($\langle v \rangle\sim33.1$\,\kms) displays a prominent linear feature associated with elevated gas temperatures and velocity dispersions. Finally, tree A ($\langle v \rangle\sim2.9$\,\kms) extends towards the north of the cloud in the direction of dust ridge cloud `b' which has a similar velocity $\sim3.4$\,\kms \ \citep{henshaw_2016}. While many of these features have been identified previously in the literature, a key and unique element of our analysis is that {\sc acorns} provides the first evidence that these features are coherent in both (projected) space and velocity. Moreover, {\sc acorns} has extracted these features blindly from the observational data. This indicates that these features were already present in data such as the HNCO emission initially presented by \citet{rathborne_2015}, but were masked by the kinematic complexity of the cloud. \\
\item We compare the trees' mean line-of-sight velocity dispersions with the fluctuations in the centroid velocity across the plane of the sky, finding $\langle \sigma_{v_{los}, {\rm 1D}} \rangle = \{5.3, 4.9, 4.0, 5.8\}$\,\kms \ and $\sigma_{v_{pos}, {\rm 1D}} = \{3.5, 5.2, 4.6, 4.5\}$\,\kms, respectively. The ratio of these two measurements yields $\langle \sigma_{v_{los}, {\rm 1D}} /\sigma_{v_{pos}, {\rm 1D}} \rangle=1.2\,\pm\,0.3$. We speculate that this isotropy in the velocity fluctuations may contain important information regarding the cloud geometry. Namely, that the line-of-sight extent of the cloud components are approximately equivalent to that in the plane of the sky.\\
\item We argue that emission from the $J=4(0,4)-3(0,3)$ transition of HNCO is (globally) optically thin, and therefore is a good tracer of the internal dynamics of the cloud overall. We disfavour the interpretation that \brick \ is a centrally-condensed molecular cloud with depletion in its cold interior, as was proposed by \citet{rathborne_2014a}, since the position-position-velocity profile would necessitate either a strong increasing gradient in density from south to the north of the cloud (or alternatively decreasing temperature), which is not observed. \\
\item We do not rule out the possibility that the merger of sub-structures within \brick \ may play an important role in producing shocked gas emission, elevating the gas temperature, and raising the velocity dispersion of the gas. However, we dispute the conclusion of \citet{higuchi_2014} that the arc emission feature is evidence that \brick \ has formed via cloud-cloud collisions. Our kinematic analysis demonstrates that emission from the arc feature is just a small fraction of the total cloud emission. Therefore it is unlikely that this is a relic signature of the formation mechanism of the cloud as a whole. \\
\item Finally, we discuss our findings in the context of the large-scale kinematics of the CMZ. \brick \ is a complex, hierarchically-structured molecular cloud exhibiting an intricate network of velocity components situated along the line-of-sight; `the Brick' is not a \emph{brick}. We argue that the morphology is most likely a product of the tangled interplay of both Galactic dynamics and feedback present in the CMZ. Recent simulations of molecular clouds orbiting galactic centres indicate that complex cloud structure is a natural outcome of the influence of the background gravitational potential and shearing motions induced by eccentric orbits (\citealp{sormani_2018, dale_2019, kruijssen_2019}). Detailed kinematic analysis of such simulations is highly promising for further constraining the physical mechanisms shaping molecular cloud structure within the CMZ. 
\end{enumerate}

In the near future, studies such as the CMZoom survey (the Sub-Millimeter Array's legacy survey of the CMZ; \citealp{battersby_2017}, Battersby et al. in preparation) as well as future ALMA surveys will facilitate a uniform description of molecular cloud dynamics throughout the CMZ. This will help to provide a statistical understanding of the earliest phases of star formation in this complex and dynamic environment. 

%% file: Tex/Acknowledgements.tex
\section*{Acknowledgements}
\addcontentsline{toc}{section}{Acknowledgements}

We thank the referee for their careful reading of the manuscript and comments. We would like to thank Jill Rathborne for making the data product used in this paper available for use. Additionally, JDH would like to thank Paola Caselli, Seamus Clarke, Christoph Federrath, Morgan Fouesneau, Iskren Georgiev, John Ilee, Dimitry Semenov, and Juan Soler, for fruitful discussions during the preparation of this paper.

ATB acknowledges funding from the European Union's Horizon 2020 research and innovation programme (grant agreement No 726384). CB gratefully acknowledges support by the National Science Foundation under Grant No. 1816715. HB acknowledges support from the European Research Council under the Horizon 2020 Framework Program via the ERC Consolidator Grant CSF-648505. TJH is funded by an Imperial College junior research fellowship. JMDK gratefully acknowledges funding from the German Research Foundation (DFG) in the form of an Emmy Noether Research Group (grant number KR4801/1-1) and from the European Research Council (ERC) under the European Union's Horizon 2020 research and innovation programme via the ERC Starting Grant MUSTANG (grant agreement number 714907). MR gratefully acknowledges funding from the European Union's Horizon 2020 research and innovation program under grant agreement No 639459 (PROMISE).

\subsection*{Facilities and Data}

This research has made use of NASA's Astrophysics Data System. This paper makes use of the following ALMA data: ADS/JAO.ALMA\#2011.0.00217.S. ALMA is a partnership of ESO (representing its member states), NSF (USA) and NINS (Japan), together with NRC (Canada) and NSC and ASIAA (Taiwan), in cooperation with the Republic of Chile. The Joint ALMA Observatory is operated by ESO, AUI/NRAO and NAOJ. 

\subsection*{Software}

This research made use of:
\begin{itemize}

\item {\sc acorns} (\url{https://github.com/jdhenshaw/acorns})
\item {\sc astropy} \citep{astropy}
\item {\sc matplotlib} \citep{Hunter:2007}
\item {\sc numpy} \citep{van2011numpy}
\item {\sc pyspeckit} (\url{http://pyspeckit.bitbucket.org})
\item {\sc scikit-learn} \citep{scikit-learn}
\item {\sc scipy} \citep{jones_scipy_2001}
\item \scouse \ (\url{https://github.com/jdhenshaw/scousepy})
\item {\sc spectralcube} (part of the radio-astro-tools package hosted at \url{http://radio-astro-tools.github.io/})
\end{itemize}

%% file: Tex/scouse_decomp.tex
\section{SCOUSEPY decomposition}\label{SCOUSE}

Here we include additional information regarding the \scouse \ fitting procedure. The left hand panel of Fig.~\ref{Figure:cov} displays the result of the new implementation for setting variable spectral averaging area (SAA) sizes based on spectral complexity. The procedure is outlined in \S~\ref{results:scouse}. Briefly however, we plot a map of $\Delta v_{\rm m}\equiv|v_{1}-v_{peak}|\sim 0$, where $v_{1}$ is the first order moment and $v_{peak}$ is the velocity of the channel containing the peak emission. We also plot a histogram of the individual pixel values. In the case of the ALMA HNCO observations of \brick \ we divide the data up into three logarithmically-spaced $\Delta v_{\rm m}$ bins, which we use to define the size of our SAAs and are overlaid on the $\Delta v_{\rm m}$ map. This enables the user to pay close attention to regions which have line profiles with a greater degree of complexity. The right hand panel of Fig.~\ref{Figure:cov} highlights the locations which have best-fitting solutions, with each pixel being colour-coded according the number of velocity components identified at that location. Table~\ref{Table:global_stats} contains the statistics of our \scouse \ decomposition. 

\input{Figures/Figure_A1_scouse.tex}
\input{Tables/Table_global_stats.tex}

%% file: Figures/Figure_A1_scouse.tex
\begin{figure*}
\begin{center}
\includegraphics[trim = 0mm 5mm 0mm 0mm, clip, width = 0.48\textwidth]{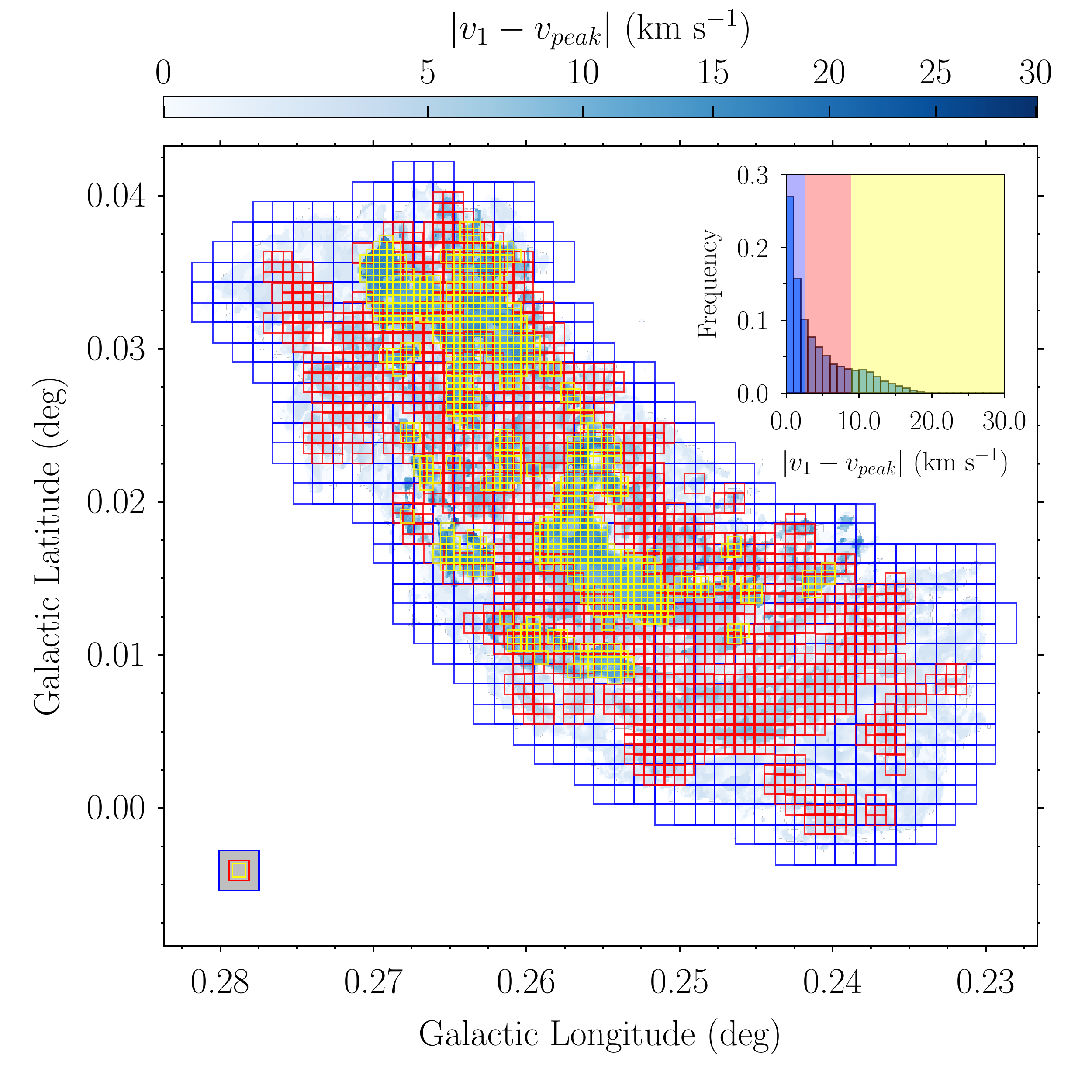}
\includegraphics[trim = 0mm 5mm 0mm 0mm, clip, width = 0.48\textwidth]{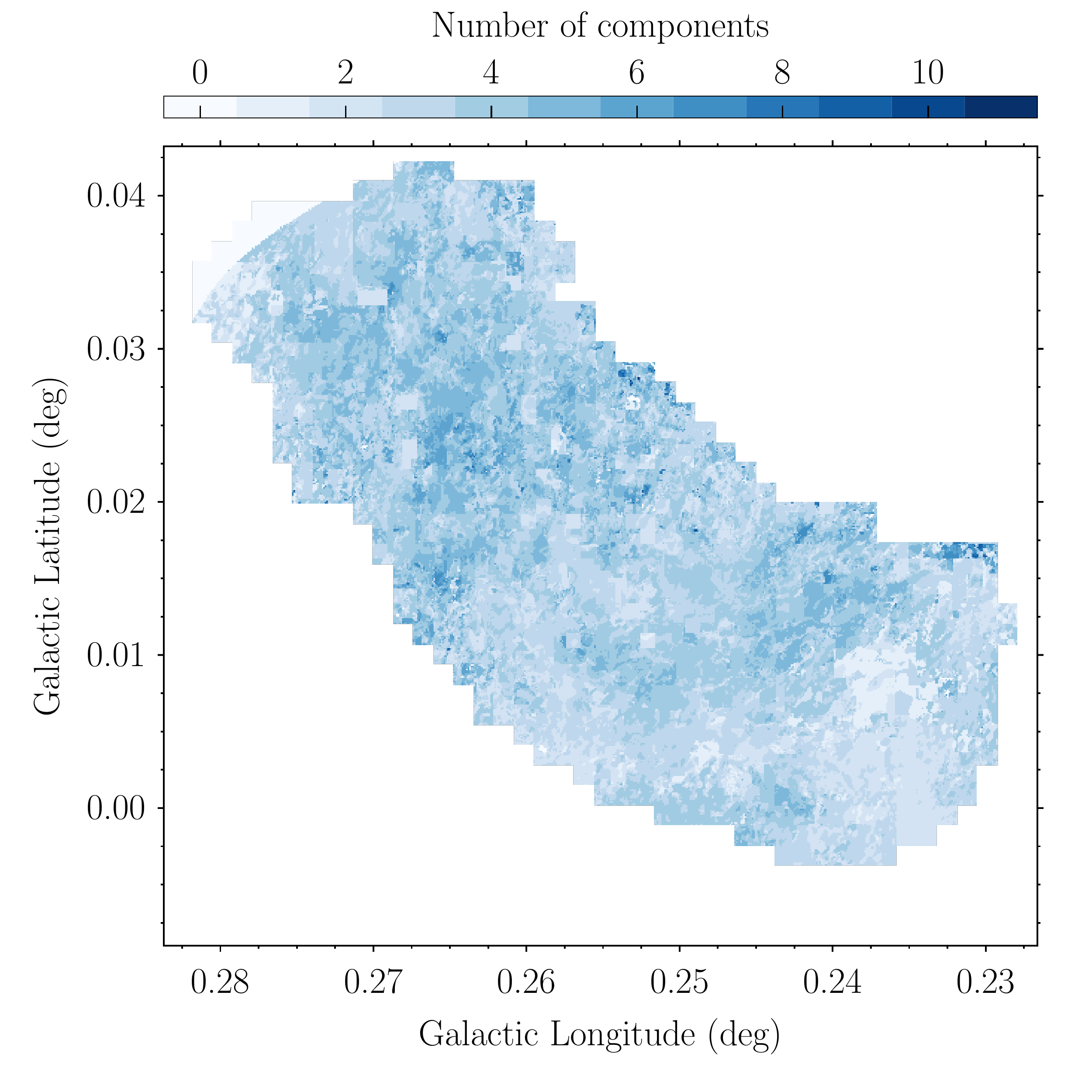}
\end{center}
\caption{Left: \scouse \ spectral averaging area (SAA) refinement based on spectral `complexity'. The background image shows a map of the absolute difference between the intensity-weighted average velocity (moment 1; $v_{1}$) and the velocity at peak emission ($v_{peak}$), which serves as a metric for spectral complexity ($\Delta v_{\rm m}=|v_{1}-v_{peak}|\sim0.0$ for a singly-peaked `simple' spectrum). The inset image is a histogram of the pixel values. \scouse \ generates $n$ logarithmically-spaced bins, based on this complexity measure. The size of the SAA is then refined according to how complex the spectra are at a given location in the map (coloured boxes). Right: The number of velocity components extracted by \scouse \ from each location in the map.}
\label{Figure:cov}
\end{figure*}

%% file: Tables/Table_global_stats.tex
\begin{table}
	\caption{\scouse: Global fitting statistics. See \S~\ref{results:scouse} for details. \vspace{0.6cm}} 
	\centering  
	\tabcolsep=0.2cm \normalsize{
	\begin{tabular}{ c  c  }
    
    \hline 
 	Output statistic &  Full resolution \\ 
	\hline  
	
	$N_{\rm tot}$ & 315219 \\ [1.0ex] 
	$N_{\rm tot,\,SAA }$ & 135020 \\ [1.0ex]
	$N_{\rm SAA }$ & 2355 \\ [1.0ex]
	$N_{\rm SAArefine}$ & [527,\,1141,\,687] \\ [1.0ex]
	$N_{\rm fit}$ & 133065 \\ [1.0ex]
	$N_{\rm comp}$ & 457264 \\ [1.0ex]
	$N_{\rm comp}$/$N_{\rm fit}$ & 3.4 \\ [1.0ex]
	$N_{\rm orig}$/$N_{\rm fit}$ (\%) & 96.4 \\ [1.0ex]
	$N_{\rm refit}$/$N_{\rm fit}$ (\%) & 2.5\\ [1.0ex]
	$N_{\rm alt}$/$N_{\rm fit}$ (\%) & 1.1 \\ [1.0ex]

	\end{tabular}
	\vspace{0.5cm}
	
\begin{minipage}{0.5\textwidth}\footnotesize{
	\centering  
	\tabcolsep=0.1cm
	\begin{tabular}{ l  l }
	$N_{\rm tot}$ & Total number of pixels in the mapped area.\\   
	$N_{\rm tot,\,SAA}$ & Total number of pixels included in the coverage.\\
	$N_{\rm SAA}$ & Total number of spectral averaging areas.\\
	$N_{\rm SAArefine}$ & Number of SAAs at each level of refinement (see text).\\
	$N_{\rm fit}$ & Total number of pixels fitted (automated). \\ 
	$N_{\rm comp}$ & Total number of components fitted. \\
	$N_{\rm comp}$/$N_{\rm fit}$ & Mean number of Gaussian components per position. \\  
	$N_{\rm orig}$/$N_{\rm fit}$ & Percentage of original fits in the final data product. \\  
	$N_{\rm refit}$/$N_{\rm fit}$ & Percentage of refitted spectra in the final data product. \\  
	$N_{\rm alt}$/$N_{\rm fit}$ & Percentage of spectra with alternative solutions selected. \\  
	\end{tabular}
}
\end{minipage}
}
\label{Table:global_stats}
\end{table}

%% file: Tex/Methodology.tex
\section{Extracting molecular gas kinematics following spectral decomposition}\label{Methodology}

As discussed in \S~\ref{Section:Introduction}, the primary aim of this study is to obtain a detailed description of the molecular gas kinematics of G0.253+0.016. To date, analyses of the gas kinematics of G0.253+0.016 have predominantly relied on techniques such as moment analysis \citep{rathborne_2015, federrath_2016}, and dendrograms \citep{kauffmann_2013}. The former technique is beneficial as it is simple and fast to implement, and it returns information on the pixel scale. However, taking an intensity-weighted average velocity along the line-of-sight results in information being lost, particularly in regions with complex LOS density and velocity structure. Conversely, the latter technique is beneficial in that complex line-of-sight structure is accounted for as the algorithm seeks to build a hierarchy of structure, which can be represented graphically in the form of a dendrogram (see e.g. \citealp{rosolowski_2008}). However, kinematic information is provided in the form of intensity-weighted average quantities relating to each structure. Further work is therefore required if one is interested in how those kinematic quantities vary with position within a given structure on the pixel scale. 

More generally, there is an array of automatic algorithms whose primary function is to parse and extract information regarding the structure of molecular clouds and their internal dynamics. These include, but are not limited to, those designed to segment and extract isolated peaks of emission for example `cores', `clumps', or `fibres' (e.g. {\sc clumpfind}, \citealp{williams_1995}; {\sc gaussclumps}, \citealp{stutzki_1990}; {\sc fellwalker}, \citealp{berry_2015}; {\sc five}, \citealp{hacar_2013}), those which target the hierarchical structure of molecular clouds (e.g. {\sc astrodendro}, \url{www.dendrograms.org}; {\sc dendrofind}/{\sc quickclump}, \citealp{wunsch_2012,sidorin_2017}; see also \citealp{miville-deschenes_2017}), those which specifically aim to extract molecular clouds (e.g. {\sc cprops}, \citealp{rosolowski_2006}; {\sc scimes}, \citealp{colombo_2015}), and those which have been used to target structure with a particular geometry, for instance filaments (e.g. {\sc disperse}, \citealp{sousbie_2011a, sousbie_2011b}; {\sc filfinder}, \citealp{koch_2015}). Despite this, there is currently no publicly available code whose primary function is to extract hierarchical structure within molecular clouds, thereby providing the important connection between cores, clumps, and clouds, but which simultaneously retains the pixel scale information needed to study variation in the kinematics throughout each member of the hierarchy. 

Our solution to this problem is the development of a new analysis tool, written in Python, named {\sc acorns} (Agglomerative Clustering for ORganising Nested Structures). The primary function of {\sc acorns} is to generate a hierarchical system of clusters within discrete data. Although {\sc acorns} was designed with the analysis of spectroscopic (position-position-velocity; PPV) data in mind, it can readily be implemented to other datasets, providing many applications.\footnote{{\sc acorns} is publicly available for download here: \url{https://github.com/jdhenshaw/acorns}.} The following section is dedicated to describing the methodology used by {\sc acorns}.

\subsection{ACORNS: Agglomerative Clustering for ORganising Nested Structures}\label{Methodology:acorns}

\subsubsection{Introduction and description of the input parameters}\label{Methodology:introduction}

{\sc acorns} follows the philosophy of hierarchical agglomerative clustering (HAC).\footnote{More information on this technique and its philosophy can be found in \citet{manning_2008}.} HAC methods fall into two main categories: `bottom-up' or `top-down'. {\sc acorns} follows the bottom-up approach in that each singleton data point begins its life as a `cluster'. Traditionally, clusters then merge until only a single cluster remains that contains all of the data. The output of this technique is often visualised graphically as a dendrogram, which have become popular in astronomy as a convenient way of representing and interpreting the hierarchical nature of molecular clouds (e.g. \citealp{houlahan_1992, rosolowski_2008}). 

Briefly, clustering in {\sc acorns} commences with the most significant data point. In the analysis presented in this work this refers to the data point with the greatest peak intensity. However, given the applicability of {\sc acorns} to different systems, this may instead refer to, for example, a density, column density, or mass. {\sc acorns} then descends in significance, merging clusters based on physically-motivated user-provided criteria, until a hierarchy is established. 

Input to {\sc acorns} is an array of $n\times m$ dimensions, where $n$ is the number of parameters, at minimum 4, but in principle has no upper limit and depends on how many parameters the user wishes to use during the clustering procedure. $m$ refers to the number of data points in the sample. As an example, in its simplest form (clustering in two spatial dimensions), this array should consist of: x position, y position, intensity (or equivalent), and an uncertainty on the intensity (or equivalent). If linking in PPV, an additional column for the velocity is a mandatory requirement. 

The linking of clusters is handled via the supply of an array containing $n-3$ elements (or $n-4$ if linking in PPP) which describes the clustering criteria (\texttt{cluster\_criteria}). Here, the user must supply the maximum spatial euclidean distance between data points, as well the maximum absolute difference in any other variable used for linking. If the separation between two data points satisfy these criteria, the data points are considered to be linked. We note however, that no two data points extracted from the same location can be linked to the same cluster. 

In addition the user must supply the following parameters:
\begin{enumerate}
\item The pixel size in equivalent units to the positional information in the input array (\texttt{pixel\_size}).
\item The radius of the smallest structures the user would like {\sc acorns} to identify (\texttt{min\_radius}). 
\item The minimum height above a merge level for a cluster to be considered as a separate structure (\texttt{min\_height}).
\item The stopping criteria, given as a multiple of the rms noise level (\texttt{stop}).
\end{enumerate}

In the following sections we provide a qualitative description of the {\sc acorns} algorithm.\footnote{Throughout this description we will use intensity as an example, however, in principle this can be exchanged for an equivalent parameter.} We begin with a description of the overall methodology before expanding on some of the individual steps.

\subsubsection{A description of the method}\label{Methodology:s1}

The main steps taken by {\sc acorns} in developing the hierarchy are as follows (these are also illustrated in the flow diagram presented in Fig.\,\ref{Figure:mainloop}):

\begin{enumerate}
\item {\sc acorns} begins by creating a catalogue of the currently unassigned data. All data whose intensity, $I$, satisfies the following criterion are added to this catalogue:
\begin{equation}
I > \texttt{stop}\times\sigma_{\rm rms} 
\end{equation}
where $\sigma_{\rm rms}$ refers to the noise level at that position. The unassigned data is then rearranged in descending order of $I$.\label{step:i}
\item These data are used to generate a k-d tree,\footnote{A data structure used to organise a number of points in a space with k-dimensions.} which can be queried to return the nearest neighbours to a given point.\label{step:ii}
\item Starting with the first data point in the unassigned catalogue, and looping over all data points in the unassigned catalogue, {\sc acorns} implements the following steps:\label{step:iii}
	\begin{enumerate}
	\item {\sc acorns} first generates a `bud cluster'. Extending the nomenclature of \citet{houlahan_1992}, a bud cluster refers to a structure which has not yet met the criteria to become a `leaf' in its own right (where leaves are the clusters situated at the top of the hierarchical system). \label{step:iiia}
	\item {\sc acorns} queries the k-d tree to find all data points which are within some maximum euclidean distance (provided in \texttt{cluster\_criteria}) from the bud cluster (see \S~\ref{Methodology:introduction}). If additional linking criteria are supplied by the user, {\sc acorns} then computes the maximum absolute difference in the desired property between the bud cluster and these data points. This is then also checked against the linking criteria supplied within \texttt{cluster\_criteria}. \label{step:iiib}
    \item All data satisfying the clustering criteria are then cross-referenced against the current cluster catalogue to see if they belong to an already established cluster within the hierarchy. If so, a link is established and the hierarchy grows (we will expand on this methodology in \S\,\ref{Methodology:s2}).\label{step:iiic}
	\end{enumerate}
\item Once {\sc acorns} has cycled through all data points in the unassigned catalogue, it begins a second loop. The cluster catalogue is first cleaned of any bud clusters and these data are used to generate a new unassigned catalogue. This step picks up any data points that were unable to be linked during the first pass of the algorithm.\label{step:iv}
\item If specified by the user (\texttt{relax}), the clustering criteria are relaxed and {\sc acorns} performs additional loops based on this new criteria. This helps further develop the hierarchy and this method is described in more detail in \S\,\ref{Methodology:s3}.\label{step:v}
\item {\sc acorns} then discards all remaining bud clusters since they did not meet the criteria to become fully-fledged clusters.
\end{enumerate}

\input{Figures/Figure_B1_flowchart_acorns.tex}
\input{Figures/Figure_B2_grow_acorns.tex}

{\sc acorns} returns a system of clusters as its output. In a given hierarchy, the antecessor is the largest common ancestor of all clusters within that hierarchy (note that for a given dataset there may be multiple antecessors and each of them may or may not have descendant substructure). Expanding the nomenclature typically used in describing dendrograms (see e.g. \citealp{houlahan_1992}), an antecessor refers to a tree in a forest of clusters. Each tree may or may not exhibit substructure, referred to as branches and leaves. 

\subsubsection{The growth of the hierarchy}\label{Methodology:s2}

\input{Figures/Figure_B3_2D_clustering.tex}

The procedure employed by {\sc acorns} during the growth of the hierarchy is described in the flow chart in Fig.~\ref{Figure:linking}. This growth strategy is developed following the methods of {\sc astrodendro} (\url{www.dendrograms.org}) and {\sc quickclump} \citep{sidorin_2017}. However, key differences in the algorithms (namely working with discrete data, rather than uniformly spaced data cubes) necessitate important differences in the details of each step. After establishing a link between the bud cluster (see \S~\ref{Methodology:s1}) and already-established clusters in the hierarchy (see step~\ref{step:iiic} in \S~\ref{Methodology:s1}), the next step depends on the number of linked clusters:

\begin{enumerate}

\item If no linked clusters are identified, the bud cluster is added to the cluster catalogue as a new cluster.
\item If only a single cluster is identified as linked, the bud cluster is merged into this already established cluster. 
\item If multiple linked clusters are identified, further decision making is required (see `Resolve ambiguity' in Fig.~\ref{Figure:linking}). {\sc acorns} first determines how many of the linked clusters are `true' clusters (i.e. not bud clusters). Once this has been determined, what happens next depends on how many fully-fledged clusters our bud cluster is linked to: 
\begin{enumerate}
\item If none, then this tells us that all of the clusters linked to our bud must also be bud clusters. We merge our bud cluster into the first of the other buds.
\item If there is only a single linked cluster we merge the bud cluster into this already established cluster. 
\item If there are multiple linked clusters, we generate a branch between these fully-fledged clusters - a new level in the hierarchy.
\end{enumerate}
All remaining bud clusters (if any) are then merged with the same cluster as our original bud cluster, be it a bud (a), a fully-fledged cluster (b), or clusters (c).
\end{enumerate}

\subsubsection{Relaxing the linking constraints}\label{Methodology:s3}

This (optional) second phase of the algorithm can be of importance when working with discrete and irregularly spaced data such as the velocity information output following the decomposition of spectroscopic data (and not, for example, regularly-spaced velocity channels within a data cube). Conceptually, the idea behind this phase is to relax the linking constraints used during stage 1 (\S~\ref{Methodology:s1}), in order to further develop the hierarchy. This can be implemented in multiple ways. The user has the option to relax the constraints either in a single step or incrementally (either interactively or non-interactively).

{\sc acorns} first generates a new catalogue of data points which were not assigned to clusters during the first and second passes of the algorithm using the initial clustering criteria.\footnote{Incidentally, the `second pass' of the algorithm (see step~\ref{step:iv} in \S~\ref{Methodology:s1}) follows the exact strategy outlined in this section, however, the linking constraints are not relaxed. } As with the implementation discussed in \S~\ref{Methodology:s1}, {\sc acorns} starts with the most significant (in this example, that with the greatest peak intensity) data point in the unassigned catalogue. Steps~\ref{step:i}-\ref{step:iii} are implemented as in Fig.~\ref{Figure:mainloop}, but this time using the new relaxed criteria. The main differences during the relax phase relate to the steps labeled `Find linked clusters' and `Link' in Fig.~\ref{Figure:mainloop}, and are outlined as follows:

\begin{enumerate}
\item During this phase, {\sc acorns} attempts to link bud clusters (Fig.~\ref{Figure:mainloop}) to an already-established forest (see \S~\ref{Methodology:s1}). It is important to ensure that any links that are created are still strong despite having relaxed the linking constraints. 

Often the user wants to link data based on more than just positional (and intensity) information. Therefore if additional properties are considered when searching for linked clusters (e.g. the centroid velocity or velocity dispersion), {\sc acorns} checks these properties against those of the linked clusters. If the properties of the bud cluster lie $>3\sigma$ away from the mean of the linked cluster properties (where $\sigma$ refers to the standard deviation of that property), then these linked clusters are prevented from creating links. This ensures that even despite relaxing the linking constraints, only strong links are forged.

\item During the relax phase, a bud cluster may be linked to multiple trees within the forest and, in some cases, it may be linked to multiple clusters belonging to the same tree. {\sc acorns} first establishes whether or not it is possible to insert the bud cluster into the correct position in the hierarchy. This is governed by the peak intensity of the bud cluster, and the minimum and maximum intensity levels of each linked cluster. If the bud cluster cannot be inserted into the linked cluster, {\sc acorns} searches downwards in the hierarchical tree (if possible) to establish a link. If the bud cluster cannot be slotted in at the correct level in any established tree, these linked clusters are ignored. {\sc acorns} returns a single linked cluster \emph{per tree} to which the bud cluster will be linked.\label{step:2iii}
\item Step~\ref{step:iiic} (Fig.~\ref{Figure:linking}) is implemented as above, with a key difference during the branching procedure. If a branch is to be created, the bud cluster is firstly merged with the closest matching cluster out of all the available linked clusters. A new branch (between multiple trees) is then created at the base of the parent hierarchies. 
\end{enumerate}

\subsection{{\sc acorns}: Clustering in 2-D}\label{methodology:pp}

\input{Figures/Figure_B4_3D_acorns.tex}

In this section we demonstrate the application of {\sc acorns} to 2-D data. The top left-hand panel of Fig.~\ref{Figure:ppclustering} depicts a clumpy `filament' from which we wish to extract structural information. The filament was generated using the clustering examples in Python's scikit-learn package.\footnote{\url{http://scikit-learn.org/stable/modules/clustering.html}.} We first generate a 2-D set of data points distributed randomly within the confines of a semi-circle with finite width. We then convert the point density into an image by convolving the point density with a Gaussian kernel.

The top central panel of Fig.~\ref{Figure:ppclustering} shows a graphical representation of the hierarchical system identified by {\sc acorns}, known as a dendrogram. {\sc acorns} picks out a total of seven `leaves', which are situated at the top of the hierarchy and highlighted in cyan, all of which belong to a single `tree' (i.e. the `filament'). The top right-hand panel highlights this information on the filament image. The leaves are indicated by cyan contours and the filament appears in dark blue. In this particular instance we chose to search for clusters using only the distance between data points as linking criteria. Consequently this solution is identical to that found with {\sc astrodendro} using equivalent input parameters. However, in principle (i.e. if available), additional constraints could be added to the {\sc acorns} linking procedure, which would result in the solutions from the two algorithms diverging. As an example, if one also had a measurement of temperature at each position, that could also be included in the clustering procedure. 

The bottom panels compare this result with other structure finding algorithms commonly used in the literature, namely {\sc clumpfind} (left; \citealp{williams_1995}), {\sc fellwalker} (centre; \citealp{berry_2015}), and \emph{gaussclumps} (right; \citealp{stutzki_1990}). Each of these algorithms seeks to identify discretised islands of emission, breaking the map up into `clumps', whereas {\sc acorns} (also {\sc astrodendro}) searches for hierarchical information within data. 

\subsection{{\sc acorns}: Clustering in 3-D}\label{methodology:ppv}

A key difference between {\sc acorns} and the algorithms mentioned in \S~\ref{Methodology:introduction}, is that {\sc acorns} is designed to work on decomposed spectroscopic data rather than data cubes. Analysis with {\sc acorns} can therefore be performed in unison with such algorithms, complementing their results by providing a detailed description of the gas kinematics. 

This is illustrated in Fig.~\ref{Figure:ppvclustering}. Here we have generated two `filaments' (one of which is identical to that shown in Fig.~\ref{Figure:ppclustering}). The intensity field of both filaments is illustrated in the top-left panels of Fig.~\ref{Figure:ppvclustering}. In this example, we also impose a velocity field. The filaments have differing velocity gradients (also shown in the top-left panels) and a uniform velocity dispersion (not shown). The intensity distribution and velocity field of the filaments are designed in such a way that the filaments overlap in PPV-space.

The top panels of Fig.~\ref{Figure:ppvclustering} display the result of applying {\sc acorns} to this configuration. {\sc acorns} identifies two clusters in the decomposed data. Importantly, the hierarchy associated with the blue filament is \emph{identical} to that found in \S~\ref{methodology:pp}. A corresponding hierarchy is identified for the green filament. The top right-hand image displays a representative dendrogram of this hierarchical system. 

The bottom left-hand image in Fig.~\ref{Figure:ppvclustering} displays structures recovered by {\sc acorns} in PPV-space. At the base of the image we demonstrate how the two filaments overlap in projection and appear as a ring. In PPV-space, we plot the velocity centroids; the data that {\sc acorns} uses for clustering. As can be seen, despite the velocity dispersion of the filaments being large enough such that they overlap in PPV-space, the two clusters are distinguishable when focusing on their centroids. 

To illustrate the difference in approach between {\sc acorns} and {\sc astrodendro}, the bottom right-hand image of Fig.~\ref{Figure:ppvclustering} displays the result of running {\sc astrodendro} on the same data cube. The light-coloured semi-transparent feature is a volume rendering of the main structure identified by {\sc astrodendro} (i.e. the trunk of the hierarchy), and the darker shaded structures refer to the leaves. Herein lies the key difference between the algorithms. Because of the blending in PPV-space between the two filaments, {\sc astrodendro}, which classifies structure as independent isosurfaces, returns a singular doughnut-shaped structure. At no point in the hierarchy are the two input filaments returned by {\sc astrodendro}. 

Encouragingly, there is close correspondence between many of the leaves identified using both algorithms. There are some very slight differences owing to the differences in the input parameters, but these are small. The key difference occurs where blending in PPV-space is observed, for example leaves \#193 and \#256. The reason these are picked out by {\sc acorns} is because these two features are identifiable (albeit blended) as multiple velocity components in the spectra, and are extracted as such during the spectral decomposition.

Consequently, the two methods complement each other nicely. While tools such as {\sc astrodendro} and {\sc scimes} (which uses {\sc astrodendro} for structure identification) are able to pick out molecular clouds as isosurfaces in spectroscopic data, {\sc acorns} can be used to search for regions of statistical similarity \emph{within} such clouds.

%% file: Figures/Figure_B1_flowchart_acorns.tex
\begin{figure}
\begin{center}
\includegraphics[trim = 95mm 60mm 85mm 50mm, clip, width = 0.48\textwidth]{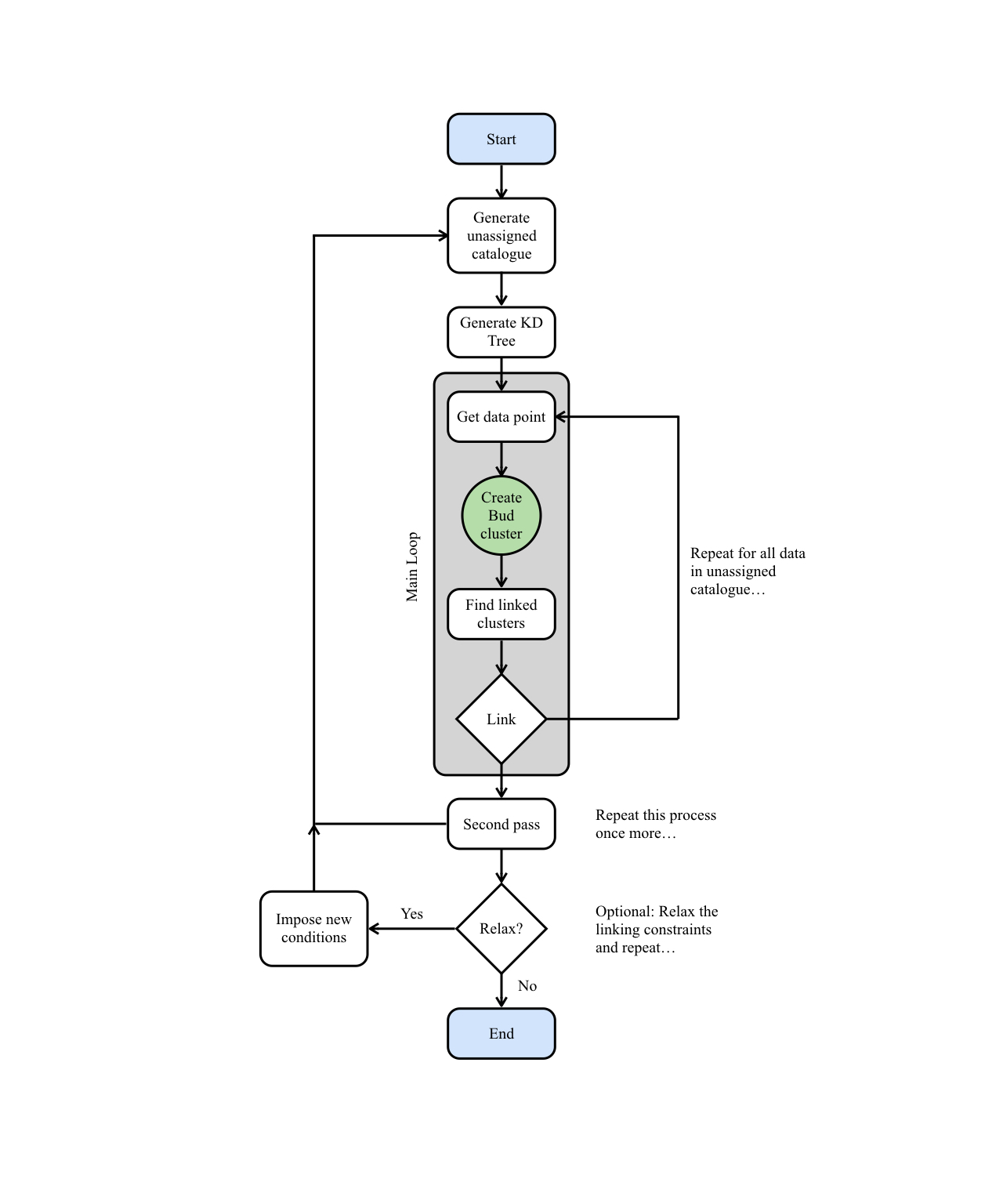}
\end{center}
\caption{A flow chart depicting the steps followed by {\sc acorns} during the clustering procedure. The main loop is indicated in dark grey. The creation of new clusters appears in green. The procedure employed during the `Link' phase is described in \S~\ref{Methodology:s2} and Fig.~\ref{Figure:linking}. }
\label{Figure:mainloop}
\end{figure}

%% file: Figures/Figure_B2_grow_acorns.tex
\begin{figure}
\begin{center}
\includegraphics[trim = 65mm 25mm 70mm 40mm, clip, width = 0.48\textwidth]{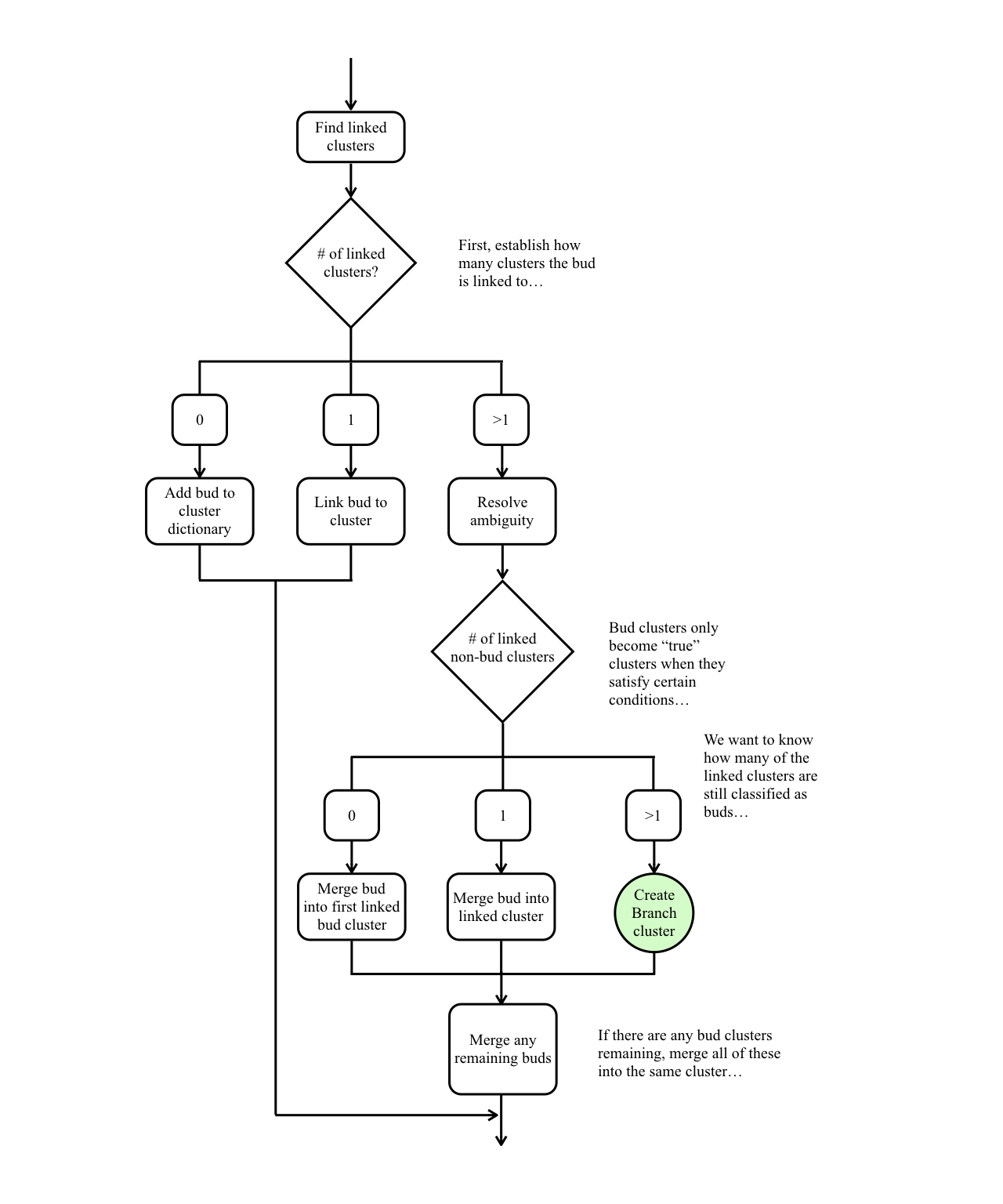}
\end{center}
\caption{A flow chart describing the growth and merging of clusters. This strategy follows the methods of {\sc astrodendro} (\url{www.dendrograms.org)} and {\sc quickclump} \citep{sidorin_2017}, see \S~\ref{Methodology:s2} for more details. }
\label{Figure:linking}
\end{figure}

%% file: Figures/Figure_B3_2D_clustering.tex
\begin{figure*}
\begin{center}
\includegraphics[trim = 0mm 0mm 0mm 0mm, clip, width = 0.33\textwidth]{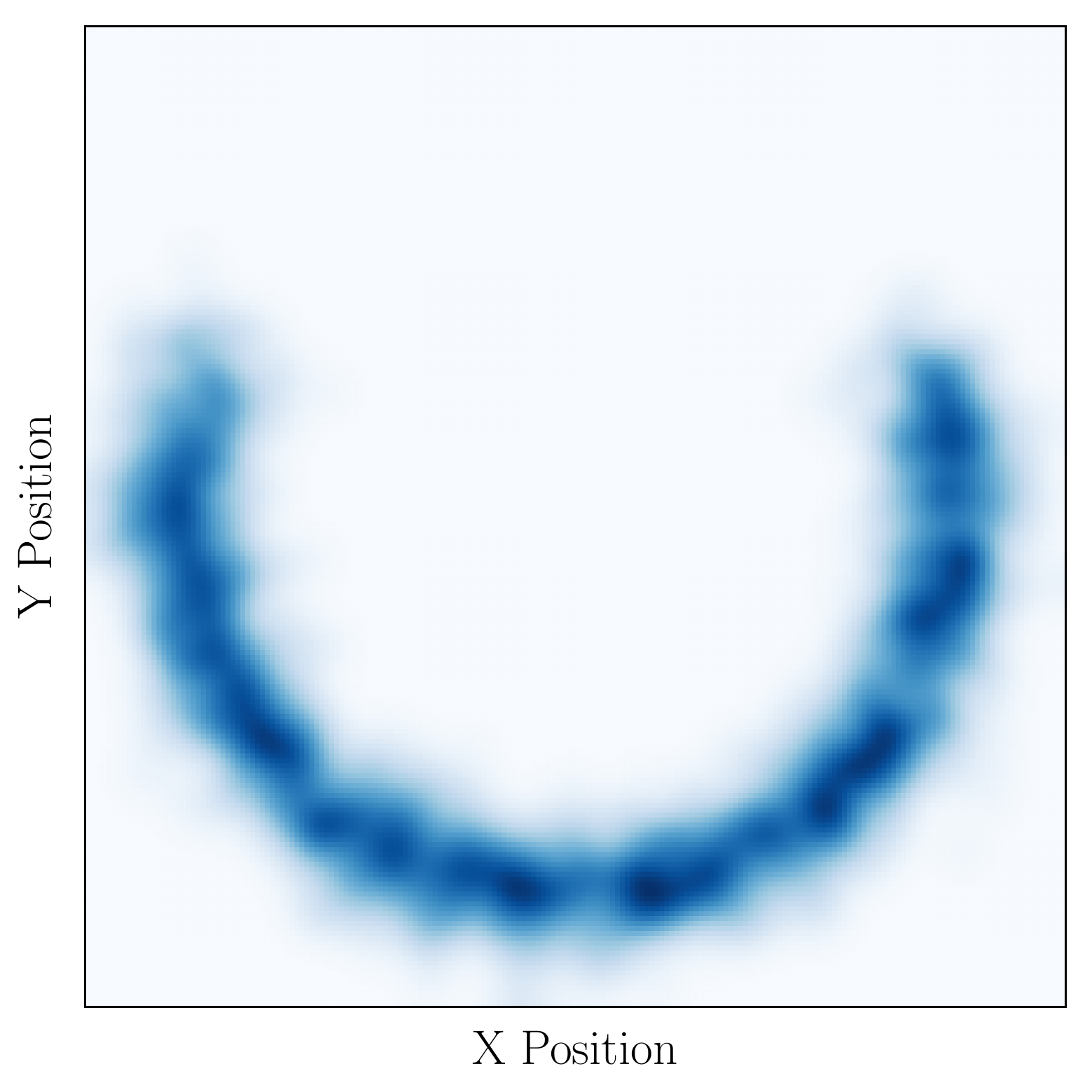}
\includegraphics[trim = 0mm 0mm 0mm 2mm, clip, width = 0.33\textwidth]{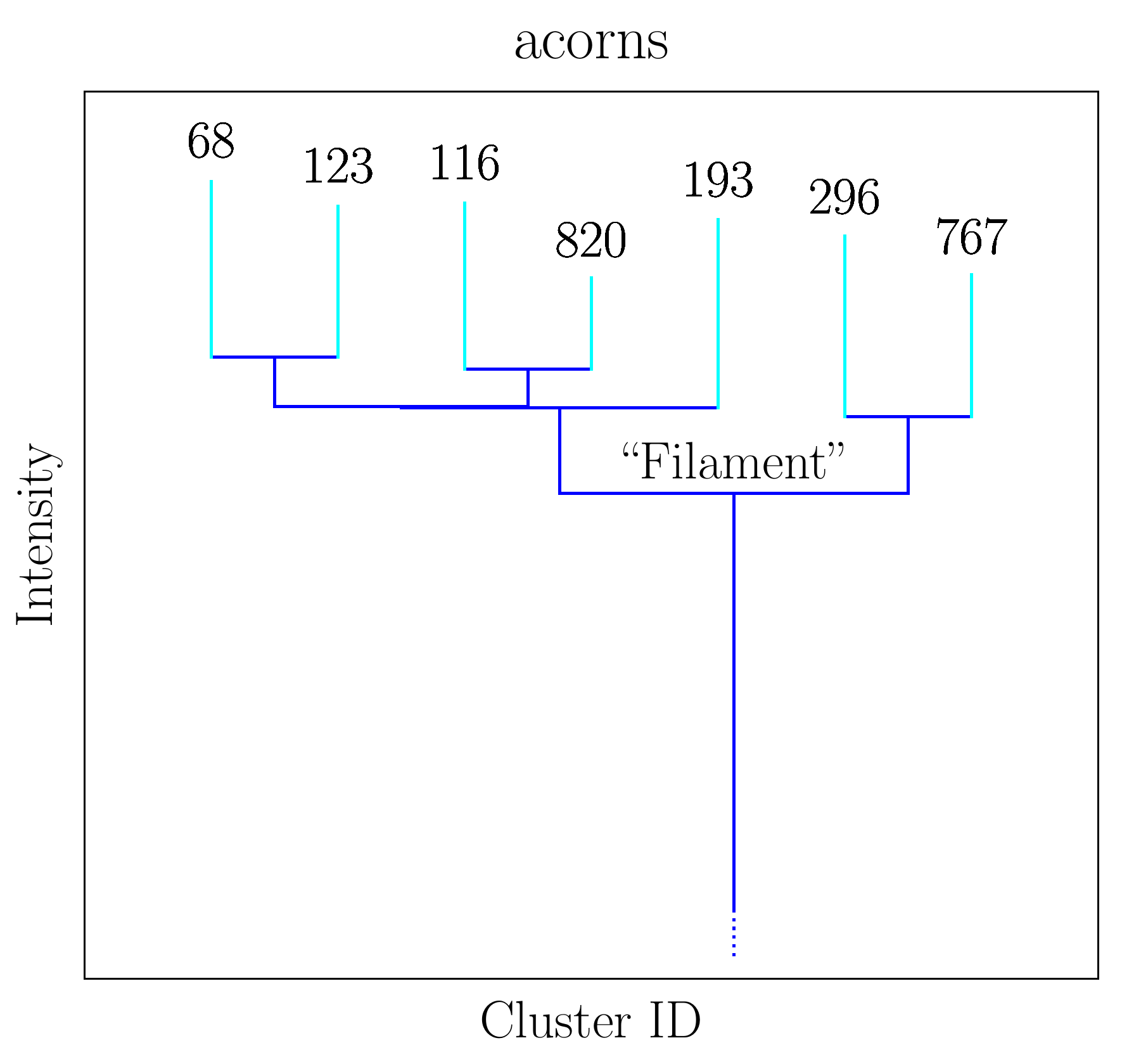}
\includegraphics[trim = 0mm 0mm 0mm 0mm, clip, width = 0.33\textwidth]{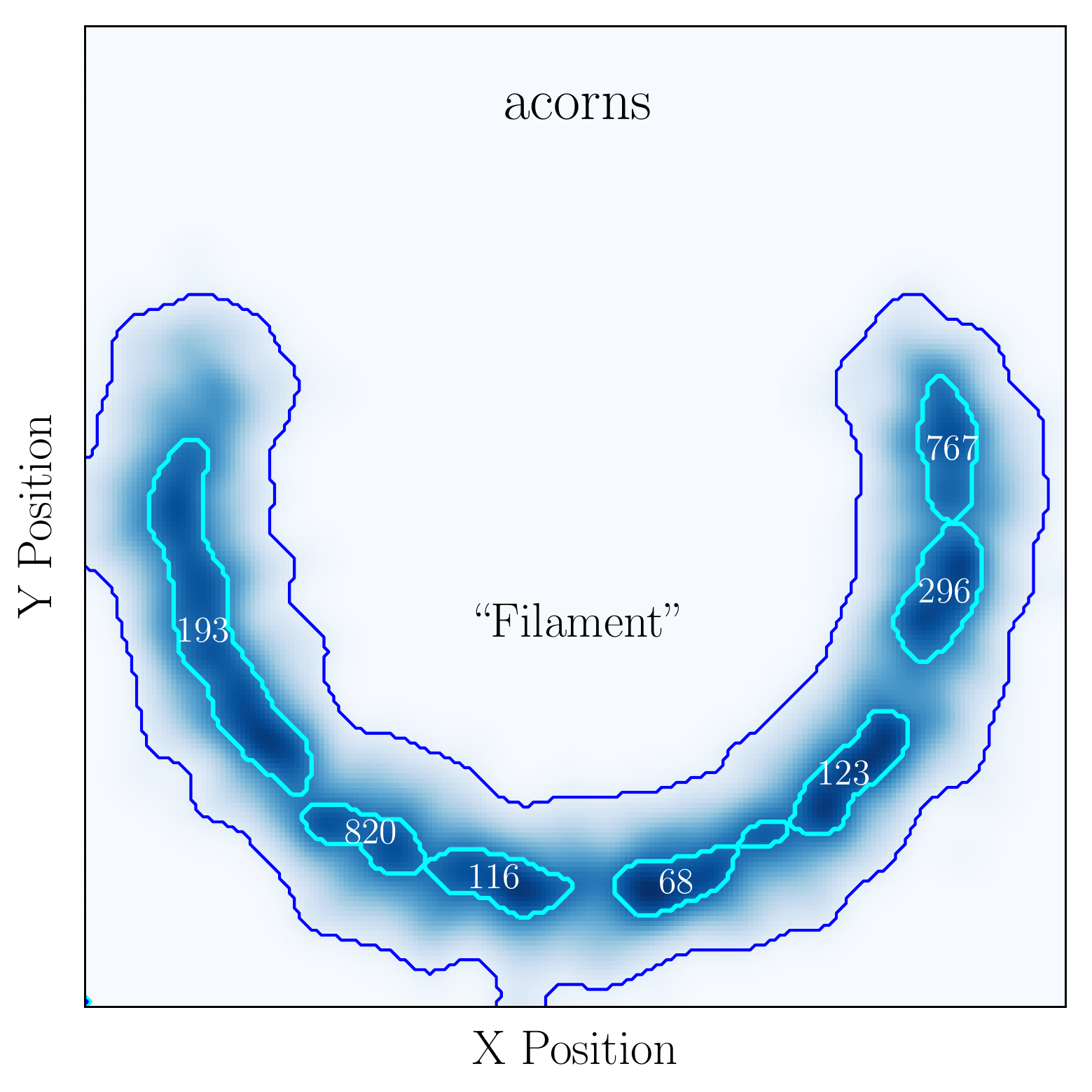}
\includegraphics[trim = 0mm 0mm 0mm 0mm, clip, width = 0.33\textwidth]{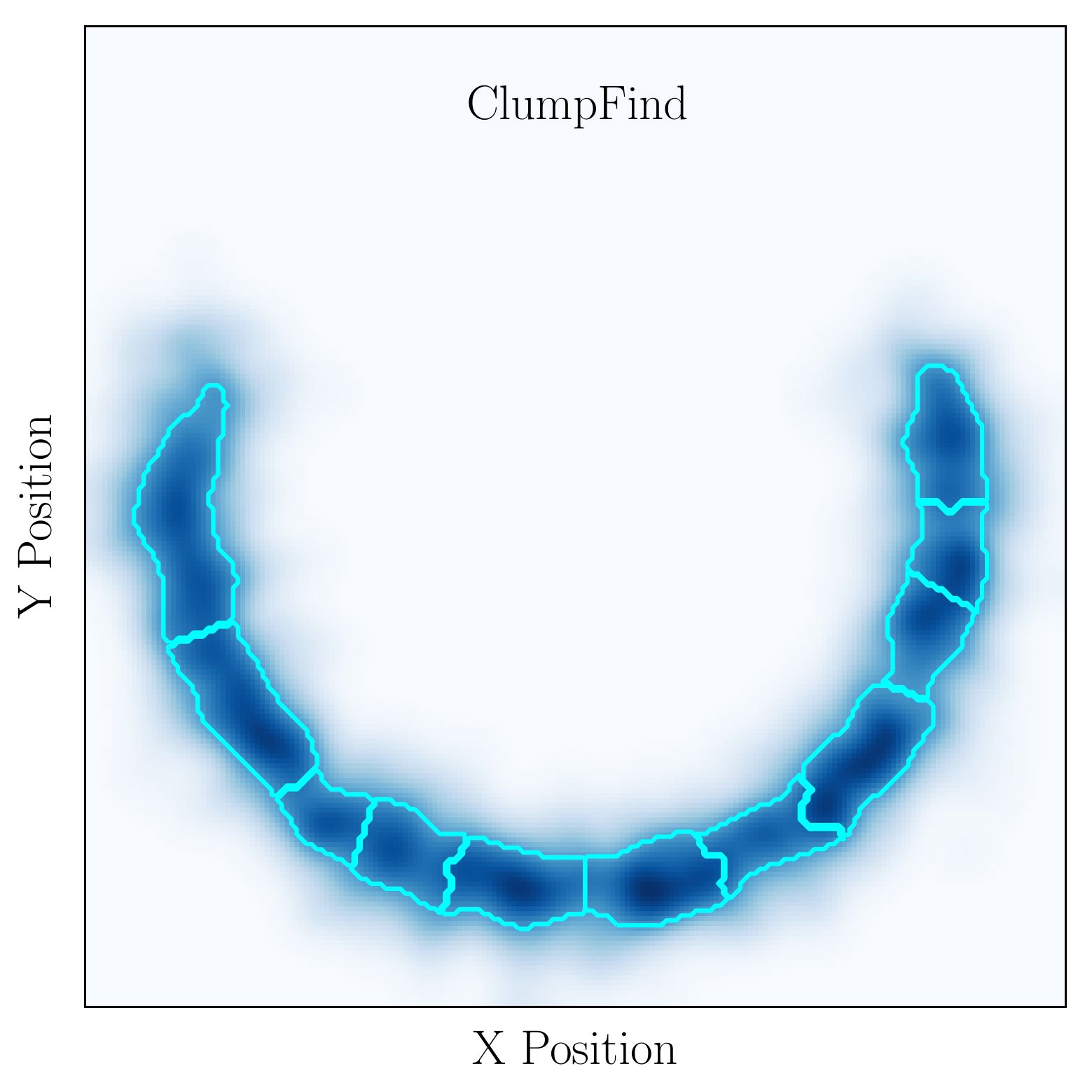}
\includegraphics[trim = 0mm 0mm 0mm 0mm, clip, width = 0.33\textwidth]{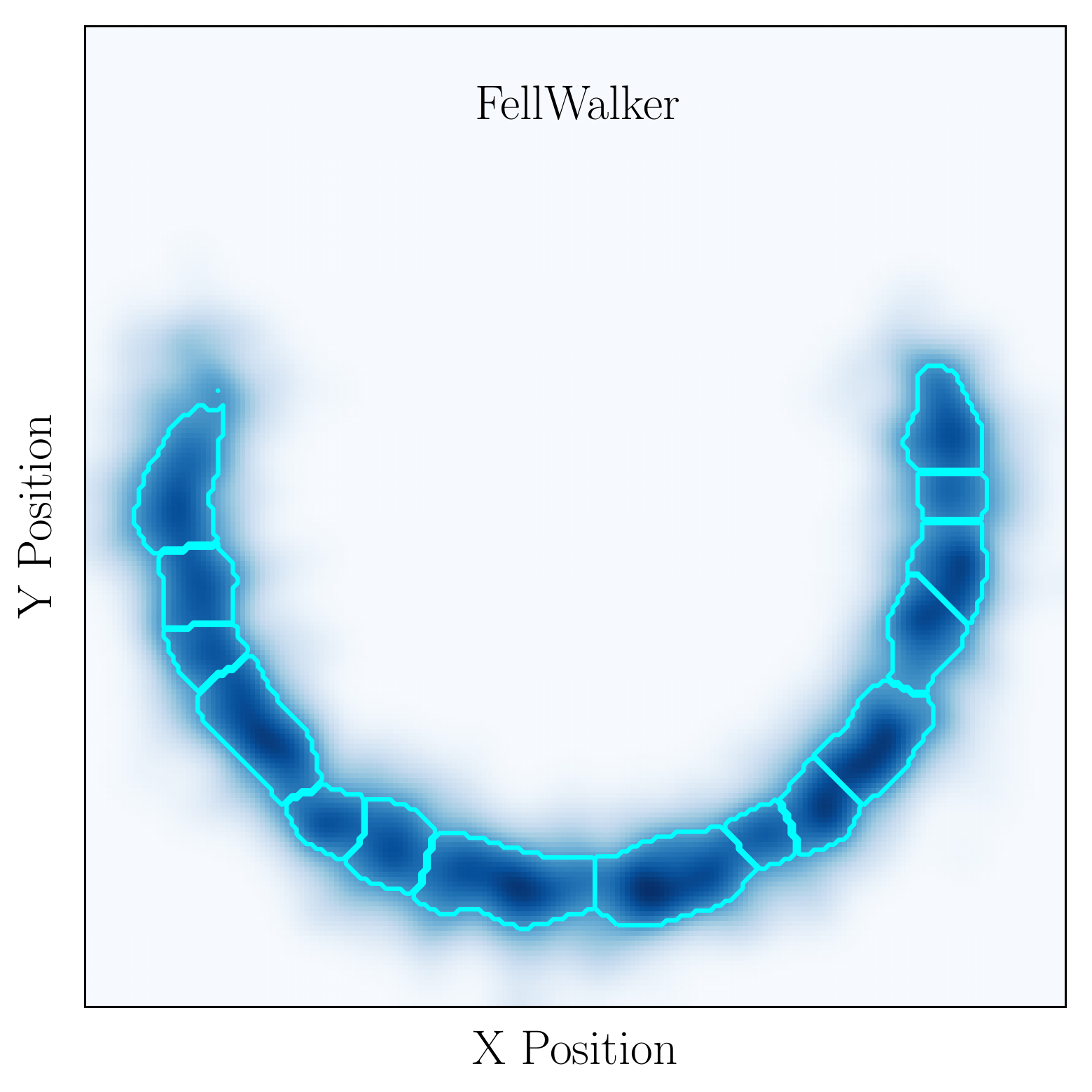}
\includegraphics[trim = 0mm 0mm 0mm 0mm, clip, width = 0.33\textwidth]{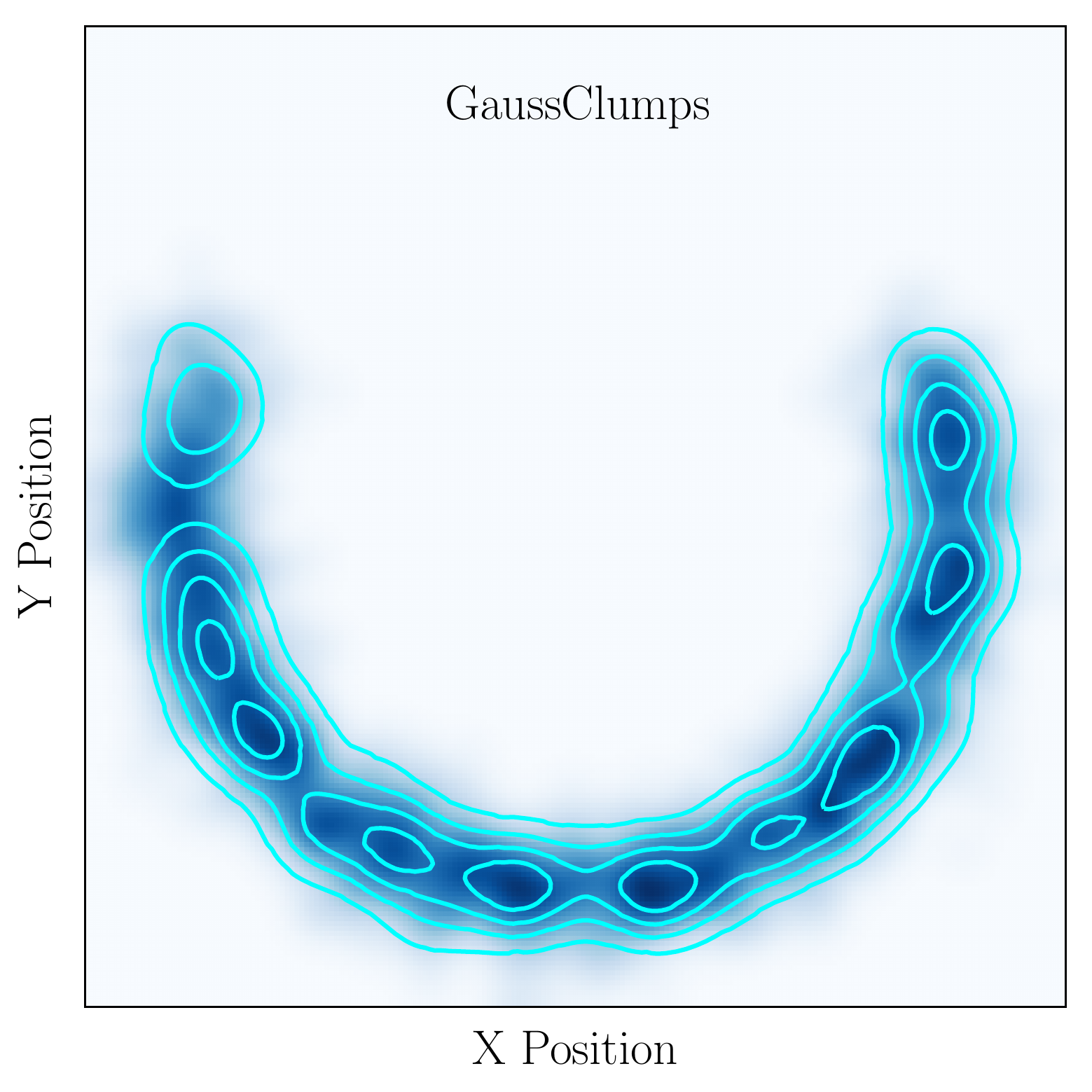}

\end{center}
\caption{Top panels: Demonstration of {\sc acorns} on 2D data. Top left panel: A fake clumpy `filament'. The colour-scale is a proxy for intensity. Top central panel: A graphical representation of the hierarchy found by {\sc acorns}. Top right panel: {\sc acorns} clusters displayed as contours. In the centre and right-hand panels, the leaves, i.e. the clusters situated at the top of the hierarchy, are displayed in cyan. The tree, corresponding to the `filament' is displayed as a dark blue contour. Bottom panels: Structure finding methods from the literature for comparison. Bottom left panel: {\sc clumpfind} \citep{williams_1995}. Bottom centre panel: {\sc fellwalker} \citep{berry_2015}. Bottom right panel: {\sc gaussclumps} \citep{stutzki_1990}. A key difference between {\sc acorns} (also {\sc astrodendro}) and the algorithms presented in the bottom panels is that the latter methods search for discretized islands of emission, whereas the former methods extract hierarchical structural information.}
\label{Figure:ppclustering}
\end{figure*}

%% file: Figures/Figure_B4_3D_acorns.tex
\begin{figure*}
\begin{center}
\includegraphics[trim = 10mm 20mm 10mm 0mm, clip, width = 0.48\textwidth]{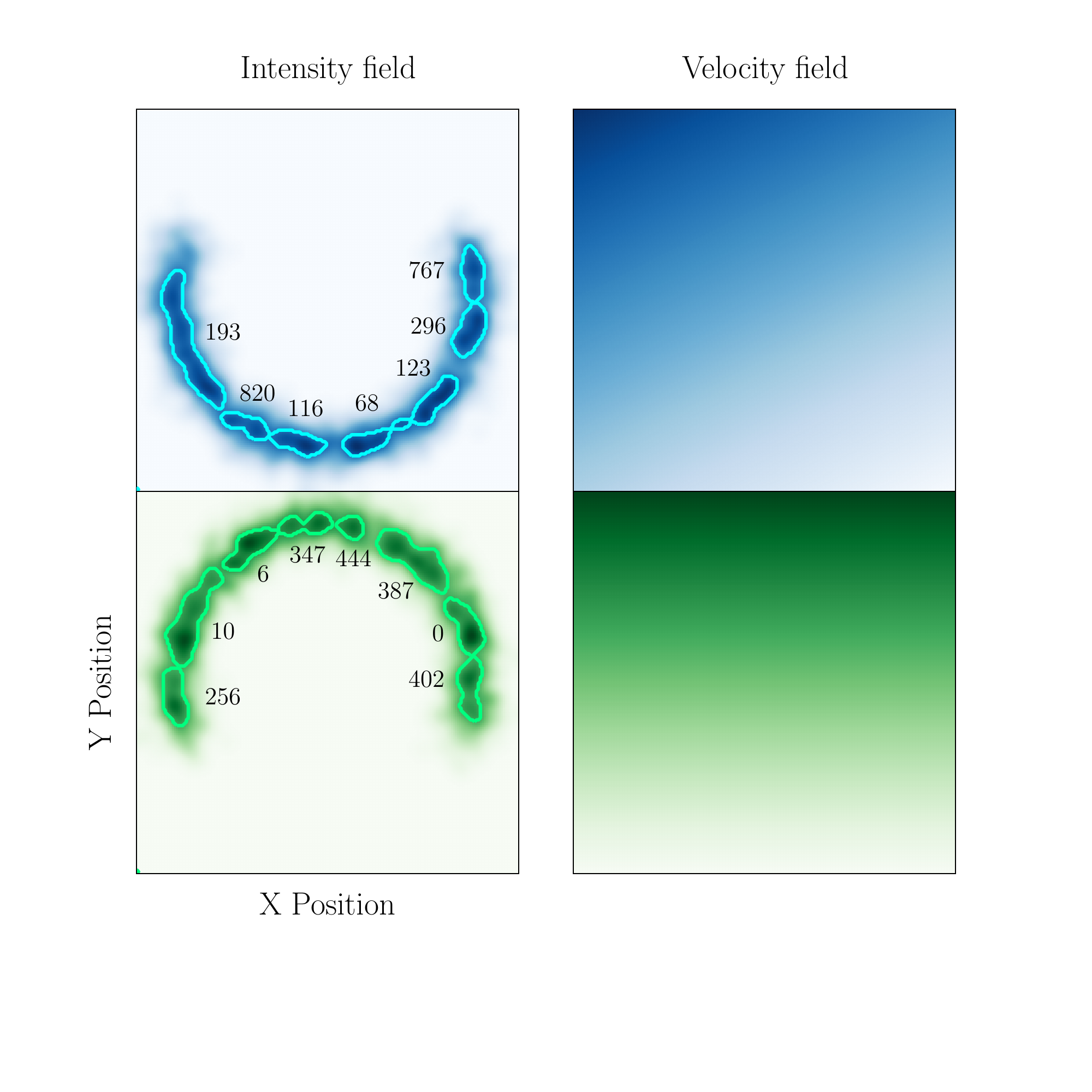}
\includegraphics[trim = 10mm 20mm 10mm 0mm, clip, width = 0.48\textwidth]{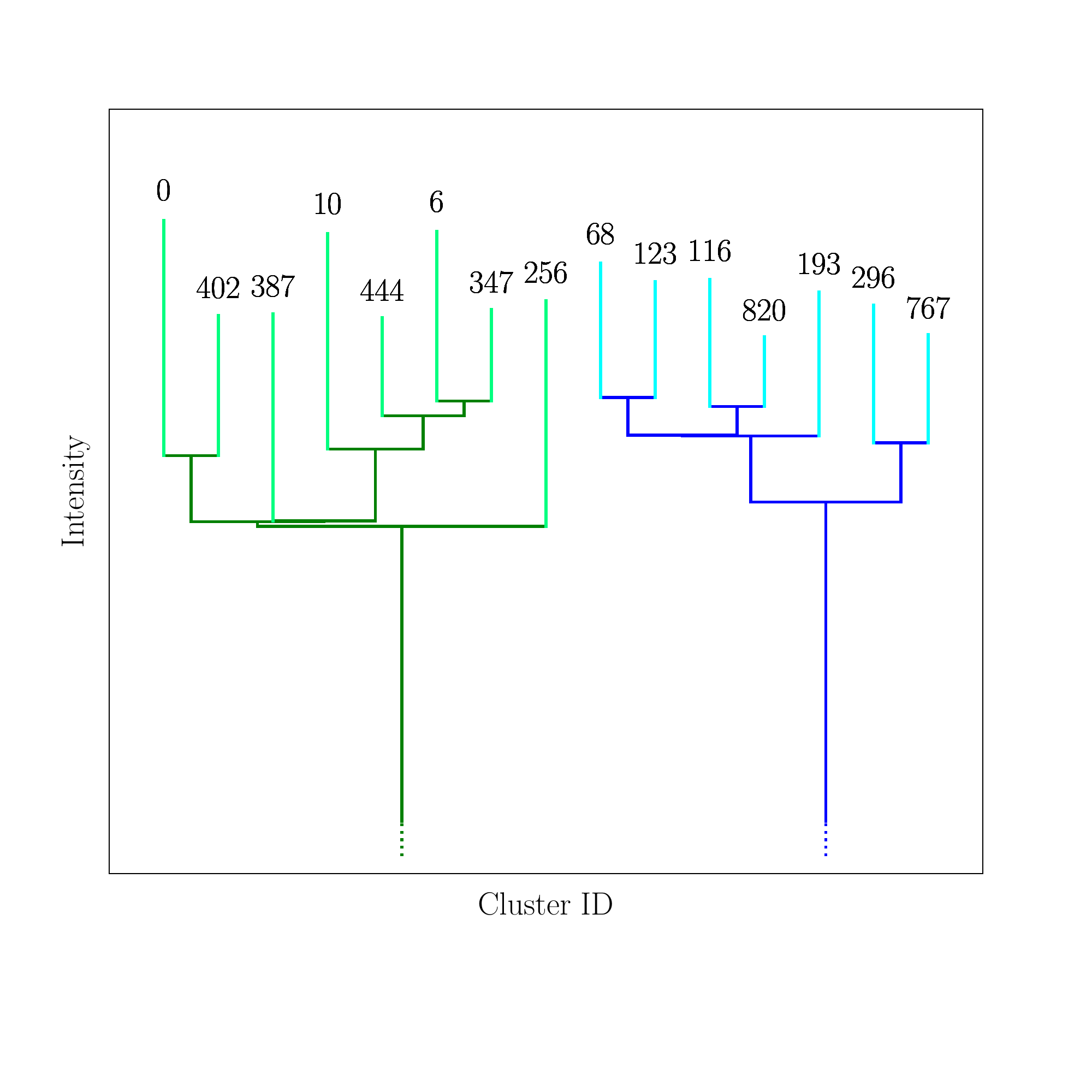}
\includegraphics[trim = 10mm 20mm 10mm 10mm, clip, width = 0.48\textwidth]{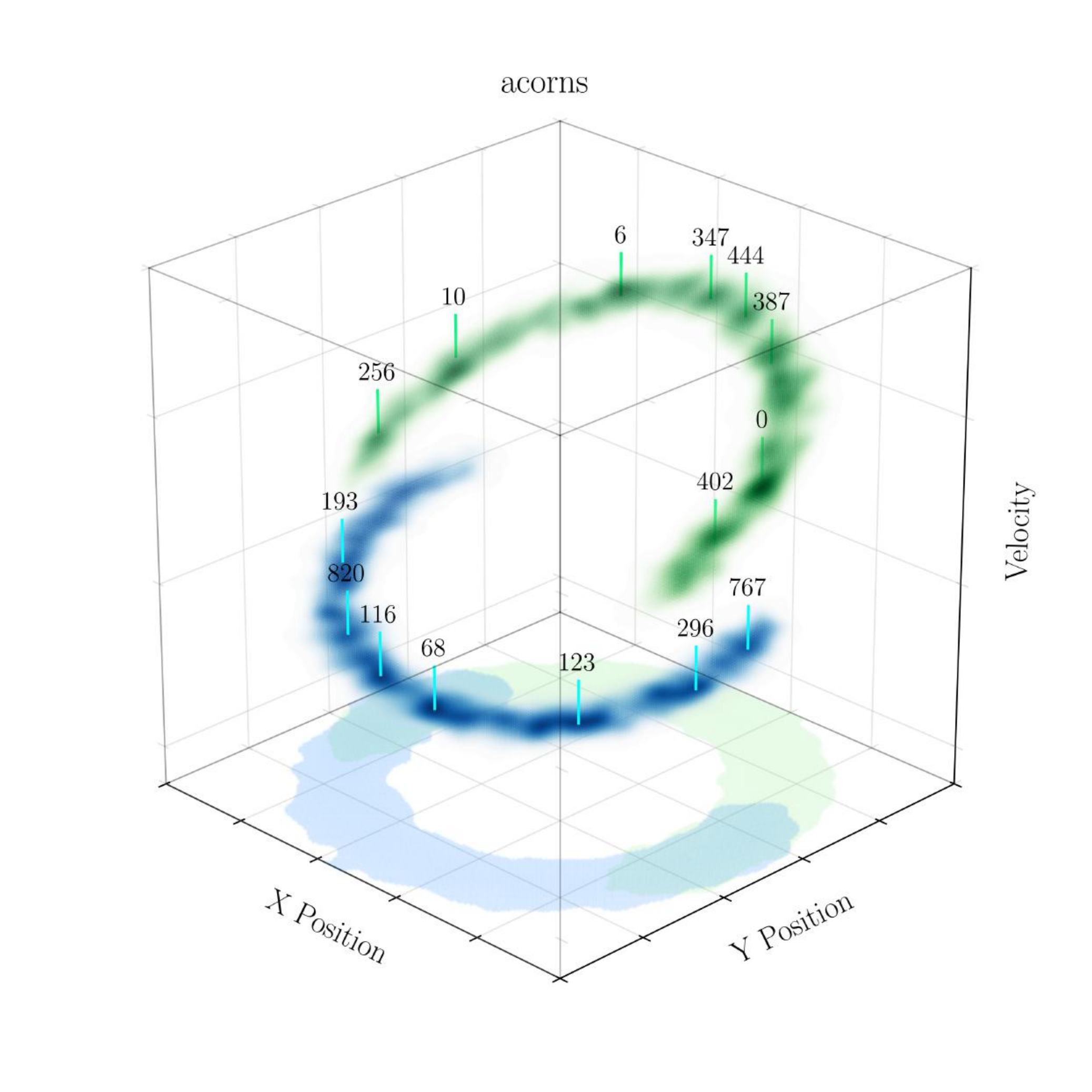}
\includegraphics[trim = 10mm 20mm 10mm 10mm, clip, width = 0.48\textwidth]{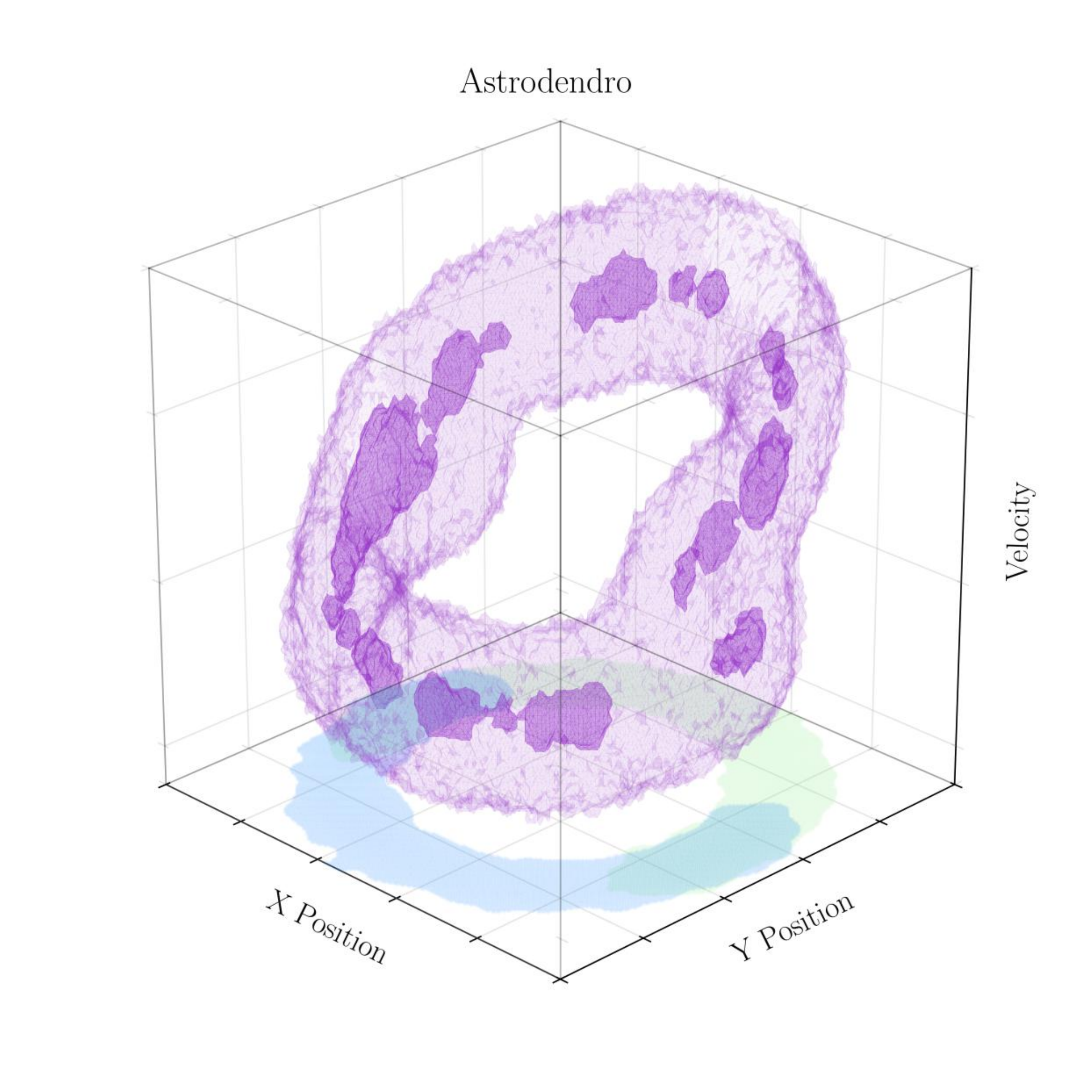}
\end{center}
\caption{A figure demonstrating the use of {\sc acorns} in position-position-velocity space. Top-left panels: two clumpy `filaments'. The colour-scale in the left-hand plots is a proxy for intensity. The right-hand plots show the corresponding velocity fields of the filaments (velocity increases from light to dark). The contours overlaid on the intensity-field highlight the leaves identified by {\sc acorns}. Top-right panel: The corresponding dendrogram. Bottom-left panel: The PPV-structure of the clusters. {\sc acorns} identified leaves are marked. Bottom-right panel: Demonstrating the use of {\sc astrodendro} on the same data. The semi-transparent purple rendering highlights the tree identified by {\sc astrodendro}. {\sc astrodendro} finds a single isosurface due to the merging of the filaments in PPV space. At no point in the {\sc astrodendro} hierarchy are the same two filamentary structures identified, in contrast to the {\sc acorns} solution. The dark purple structures highlight the leaves. Encouragingly, there is close correspondence between many of the leaves identified by both algorithms. However, a crucial difference is that leaves \#193 and \#256, identified with {\sc acorns}, merge in PPV-space and therefore are also merged in the {\sc astrodendro} solution. }
\label{Figure:ppvclustering}
\end{figure*}